\documentclass[10pt,twoside]{amundi_article}

\usepackage[table]{xcolor}
\usepackage[latin9]{inputenc}
\usepackage{amssymb}
\usepackage{amsfonts}
\usepackage{amsmath}
\usepackage{fancybox}
\usepackage{fancyhdr}
\usepackage{graphicx}
\usepackage[hyphens]{url}
\usepackage{arydshln}
\usepackage{multirow}
\usepackage{pdflscape}
\usepackage{tikz}
\usepackage{comment}
\usepackage{nicefrac}
\usepackage{dsfont}

\newlength{\figurewidth}
\newlength{\figureheight}
\def\tableskip{\vskip 10pt plus 2pt minus 2pt\relax}
\def\figureskip{\vskip 10pt plus 2pt minus 2pt\relax}

\newtheorem{remark}{Remark}
\def\limfunc#1{\mathop{\rm #1}}
\def\func#1{\mathop{\rm #1}}

\newcommand{\TsIII}{\hspace{3pt}}
\newcommand{\TsV}{\hspace{5pt}}
\newcommand{\TsVIII}{\hspace{8pt}}
\newcommand{\TsX}{\hspace{10pt}}
\newcommand{\TsXIII}{\hspace{13pt}}

\begin{document}

\setcounter{page}{1}

\title{\textbf{\color{amundi_blue}Robust Asset Allocation for Robo-Advisors}%
\footnote{The authors are very grateful to Silvia Bocchiotti, Arnaud
Gamain, Patrick Herfroy, Matthieu Keip, Didier Maillard, Hassan Malongo, Binh Phung-Que,
Christophe Romero and Takaya Sekine for their helpful comments.}}
\author{
{\color{amundi_dark_blue} Thibault Bourgeron} \\
Quantitative Research \\
Amundi Asset Management, Paris \\
\texttt{thibault.bourgeron@amundi.com} \and
{\color{amundi_dark_blue} Edmond Lezmi} \\
Quantitative Research \\
Amundi Asset Management, Paris \\
\texttt{edmond.lezmi@amundi.com} \and
{\color{amundi_dark_blue} Thierry Roncalli} \\
Quantitative Research \\
Amundi Asset Management, Paris \\
\texttt{thierry.roncalli@amundi.com}}

\date{\color{amundi_dark_blue}September 2018}

\maketitle

\begin{abstract}
In the last few years, the financial advisory industry has been
impacted by the emergence of digitalization and robo-advisors. This
phenomenon affects major financial services, including wealth
management, employee savings plans, asset managers, private banks,
pension funds, banking services, etc. Since the robo-advisory model
is in its early stages, we estimate that robo-advisors will
help to manage around \$1 trillion of assets in 2020 (OECD, 2017).
And this trend is not going to stop with future generations, who
will live in a technology-driven and social media-based
world.

In the investment industry, robo-advisors face different challenges:
client profiling, customization, asset pooling, liability
constraints, etc. In its primary sense, robo-advisory is a term for
defining automated portfolio management. This includes automated
trading and rebalancing, but also automated portfolio allocation.
And this last issue is certainly the most important challenge for
robo-advisory over the next five years. Today, in many robo-advisors,
asset allocation is rather human-based and very far from being
computer-based. The reason is that portfolio optimization is a very
difficult task, and can lead to optimized mathematical solutions
that are not optimal from a financial point of view (Michaud, 1989).
The big challenge for robo-advisors is therefore to be able to optimize
and rebalance hundreds of optimal portfolios without human
intervention.

In this paper, we show that the mean-variance optimization approach
is mainly driven by arbitrage factors that are related to the
concept of hedging portfolios. This is why regularization and
sparsity are necessary to define robust asset allocation. However,
this mathematical framework is more complex and requires
understanding how norm penalties impacts portfolio optimization. From a
numerical point of view, it also requires the implementation of
non-traditional algorithms based on ADMM methods and proximal
operators.
\end{abstract}

\noindent \textbf{Keywords:} Robo-advisor, asset allocation, active
management, portfolio optimization, Black-Litterman model, spectral
filtering, machine learning, Tikhonov regularization, mixed penalty,
ridge regression, lasso method, sparsity, ADMM algorithm, proximal
operator.\medskip

\noindent \textbf{JEL classification:} C61, C63, G11.

\clearpage

\section{Introduction}

The concept of portfolio optimization has a long history and dates back to the
seminal work of Markowitz (1952). In this paper, Markowitz defined precisely
what \textit{portfolio selection} means: \textquotedblleft \textsl{the investor does
(or should) consider expected return a desirable thing and variance of return
an undesirable thing}\textquotedblright. This was the starting point of
mean-variance optimization and portfolio allocation based on quantitative
models. In particular, the Markowitz approach became the standard model for
strategic asset allocation until the end of the 2000s.\smallskip

Since the financial crisis of 2008, another model has emerged and is now a very
serious contender for asset allocation (Roncalli, 2013). The risk budgeting
approach is successfully used for managing multi-asset portfolios, equity risk
factors or alternative risk premia. The main difference with mean-variance
optimization is the objective function. The Markowitz approach mainly focuses
on expected returns and exploits the trade-off between performance and
volatility. The risk budgeting approach is based on the risk allocation of the
portfolio, and does not take into account expected returns of assets.\smallskip

The advantage of the risk budgeting approach is that it produces stable and
robust portfolios. On the contrary, mean-variance optimization is very
sensitive to input parameters. These stability issues make the practice of
portfolio optimization less attractive than the theory (Michaud, 1989). Even
for strategic asset allocation, many weight constraints need to be introduced in
order to regularize the mathematical solution and obtain an acceptable
financial solution. In the case of tactical asset allocation, professionals
generally prefer to implement the model of Black and Litterman (1991, 1992),
because the optimized portfolio depends on the current allocation. Therefore,
the Black-Litterman model appears to be slightly more robust than the Markowitz
model because having a benchmark or introducing a tracking error constraint is
already a form of portfolio regularization. However, since the Black-Litterman
model is a slight modification of the Markowitz model, it suffers from the same
drawbacks.\smallskip

Since the 1990s, academics have explored how to robustify portfolio
optimization in two different directions. The first one deals with the
estimation of the input parameters. For instance, we can use de-noising methods
(Laloux \textsl{et al.}, 1999) or shrinkage approaches (Ledoit and Wolf, 2004)
to reduce estimation errors of the covariance matrix. The second one
deals with the objective function. As explained by Roncalli (2013), the
Markowitz model is an aggressive model of active management due to the
mean-variance objective function. Academics have suggested regularizing
the optimization problem by adding penalization functions. For instance, it is
common to include a $L_1$ or $L_2$ norm loss function. The advantage of this is to
obtain a \textquotedblleft sparser\textquotedblright\ or \textquotedblleft
smoother\textquotedblright\ solution.\smallskip

The success of risk parity, equal risk contribution (ERC) and risk budgeting
portfolios has put these new developments in second place. However, the rise of
robo-advisors is changing the current trend and highlights the need for active
allocation models that are focused on expected returns. Indeed, the challenge
of robo-advice concerns tactical asset allocation and not the portfolio
construction of strategic asset allocation. Building a defensive, balance or
dynamic portfolio profile is not an issue, because they are defined from an
ex-ante point of view. Quantitative models can be used to define this step, but
they are not necessarily required. For example, this step can also be done
using a discretionary approach, since portfolio profiles are revised once and
for all. The difficulty lies with the life of the invested portfolio and the
dynamic allocation. A robo-advisor that would consist in rebalancing a
constant-mix allocation is not a true robo-advisor, since it is reduced to the
profiling of clients. The main advantage of robo-advisors is to perform dynamic
allocation by including investment views, side assets or the client's dynamic constraints,
or some alpha engines provided by the robo-advisor's manager or distributor.\smallskip

The challenge for a robo-advisor is therefore to perform dynamic allocation or
tactical asset allocation in a systematic way without human
interventions. In this case, expected returns or trading signals
must be taken into account. One idea is to consider an extension
of the ERC portfolio by using a risk measure that depends on
expected returns (Roncalli, 2015). However, this approach is not
always suitable when we target a high tracking error. Otherwise, it makes a lot of sense
for the mean-variance optimization to be the allocation
engine of robo-advisors. As said previously, the challenge is
to develop a robust asset allocation model. The purpose of this
research is to provide a practical solution that does not require
human interventions.\smallskip

This paper is organized as follows. Section Two illustrates the
practice of mean-variance optimization and highlights the limits of
such models. In Section Three, we apply the theory of regularization
to asset allocation. In particular, we point out the calibration
procedure of the Lagrange coefficients of norm functions. In Section
Four, we consider application to robo-advisory. Finally, Section
Five offers some concluding remarks.

\section{Practice and limits of mean-variance optimization}

\subsection{The mean-variance optimization framework}

We follow the presentation of Roncalli (2013). We consider a universe of $n$
assets. Let $x=\left( x_{1},\ldots ,x_{n}\right) $ be the vector of weights in
the portfolio. We denote by $\mu $ and $\Sigma $ the vector of expected returns
and the covariance matrix of asset returns. It follows that the expected return
and the volatility of the portfolio are equal to $\mu \left( x\right) =x^{\top
}\mu $ and $\sigma \left( x\right) =\sqrt{x^{\top }\Sigma x}$. The Markowitz
approach consists in maximizing the expected return of the portfolio under a
volatility constraint ($\sigma $-problem):
\begin{equation}
x^{\star }=\arg \max \mu \left( x\right) \text{ s.t. }\sigma \left( x\right)
\leqslant \sigma ^{\star }  \label{eq:sigma-problem}
\end{equation}
or minimizing the volatility of the portfolio under a return constraint ($\mu
$-problem):
\begin{equation}
x^{\star }=\arg \min \sigma \left( x\right) \text{ s.t. }\mu \left( x\right)
\geqslant \mu ^{\star }  \label{eq:mu-problem}
\end{equation}%
Replacing the volatility by the variance scaled with the factor
$\nicefrac{1}{2}$ does not change the solution. Therefore, we deduce that the
Lagrange functions associated with Problems (\ref{eq:sigma-problem}) and
(\ref{eq:mu-problem}) are:
\begin{equation*}
\mathcal{L}_{1}\left( x,\lambda _{1},\sigma ^{\star }\right) =x^{\top }\mu
-\lambda _{1}\left( \frac{1}{2}\sigma ^{2}\left( x\right) -\frac{1}{2}\sigma
^{\star ^{2}}\right)
\end{equation*}%
and:%
\begin{equation*}
\mathcal{L}_{2}\left( x,\lambda _{2},\mu ^{\star }\right) =\frac{1}{2}\sigma
^{2}\left( x\right) -\lambda _{2}\left( \mu \left( x\right) -\mu ^{\star
}\right)
\end{equation*}%
They satisfy $\mathcal{L}_{2}\left( x,\lambda _{2},0\right) =-\lambda _{2}%
\mathcal{L}_{1}\left( x,\theta ,0\right) $ where $\theta =\lambda _{2}^{-1}$ is
the risk aversion of the quadratic utility function. As strong duality holds,
these two problems are equivalent. Moreover, we can show that they can be
written as a standard quadratic programming problem (Markowitz, 1956):
\begin{equation}
x^{\star }\left( \gamma \right) =\arg \min \frac{1}{2}x^{\top }\Sigma
x-\gamma x^{\top }\mu \label{eq:qp1}
\end{equation}%
where $\gamma$ is the risk/return trade-off parameter. Since the problem is
strongly convex and the solution is $x^{\star }\left( \gamma \right) =\gamma
\Sigma ^{-1}\mu $, we deduce that the solution of the $\mu $-problem is given
by:
\begin{equation*}
\gamma =\frac{\mu ^{\star }}{\mu ^{\top }\Sigma ^{-1}\mu }
\end{equation*}%
whereas the solution of the $\sigma $-problem is obtained for the following value
of $\gamma $:
\begin{equation*}
\gamma =\frac{\sigma ^{\star }}{\sqrt{\mu ^{\top }\Sigma ^{-1}\mu }}
\end{equation*}
\smallskip

The previous framework can be extended by considering a risk-free asset and
portfolio constraints:%
\begin{eqnarray}
x^{\star }\left( \gamma \right)  &=&\arg \min \frac{1}{2}x^{\top }\Sigma
x-\gamma x^{\top }\left( \mu -r\mathbf{1}\right)   \label{eq:qp2} \\
&\text{s.t.}&x\in \Omega   \notag
\end{eqnarray}%
where $r$ is the risk-free rate and $\Omega $ is the set of restrictions. Let
$\mu ^{-}\leqslant \mu \left( x\right) \leqslant \mu ^{+}$ and $\sigma ^{-}\leqslant \sigma
\left( x\right) \leqslant \sigma ^{+}$ be the bounds of the expected return and the
volatility such that $x\in \Omega $. It follows that there is a solution to the
$\sigma $-problem and the $\mu $-problem if $\sigma ^{\star }\geqslant \sigma ^{-}$
and $\mu ^{\star }\leqslant \mu ^{+}$.

\begin{remark}
The Sharpe ratio is the standard risk/return measure used in finance, and
corresponds to the zero-homogeneous quantity:
\begin{eqnarray*}
\func{SR}\left(x \mid r\right) & = & \frac{\mu\left(x\right) - r}{\sigma\left(x\right)} \\
& = & \frac{x^{\top} \mu - r}{\sqrt{x^{\top} \Sigma x}}
\end{eqnarray*}
The capital asset pricing model (CAPM) defines the tangency portfolio as the
optimized portfolio that has the maximum Sharpe ratio. When the capital budget
is reached (meaning that $\sum_{i=1}^{n} x_i = 1$), the solution of Problem
(\ref{eq:qp2}) is equal to $x^{\star} = \gamma \Sigma^{-1} \left(\mu - r
\mathbf{1}\right)$ where $\gamma^{\star} = \left( \mathbf{1}^{\top} \Sigma^{-1}
\left(\mu - r \mathbf{1}\right) \right) ^{-1}$. Since the matrix $\Sigma$ has a
unique symmetric positive definite square root denoted by $\Sigma^{1/2}$, the
Cauchy-Schwarz inequality yields:
\begin{equation*}
\left(x^{\top} \left(\mu - r \mathbf{1}\right)\right)^2 =
\left(x^{\top} \Sigma^{1/2} \Sigma^{-1/2} \left(\mu - r \mathbf{1}\right)\right)^2
\leqslant \left( x^{\top} \Sigma x \right) \left( \left(\mu - r \mathbf{1}\right)^{\top} \Sigma^{-1}
\left(\mu - r \mathbf{1}\right) \right)
\end{equation*}
The equality holds if and only if there exists a scalar $\gamma \in \mathbb{R}$
such that $\Sigma^{1/2} x = \gamma \Sigma^{-1/2} \left(\mu - r
\mathbf{1}\right)$. It follows that:
\begin{equation}
\forall\, x \in \mathbb{R}^n \quad \func{SR}\left(x \mid r\right) \leqslant \sqrt{\left(\mu - r
\mathbf{1}\right)^{\top} \Sigma^{-1} \left(\mu - r
\mathbf{1}\right)} \label{eq:sharpe-bound}
\end{equation}
We deduce that the set of portfolios maximizing the Sharpe ratio is the
one-dimensional vector space defined by $x \in \Sigma^{-1} \left(\mu - r
\mathbf{1}\right)$. This means that unconstrained and constrained portfolio
optimizations are related when we impose only one simple constraint like the
capital budget restriction. In more complex cases, the constrained solution is
not necessarily related to the unconstrained solution. However, the bound
remains valid, because it only depends on the Cauchy-Schwarz inequality.
\end{remark}

The previous result highlights the importance of constraints in portfolio
optimization. A portfolio is long-only if $\forall\, i \in \{1,\ldots,n\}\ x_i
\geqslant 0$ whereas it is long-short if $\exists\, \left(i,j\right) \in
\{1,\ldots,n\}$ such that $x_i > 0$ and $x_j < 0$. For long-only portfolios, a
capital budget is usually assumed, meaning that the portfolio is fully invested
($\sum_{i=1}^{n} x_i = 1$). For long-short portfolios, professionals sometimes
impose a neutral or zero-capital budget, implying that the long exposure is
financed by the short exposure ($\sum_{i=1}^{n} x_i = 0$). They can also impose
leverage constraints ($\sum_{i=1}^{n} \left\vert x_i\right\vert \leqslant c$),
while risk-budgeting portfolios require adding a logarithmic barrier constraint
($\sum_{i=1}^{n} \omega_i \ln x_i \geqslant c$).\smallskip

In practice, the quantities $\mu $ and $\Sigma $ are unknown and must be
specified. We can assume that they are estimated using an historical sample
$\left\{ R_{1},\ldots ,R_{T}\right\} $ where $R_{t}$ is the vector of asset
returns at time $t$. Let $\hat{\mu}$ and $\hat{\Sigma}$ be the
corresponding estimators. We have:%
\begin{equation*}
\hat{\mu}=\sum_{t=1}^{T}w_{t}R_{t}
\end{equation*}%
and:%
\begin{equation*}
\hat{\Sigma}=\sum_{t=1}^{T}w_{t}\left( R_{t}-\hat{\mu}\right) \left( R_{t}-%
\hat{\mu}\right) ^{\top }
\end{equation*}%
where $w_{t}$ is the weighting scheme such that $\sum_{t=1}^{T}w_{t}=1$.
In Appendix \ref{appendix:matrix-form} on page \pageref{appendix:matrix-form},
we show that Problem (\ref{eq:qp2}) can be written as follows\footnote{%
The norm $\left\Vert x\right\Vert _{A}$ is equal to $\left( x^{\top
}Ax\right) ^{1/2}$. All the notations are defined in Appendix \ref{appendix:notations}
on page \pageref{appendix:notations}.}:%
\begin{eqnarray}
x^{\star }\left( \gamma \right)  &=&\arg \min \frac{1}{2}\left\Vert
Rx\right\Vert_W^{2}-\gamma x^{\top }\left( R^{\top }w-r\mathbf{%
1}\right)   \label{eq:qp3} \\
&\text{s.t.}&x\in \Omega   \notag
\end{eqnarray}%
where $w=\left( w_{1},\ldots ,w_{T}\right) \in \mathbb{R}^{T}$, $R=\left(
R_{1},\ldots ,R_{T}\right) \in \mathbb{R}^{T\times n}$ and $W =
\func{diag}\left( w\right) - ww^{\top }$. In this case, the Markowitz solution
is the portfolio that maximizes the backtest for a given volatility. When
$w_{t+1}\geqslant w_{t}$, we conclude that Problem (\ref{eq:qp2}) is a
trend-following optimization program, whose moving average is defined by the
weighting scheme $w$. In order not to be trend-following, we have to use a
vector of expected returns $\mu $ that does not satisfy $w_{t+1}\geqslant w_{t}$ or
that does not depend on the sample of asset returns.

\subsection{Stability issues}

According to Hadamard (1902), a well-posed problem must satisfy three
properties:
\begin{enumerate}
\item a solution exists;
\item the solution is unique;
\item the solution's behavior changes continuously with the initial
    conditions.
\end{enumerate}
We recall that the solution to Problem (\ref{eq:qp1}) is $x^{\star }\left(
\gamma \right) =\gamma \Sigma ^{-1}\mu $. If $\Sigma $ has no zero eigenvalues,
it follows that the existence and uniqueness is ensured, but not necessarily
the stability. Indeed, this third property implies that $\Sigma $ has no
\textquotedblleft small\textquotedblright\ eigenvalues. This problem is
extensively illustrated by Bruder \textsl{et al.} (2013) and Roncalli (2013).
If we consider the eigendecomposition $\Sigma =V\Lambda V^{\top }$, we have
$\Sigma ^{-1}=V\Lambda ^{-1}V^{\top }$ and $x^{\star }\left( \gamma \right)
=\gamma V\Lambda ^{-1}V^{\top }\mu $. It follows that $V^{\top }x^{\star
}\left( \gamma \right) =\gamma \Lambda ^{-1}V^{\top }\mu $ or:
\begin{equation}
\tilde{x}^{\star }\propto \Lambda ^{-1}\tilde{\mu} \label{eq:eig1}
\end{equation}%
where $\tilde{x}^{\star }=V^{\top }x^{\star }\left( \gamma \right) $ and $%
\tilde{\mu}=V^{\top }\mu $. By applying the change of basis $V^{-1}$, we notice
that the Markowitz solution is proportional to the vector of return and
inversely proportional to the eigenvectors. We conclude that the mean-variance
optimization problem mainly focuses on the small eigenvalues. This is why the
stability property is lacking in the original portfolio optimization
problem.\smallskip

Let us consider an example\label{ex:example1} to illustrate this problem. The investment universe
is composed of 4 assets. The expected returns are equal to $\mu_1 = 7\%$,
$\mu_2 = 8\%$, $\mu_3= 9\%$ and $\mu_4 = 10\%$ whereas the volatilities are
equal to $\sigma_1 = 15\%$, $\sigma_2 = 18\%$, $\sigma_3 = 20\%$ and $\sigma_4
= 25\%$. The correlation matrix is the following:
\begin{equation*}
\mathcal{C} = \left(
\begin{array}{rrrr}
 1.00 &       &       &      \\
 0.50 &  1.00 &       &      \\
 0.50 &  0.50 &  1.00 &      \\
 0.60 &  0.50 &  0.40 & 1.00
\end{array}
\right)
\end{equation*}
The portfolio manager's objective is to maximize the expected return for a
$15\%$ volatility target and a full investment\footnote{We only impose that the
sum of the weights is equal to $100\%$.}. The optimal portfolio $x^{\star}$ is
$\left(26.3\%,25.5\%,32.3\%,15.9\%\right)$. In Table \ref{tab:mvo1}, we
indicate how this solution differs when we slightly change the value of input
parameters. For example, if the volatility of the third asset is equal to
$19\%$, the weight of the third asset becomes $39.1\%$ instead of $32.3\%$. In
real life, we know exactly the true parameters. For instance, there is a low
probability that the realized correlation matrix is exactly the one specified
above. If we consider a uniform correlation matrix of $70\%$, we observe
significant differences in terms of allocation.

\begin{table}[tbph]
\centering
\caption{Sensitivity of the MVO portfolio to input parameters}
\label{tab:mvo1}
\tableskip
\begin{tabular}{c:ccccccc}
\hline
$\sigma_3$ &         &  $19\%$ &  $21\%$ &              &                 &                 &          $21\%$ \\
$C$        &         & & & $\mathcal{C}_4\left(30\%\right)$ & $\mathcal{C}_4\left(70\%\right)$ & & $\mathcal{C}_4\left(70\%\right)$ \\
$\mu_2$    &         &         &         &              &                 &           $5\%$ &           $7\%$ \\ \hline
$x_1$      & $26.30$ & $21.48$ & $30.20$ & ${\TsV}7.03$ & ${\TsIII}54.59$ & ${\TsIII}54.72$ & ${\TsIII}70.75$ \\
$x_2$      & $25.52$ & $22.90$ & $27.79$ &      $24.23$ & ${\TsIII}26.81$ &         $-2.43$ & ${\TsIII}13.95$ \\
$x_3$      & $32.28$ & $39.10$ & $26.48$ &      $37.53$ & ${\TsIII}22.38$ & ${\TsIII}35.38$ & ${\TsIII}16.57$ \\
$x_4$      & $15.90$ & $16.52$ & $15.53$ &      $31.21$ &         $-3.78$ & ${\TsIII}12.34$ &         $-1.27$ \\
\hline
\end{tabular}
\end{table}

We have seen that the lack of stability is due to the small eigenvalues of the
covariance matrix. More specifically, we notice that the important quantity in
mean-variance optimization is not the covariance matrix itself, but the
precision matrix, which is the inverse of the covariance matrix. In Tables
\ref{tab:mvo2-1} and \ref{tab:mvo2-2}, we have reported the eigendecomposition of $\Sigma$ and $\mathcal{I} =
\Sigma^{-1}$. We verify that the eigenvectors of the precision matrix are the
same as those of the covariance matrix, but the eigenvalues of the precision
matrix are the inverse of the eigenvalues of the covariance matrix.
\begin{table}[tbph]
\centering
\caption{Principal component analysis of the covariance matrix $\Sigma$}
\label{tab:mvo2-1}
\tableskip
\begin{tabular}{cc:rrrr}
\hline
\multicolumn{2}{c:}{Factor} & \multicolumn{1}{c}{1} &
\multicolumn{1}{c}{2} & \multicolumn{1}{c}{3} & \multicolumn{1}{c}{4} \\ \hline
\multirow{4}{*}{Asset}
& $1$                               & $36.16\%$ & $  2.44\%$ & $  5.72\%$ & $-93.03\%$ \\
& $2$                               & $42.19\%$ & $ 25.48\%$ & $-86.21\%$ & $ 11.76\%$ \\
& $3$                               & $44.74\%$ & $ 73.10\%$ & $ 46.52\%$ & $ 22.16\%$ \\
& $4$                               & $70.08\%$ & $-63.26\%$ & $ 19.25\%$ & $ 26.76\%$ \\ \hline
\multicolumn{2}{c:}{Eigenvalue}     & $ 0.10\%$ & $  0.03\%$ & $  0.02\%$ & $  0.01\%$ \\
\multicolumn{2}{c:}{$\%$ cumulated} & $63.80\%$ & $ 18.72\%$ & $ 10.65\%$ & $  6.83\%$ \\ \hline
\end{tabular}
\end{table}
\begin{table}[tbph]
\centering
\caption{Eigendecomposition of the precision matrix $\mathcal{I}$}
\label{tab:mvo2-2}
\tableskip
\begin{tabular}{cc:rrrr}
\hline
\multicolumn{2}{c:}{Factor} & \multicolumn{1}{c}{1} &
\multicolumn{1}{c}{2} & \multicolumn{1}{c}{3} & \multicolumn{1}{c}{4} \\ \hline
\multirow{4}{*}{Asset}
& $1$                               & $-93.03\%$ & $  5.72\%$ & $  2.44\%$ & $36.16\%$ \\
& $2$                               & $ 11.76\%$ & $-86.21\%$ & $ 25.48\%$ & $42.19\%$ \\
& $3$                               & $ 22.16\%$ & $ 46.52\%$ & $ 73.10\%$ & $44.74\%$ \\
& $4$                               & $ 26.76\%$ & $ 19.25\%$ & $-63.26\%$ & $70.08\%$ \\ \hline
\multicolumn{2}{c:}{Eigenvalue}     & $ 93.06\%$ & $ 59.65\%$ & $ 33.94\%$ & $ 9.96\%$ \\
\multicolumn{2}{c:}{$\%$ cumulated} & $ 47.33\%$ & $ 30.34\%$ & $ 17.26\%$ & $ 5.06\%$ \\ \hline
\end{tabular}
\end{table}
This means that the risk factors are the same, but they are in reverse order.
We see that the most important risk factor for portfolio optimization is
a long/short portfolio, which is short on the first asset and long on the other assets.
The second most important risk factor is another long/short portfolio,
which is short on the second asset and long on the third asset%
\footnote{On Page \pageref{tab:mvo2-3}, we have reported the representation
quality and the contribution of each variable for the PCA factors of $\Sigma$.
Since the second risk factor of $\mathcal{I}$ is the third risk factor of
$\Sigma$, we deduce that the first and fourth assets have a very small
contribution (respectively $0.33\%$ and $3.71\%$).}. Any changes in the
covariance matrix then impacts the largest eigenvalues of $\mathcal{I}$ and the
long/short risk factors.\smallskip

\subsection{Which risk factors are important?}

The previous eigendecomposition analysis is the traditional way to illustrate
the stability issue (Roncalli, 2017). However, the corresponding arbitrage
factors are difficult to interpret and, moreover, they do not fully help
understand the Markowitz machinery, in particular how mean-variance
portfolios are built. In this section, we use the method developed by Stevens
(1998) in order to better characterize the underlying mechanism.\smallskip

We have seen that the solution is $x^{\star }\left( \gamma \right) =\gamma
\Sigma ^{-1}\mu $. If we assume that asset returns are independent --
$\mathcal{C}=I_{n}$, we obtain the famous result:
\begin{equation*}
x_{i}^{\star }\left( \gamma \right) =\gamma \frac{\mu _{i}}{\sigma _{i}^{2}}
\end{equation*}%
The optimal weights are proportional to expected returns and inversely
proportional to variances of asset returns. In the general case --
$\mathcal{C}\neq I_{n}$, Stevens (1998) shows that the optimal portfolio
$x^{\star }$ is connected to the linear regression\footnote{This means that:
\begin{equation*}
R_{i,t}=\alpha _{i}+\sum_{j\neq i}\beta _{i,j}R_{t,j}+\varepsilon _{i,t}
\end{equation*}%
}:
\begin{equation}
R_{i,t}=\alpha _{i}+\beta _{i}^{\top }R_{t}^{\left( -i\right) }+\varepsilon
_{i,t} \label{eq:hedging1}
\end{equation}%
where $R_{t}^{\left( -i\right) }$ denotes the vector of asset returns excluding
the $i^{\mathrm{th}}$ asset. By noting $\mathfrak{R}_{i}^{2}$ the coefficient
of determination and $s_{i}^{2}$ the variance of $\varepsilon _{i,t}$, we have:
\begin{equation*}
\left[ \Sigma ^{-1}\right] _{i,i}=\frac{1}{\sigma _{i}^{2}\left( 1-\mathfrak{%
R}_{i}^{2}\right) }
\end{equation*}%
and:%
\begin{equation*}
\left[ \Sigma ^{-1}\right] _{i,j}=-\frac{\beta _{i,j}}{\sigma _{i}^{2}\left(
1-\mathfrak{R}_{i}^{2}\right) }=-\frac{\beta _{j,i}}{\sigma _{j}^{2}\left( 1-%
\mathfrak{R}_{j}^{2}\right) }
\end{equation*}%
We deduce that:%
\begin{eqnarray*}
x_{i}^{\star }\left( \gamma \right)  &=&\gamma \left[ \Sigma ^{-1}\mu \right] _{i} \\
&=&\gamma \frac{\mu _{i}-\beta _{i}^{\top }\mu ^{\left( -i\right) }}{\sigma
_{i}^{2}\left( 1-\mathfrak{R}_{i}^{2}\right) }
\end{eqnarray*}%
where $\mu ^{\left( -i\right) }$ is the vector of expected returns excluding
the $i^{\mathrm{th}}$ asset. Since we have\footnote{See Appendix
\ref{appendix:section-conditional-expectation} on page
\pageref{appendix:section-conditional-expectation}.} $s_{i}^{2}=\sigma
_{i}^{2}\left( 1-\mathfrak{R}_{i}^{2}\right) $ and $\alpha _{i}=\mu _{i}-\beta
_{i}^{\top }\mu ^{\left( -i\right) }$, we obtain:
\begin{equation*}
x_{i}^{\star }\left( \gamma \right) =\gamma \frac{\alpha _{i}}{s_{i}^{2}}
\end{equation*}%
In the general case, the optimal weights are proportional to idiosyncratic
returns $\alpha _{i}$ and inversely proportional to idiosyncratic variances
$s_{i}^{2}$.\smallskip

We notice that $\beta _{i}$ represents the best portfolio for replicating the
returns of Asset $i$. This is why it is called the hedging (or tracking)
portfolio of Asset $i$. The idiosyncratic return $\alpha _{i}$ is the
difference between the expected return $\mu _{i}$ of Asset $i$ and the expected
return $\beta _{i}^{\top }\mu ^{\left( -i\right) }$ of its hedging portfolio.
The idiosyncratic volatility $s_{i}$ is the standard deviation of residuals
$\varepsilon _{i,t}$. It is also equal to the volatility of the tracking errors
$e_{i,t}=R_{i,t}-\hat{R}_{i,t}$ where $\hat{R}_{i,t}$ is the return of the
hedging portfolio. The hedging portfolio concept is at the core of the
Markowitz optimization. Indeed, the Markowitz framework consists in estimating
the hedging strategy $\beta _{i}$ for each asset, and in forming two portfolios:

\begin{enumerate}
\item the first portfolio $y^{\star }$ is the optimal portfolio of assets
    assuming that assets are not correlated:
\begin{equation*}
y_{i}^{\star }=\gamma \frac{\mu _{i}}{\sigma _{i}^{2}}
\end{equation*}

\item the second portfolio $z^{\star }$ is the optimal portfolio of the
    hedging strategies:\footnote{Because the hedging strategies are
    independent and we have $\limfunc{var}\left( \beta _{i}^{\top
    }R_{t}^{\left(
-i\right) }\right) =\limfunc{var}\left( R_{i,t}\right) -\limfunc{var}\left(
\varepsilon _{i,t}\right) =\sigma _{i}^{2}-s_{i}^{2}$.}:%
\begin{equation*}
z_{i}^{\star }=\gamma \frac{\beta _{i}^{\top }\mu ^{\left( -i\right) }}{%
\sigma _{i}^{2}-s_{i}^{2}}
\end{equation*}
\end{enumerate}
We deduce that:%
\begin{eqnarray*}
x_{i}^{\star }\left( \gamma \right)  &=&\left( \gamma \frac{\mu _{i}}{\sigma
_{i}^{2}\left( 1-\mathfrak{R}_{i}^{2}\right) }\right) -\left( \gamma \frac{%
\beta _{i}^{\top }\mu ^{\left( -i\right) }}{\sigma _{i}^{2}\left( 1-%
\mathfrak{R}_{i}^{2}\right) }\right)  \\
&=&\left( \frac{1}{\left( 1-\mathfrak{R}_{i}^{2}\right) }\cdot \frac{\gamma
\mu _{i}}{\sigma _{i}^{2}}\right) -\left( \frac{\sigma _{i}^{2}-s_{i}^{2}}{%
\sigma _{i}^{2}\left( 1-\mathfrak{R}_{i}^{2}\right) }\cdot \frac{\gamma
\beta _{i}^{\top }\mu ^{\left( -i\right) }}{\sigma _{i}^{2}-s_{i}^{2}}%
\right)  \\
&=&\left( 1+\omega _{i}\right) \left( \phi ^{-1}\frac{\mu _{i}}{\sigma
_{i}^{2}}\right) -\omega _{i}\left( \phi ^{-1}\frac{\beta _{i}^{\top }\mu
^{\left( -i\right) }}{\sigma _{i}^{2}-s_{i}^{2}}\right)  \\
&=&y_{i}^{\star }+\omega _{i}\left( y_{i}^{\star }-z_{i}^{\star }\right)
\end{eqnarray*}%
where:
\begin{equation*}
\omega _{i}=\frac{\mathfrak{R}_{i}^{2}}{1-\mathfrak{R}_{i}^{2}}=\frac{\sigma
_{i}^{2}-s_{i}^{2}}{s_{i}^{2}}
\end{equation*}%
To take into account the correlation diversification, the optimal portfolio $%
x^{\star }$ adds to the portfolio $y^{\star }$ a long/short exposure between
$y^{\star }$ and $z^{\star }$ with a leverage that depends on the quality of
the hedge.
\smallskip

Let us consider the previous example. In Table \ref{tab:hedging1a}, we have
reported the linear regressions between the four assets, which are the hedging
portfolios of each asset. We observe that the coefficient of determination lies
between $33.5\%$ and $45.8\%$. $\mathfrak{R}^2_i$ is the highest for the first
asset, because it exhibits the largest cross-correlations. Therefore, it is the
lowest contributor to the diversification whereas the third asset is the
highest contributor to the diversification.

\begin{table}[tbph]
\centering
\caption{Linear dependence between the four assets (hedging portfolios)}
\label{tab:hedging1a}
\tableskip
\begin{tabular}{c:c:cccc:c} \hline
Asset  & $\alpha_{i}$  & \multicolumn{4}{c:}{$\beta_i$}        & $\mathfrak{R}^2_i$   \\ \hline
1      & $1.70\%$      & $     $ & $0.139$ & $0.187$ & $0.250$ & $45.83\%$            \\
2      & $2.06\%$      & $0.230$ & $     $ & $0.268$ & $0.191$ & $37.77\%$            \\
3      & $2.85\%$      & $0.409$ & $0.354$ & $     $ & $0.045$ & $33.52\%$            \\
4      & $1.41\%$      & $0.750$ & $0.347$ & $0.063$ & $     $ & $41.50\%$            \\ \hline
\end{tabular}
\medskip

\centering
\caption{Risk/return analysis of hedging portfolios}
\label{tab:hedging1b}
\tableskip
\begin{tabular}{c:ccc:ccc:c} \hline
Asset  & $\mu_{i}$ & $\hat{\mu}_i$ & $\alpha_i$
       & $\sigma_{i}$ & $\hat{\sigma}_i$ & $s_i$ & $\mathfrak{R}^2_{i}$   \\ \hline
1      & ${\TsV}7.00\%$ & $5.30\%$ & $1.70\%$ & $15.00\%$ & $10.16\%$ & $11.04\%$ & $45.83\%$ \\
2      & ${\TsV}8.00\%$ & $5.94\%$ & $2.06\%$ & $18.00\%$ & $11.06\%$ & $14.20\%$ & $37.77\%$ \\
3      & ${\TsV}9.00\%$ & $6.15\%$ & $2.85\%$ & $20.00\%$ & $11.58\%$ & $16.31\%$ & $33.52\%$ \\
4      &      $10.00\%$ & $8.59\%$ & $1.41\%$ & $25.00\%$ & $16.11\%$ & $19.12\%$ & $41.50\%$ \\ \hline
\end{tabular}
\medskip

\centering \caption{Optimal portfolio} \label{tab:hedging1c} \tableskip
\begin{tabular}{c:cccc} \hline
Asset  & $\omega_{i}$ & $y^{\star}_i$ & $z^{\star}_i$ & $x^{\star}_i$  \\ \hline
1      & $84.62\%$ & $80.22\%$ &      $132.48\%$ &      $36.00\%$ \\
2      & $60.68\%$ & $63.67\%$ &      $125.09\%$ &      $26.39\%$ \\
3      & $50.43\%$ & $58.02\%$ &      $118.19\%$ &      $27.67\%$ \\
4      & $70.94\%$ & $41.26\%$ & ${\TsV}85.40\%$ & ${\TsV}9.94\%$ \\ \hline
\end{tabular}
\end{table}

We then calculate the risk/return statistics of hedging portfolios in Table
\ref{tab:hedging1b}. We verify that the following equalities hold%
\footnote{We have:
\begin{equation*}
\hat{\mu}_{i} = \mathbb{E}\left[ \hat{R}_{i,t} \right] =
\mathbb{E}\left[ \beta _{i}^{\top }R_{t}^{\left( -i\right) } \right] =
\beta _{i}^{\top }\mu ^{\left( -i\right) }
\end{equation*}%
and:%
\begin{equation*}
\hat{\sigma}_{i}^{2} = \func{var}\left( \hat{R}_{i,t}\right) =
\func{var}\left( \beta _{i}^{\top }R_{t}^{\left(-i\right) }\right) =
\sigma _{i}^{2}\mathfrak{R}_{i}^{2}
\end{equation*}
}: $\mu _{i}=\hat{\mu}_{i} + \alpha _{i}$ and $\sigma
_{i}^{2}=\hat{\sigma}_{i}^{2}+s_{i}^{2}$. Finally, we obtain the optimal
portfolio given in Table \ref{tab:hedging1c}. $\gamma $ is set to $0.2578$ in
order to obtain a $100\%$ exposure. In this example, the optimal portfolio is:
$x_{1}^{\star }=36\%$, $x_{2}^{\star }=26.39\%$, $x_{3}^{\star }=27.67\%$ and
$x_{4}^{\star }=9.94\%$. There is no short position, because the alpha $\alpha
_{i}$ is positive for all the assets, meaning that hedging portfolios are not
able to produce a better expected return than the corresponding
assets.\smallskip

We now modify the correlation between the third and fourth assets, and set
$\rho_{3,4} = 95\%$. This high correlation changes the results of the linear
regression (see Tables \ref{tab:hedging2a} and \ref{tab:hedging2b}). Indeed, the coefficient of
determination for Assets $3$ and $4$ is larger than $90\%$, and the fourth
hedging portfolio has an expected return that is higher than that of the fourth
asset. Since $\alpha_4$ is the only negative alpha, the optimal portfolio is
short on the fourth asset and long on the other assets (see Table
\ref{tab:hedging2c}). Another important factor is the impact of
$\mathfrak{R}^2_i$ on the weights $\omega_i$. Thus, $\omega_3$ and $\omega_4$
are larger than $10$ whereas $\omega_1$ and $\omega_2$ are smaller than $1$.
Even if the difference between $y_i^{\star}$ and $z_i^{\star}$ is the smallest
for Assets 3 and 4, the leverage effect largely compensates the long/short
effect, and explains why the optimal portfolio has a large exposure on Assets 3
and 4.

\begin{table}[tbph]
\centering
\caption{Linear dependence between the four assets ($\rho_{3,4} = 95\%$)}
\label{tab:hedging2a}
\tableskip
\begin{tabular}{c:c:cccc:c} \hline
Asset  & $\alpha_{i}$  & \multicolumn{4}{c:}{$\beta_i$}        & $\mathfrak{R}^2_i$   \\ \hline
1 & ${\TsVIII}3.16\%$ &         $      $ & ${\TsVIII}0.244$ &         $-0.595$ & ${\TsVIII}0.724$ & $47.41\%$ \\
2 & ${\TsVIII}2.23\%$ & ${\TsVIII}0.443$ &         $      $ & ${\TsVIII}0.470$ &         $-0.157$ & $33.70\%$ \\
3 & ${\TsVIII}1.66\%$ &         $-0.174$ & ${\TsVIII}0.076$ &         $      $ & ${\TsVIII}0.795$ & $91.34\%$ \\
4 &         $-1.61\%$ & ${\TsVIII}0.292$ &         $-0.035$ & ${\TsVIII}1.094$ &         $      $ & $92.37\%$ \\ \hline
\end{tabular}
\medskip

\centering
\caption{Risk/return analysis of hedging portfolios ($\rho_{3,4} = 95\%$)}
\label{tab:hedging2b}
\tableskip
\begin{tabular}{c:ccc:ccc:c} \hline
Asset  & $\mu_{i}$ & $\hat{\mu}_i$ & $\alpha_i$
       & $\sigma_{i}$ & $\hat{\sigma}_i$ & $s_i$ & $\mathfrak{R}^2_{i}$   \\ \hline
1      & ${\TsV}7.00\%$ & ${\TsV}3.84\%$ & ${\TsVIII}3.16\%$ & $15.00\%$ & $10.33\%$ & $10.88\%$ & $47.41\%$ \\
2      & ${\TsV}8.00\%$ & ${\TsV}5.77\%$ & ${\TsVIII}2.23\%$ & $18.00\%$ & $10.45\%$ & $14.66\%$ & $33.70\%$ \\
3      & ${\TsV}9.00\%$ & ${\TsV}7.34\%$ & ${\TsVIII}1.66\%$ & $20.00\%$ & $19.11\%$ & $ 5.89\%$ & $91.34\%$ \\
4      &      $10.00\%$ &      $11.61\%$ &         $-1.61\%$ & $25.00\%$ & $24.03\%$ & $ 6.90\%$ & $92.37\%$ \\ \hline
\end{tabular}
\medskip

\centering
\caption{Optimal portfolio ($\rho_{3,4} = 95\%$)}
\label{tab:hedging2c}
\tableskip
\begin{tabular}{c:cccc} \hline
Asset  & $\omega_{i}$ & $y^{\star}_i$ & $z^{\star}_i$ & $x^{\star}_i$  \\ \hline
1      & ${\TsX}90.16\%$ & $60.73\%$ & ${\TsV}70.30\%$ & ${\TsVIII}52.10\%$ \\
2      & ${\TsX}50.82\%$ & $48.20\%$ &      $103.08\%$ & ${\TsVIII}20.31\%$ \\
3      &     $1054.10\%$ & $43.92\%$ & ${\TsV}39.22\%$ & ${\TsVIII}93.44\%$ \\
4      &     $1211.48\%$ & $31.23\%$ & ${\TsV}39.25\%$ &         $-65.85\%$ \\ \hline
\end{tabular}
\end{table}

The theoretical analysis presented in this paragraph also highlights the
importance of the expected returns. Indeed, even if they do not change the
composition and the risk analysis of hedging portfolios, they impact the
return analysis. An example is provided in Appendix
\ref{appendix:section-tables} on page \pageref{tab:hedging3a}. We change the
expected return of the first asset and set $\mu_1 = 3\%$. In this case, the
expected return of the first asset is largely smaller than the expected return
of the corresponding hedging portfolio. At the same time, the alpha of the
other three assets increases sharply. This is why Markowitz optimization
increases the allocation in the third asset and takes a short position on
the first asset.\smallskip

Let us write Equation (\ref{eq:hedging1}) as follows:%
\begin{equation*}
\frac{R_{i,t}-\mu _{i}}{\sigma _{i}}=\sum_{j\neq i}\tilde{\beta}_{i,j}\left(
\frac{R_{j,t}-\mu _{j}}{\sigma _{j}}\right) +\varepsilon _{i,t}
\end{equation*}%
where the coefficients $\tilde{\beta}_{i,j}$ only depend on the correlation
matrix $\mathcal{C}$. We have the following correspondence:%
\begin{equation*}
\alpha _{i}=\mu _{i}-\sigma _{i}\sum_{j\neq i}\tilde{\beta}_{i,j}\left(
\frac{\mu _{j}}{\sigma _{j}}\right)
\end{equation*}%
and:%
\begin{equation*}
\beta _{i,j}=\tilde{\beta}_{i,j}\left( \frac{\sigma _{i}}{\sigma _{j}}%
\right)
\end{equation*}%
Moreover, we notice that:%
\begin{equation*}
s_{i}^{2}=\sigma _{i}^{2}\left( 1-\mathfrak{R}_{i}^{2}\right)
\end{equation*}%
and:%
\begin{equation*}
\mathfrak{R}_{i}^{2}=\left( e_{i}^{\top }\mathcal{C}^{\left( -i\right) }\right) \left(
\mathcal{C}^{\left( -i\right) }\right) ^{-1}\left( \mathcal{C}^{\left( -i\right)
}e_{i}\right)
\end{equation*}%
where $\mathcal{C}^{\left( -i\right) }$ is the correlation matrix excluding the $i^{%
\mathrm{th}}$ asset. We obtain the following effects:

\begin{itemize}
\item A change in the expected return $\mu _{i}$ impacts the alpha $\alpha_i$ of the
    hedging portfolios. It does not change the composition $\beta_i$ of hedging portfolios
    or the weights $\omega_i$;

\item A change in the volatility $\sigma _{i}$ impacts the exposures
    $\beta_i$ of the hedging portfolios. It does not change the weights
    $\omega_i$, but modifies the value of alphas. As such, the composition of
    the portfolio $z_i$ changes;

\item A change in the correlation $\rho _{i,j}$ impacts all the parameters
    ($\alpha_i$, $\beta_i$ and $w_i$).
\end{itemize}
We also notice that the correlations are the only parameters that are used for
calculating the coefficient of determination $\mathfrak{R}_{i}^{2}$. Therefore,
correlations are the key parameters for understanding the leverage effects in
the Markowitz model. Indeed, they impact both the tracking error volatilities
$s_i$ and the weights $\omega_i$. The main effect of the volatility $\sigma_i$
concerns the tracking error, because $s_i$ is an increasing function of
$\sigma_i$. A high volatility $\sigma_i$ therefore negatively impacts the allocation
$y_i$ and $z_i$.

\section{Theory of regularization}

The stability issue has been considered by Michaud (1989) in a very famous
publication \textquotedblleft The Makowitz Optimization Enigma: Is Optimized
Optimal?\textquotedblright. In his works, Michaud clearly makes the distinction
between mathematical optimization and financial optimality. For instance, if we
consider two assets that are highly similar in terms of risk and return, a fund
manager will most likely spread a long exposure into these two assets, whereas
Markowitz will play an arbitrage between them. Academics have proposed several
approaches to make Markowitz's solutions more robust. Two main directions have
been explored. The first one concerns the regularization of the covariance
matrix. As seen in Equation (\ref{eq:eig1}), the problem is ill-conditioned
because of the magnitude of eigenvectors. One solution is therefore to change
the eigenvalues of $\Sigma$. For instance, the direct approach consists in
deleting the lowest eigenvalues (Laloux \textsl{et al.}, 1999). The indirect
approach mixes different covariance matrices in order to obtain a more robust
estimator, and is called the shrinkage method (Ledoit and Wolf, 2003). The
second direction concerns the regularization of the optimization problem (e.g.
adding $L_2$ penalty) or the sparsity of the solution (e.g. adding $L_1$
penalty). The simplest way is to add some weight constraints. For instance, we
can impose that the sum of weights is equal to one, the weights are positive,
etc. Another approach consists in modifying the objective function by adding
some penalties, such as ridge or lasso norms.

\subsection{Adding constraints}

Let us specify the Markowitz problem in the following way:
\begin{eqnarray*}
&\min &\frac{1}{2}x^{\top }\Sigma x \\
&\text{s.t.}&\left\{
\begin{array}{l}
\mathbf{1}^{\top }x=1 \\
x ^{\top }\mu \geqslant \mu ^{\star } \\
x\in \Omega%
\end{array}%
\right.
\end{eqnarray*}%
where $\Omega$ is the set of weight constraints. This is a variant of the
$\mu$-problem (\ref{eq:mu-problem}) described on page \pageref{eq:mu-problem}.
We consider two optimized portfolios:

\begin{itemize}
\item The first one is the unconstrained portfolio $x^{\star }\left( \mu
    ,\Sigma \right) $ with $\Omega =\mathbb{R}^{n}$.

\item The second one is the constrained portfolio $\tilde{x}\left( \mu
    ,\Sigma \right) $ with some constraints added.
\end{itemize}

\noindent Jagannathan and Ma (2003) assume that the weight of asset $i$ is
between a lower bound $x_{i}^{-}$ and an upper bound $x_{i}^{+}$:%
\begin{equation*}
\Omega=\left\{ x\in \mathbb{R}^{n}:x_{i}^{-}\leqslant x_{i}\leq
x_{i}^{+}\right\}
\end{equation*}%
They show that the constrained optimal portfolio is the solution of the
unconstrained problem:%
\begin{equation*}
\tilde{x}\left( \mu ,\Sigma \right) =x^{\star }\left( \tilde{\mu},\tilde{%
\Sigma}\right)
\end{equation*}%
with:%
\begin{equation*}
\left\{
\begin{array}{l}
\tilde{\mu}=\mu  \\
\tilde{\Sigma}=\Sigma +\left( \lambda ^{+}-\lambda ^{-}\right) \mathbf{1}%
^{\top }+\mathbf{1}\left( \lambda ^{+}-\lambda ^{-}\right) ^{\top }%
\end{array}%
\right.
\end{equation*}%
where $\lambda ^{-}$ and $\lambda ^{+}$ are the Lagrange coefficients vectors
associated with the lower and upper bounds. Introducing weight
constraints is then equivalent to using another covariance matrix $\tilde{%
\Sigma}$, or shrinking the covariance matrix. More generally, if we
introduce linear inequality constraints:%
\begin{equation*}
\Omega = \left\{ x\in \mathbb{R}^{n}:Cx\geqslant d\right\}
\end{equation*}%
we obtain a similar result. The covariance matrix is shrunk as follows%
\footnote{The shrinkage covariance matrix is not necessarily positive definite (Roncalli, 2013).}:%
\begin{equation*}
\tilde{\Sigma}=\Sigma -\left( C^{\top }\lambda \mathbf{1}^{\top }+\mathbf{1}%
\lambda ^{\top }C\right)
\end{equation*}%
where $\lambda $ is the vector of Lagrange coefficients associated with the
constraints $Cx\geqslant d$.\smallskip

We again consider the previous example given on page \pageref{ex:example1}. If
we compute the global minimum variable, the solution $x^{\star}$ is equal to
$65.57\%$, $29.06\%$, $13.61\%$ and $-8.24\%$. Let us suppose that the portfolio
manager is not satisfied with this optimized portfolio and decides to impose some
constraints. For instance, he could decide that the portfolio must contain at
least $10\%$ of all assets. In order to achieve a certain degree of
diversification, he could also decide to impose an upper bound of $40\%$. With
these constraints $x_i^{-} = 10\%$ and $x_i^{+} = 40\%$, the solution becomes
$40.00\%$, $31.18\%$, $18.82\%$ and $10.00\%$. Thanks to the Jagannathan-Ma
framework, we can compute the shrinkage covariance matrix%
\footnote{We have $\lambda_4^{-} = 48.89$ bps and $\lambda_1^{+} = 28.58$ bps.
The other Lagrange coefficients are equal to zero.}, and deduce the shrinkage
volatilities $\tilde{\sigma}_i$ and correlation matrix $\tilde{\mathcal{C}}$,
which are reported in Table \ref{tab:shrinkage1a}. To obtain this new solution,
one must increase (implicitly) the volatility of the first asset, and decrease
(implicitly) the volatility of the fourth asset. Concerning the correlations,
we also notice that they have changed. In Table \ref{tab:shrinkage1b}, we
report the results when the objective function is to target an expected return
of $9\%$. In this case, we notice that introducing constraints is equivalent to
introducing some views on the first asset. Indeed, this allows us to impose a better
Sharpe ratio and a lower correlation with the second asset.\smallskip

\begin{table}[tbph]
\centering
\caption{Jagannathan-Ma shrinkage of the GMV portfolio}
\label{tab:shrinkage1a}
\tableskip
\begin{tabular}{c:cc:c:cccc}
\hline
       &                 &                 &                    &                 &                 &            \\[-1em]
Asset  & $x^{\star}_{i}$ & $\tilde{x}_{i}$ & $\tilde{\sigma}_i$ & \multicolumn{4}{c}{$\tilde{\mathcal{C}}$}  \\ \hline
1 & ${\TsIII}65.57\%$ & $40.00\%$ & $16.80\%$ & $100.00\%$      &                 &                 &            \\
2 & ${\TsIII}29.06\%$ & $31.18\%$ & $18.00\%$ & ${\TsV}54.10\%$ &      $100.00\%$ &                 &            \\
3 & ${\TsIII}13.61\%$ & $18.82\%$ & $20.00\%$ & ${\TsV}53.16\%$ & ${\TsV}50.00\%$ &      $100.00\%$ &            \\
4 &         $-8.24\%$ & $10.00\%$ & $22.96\%$ & ${\TsV}53.07\%$ & ${\TsV}42.61\%$ & ${\TsV}32.90\%$ & $100.00\%$ \\ \hline
\end{tabular}
\end{table}

\begin{table}[tbph]
\centering
\caption{Jagannathan-Ma shrinkage of the MVO portfolio ($\mu^{\star} = 9\%$)}
\label{tab:shrinkage1b}
\tableskip
\begin{tabular}{c:cc:c:cccc}
\hline
       &                 &                 &                    &                 &                 &            \\[-1em]
Asset  & $x^{\star}_{i}$ & $\tilde{x}_{i}$ & $\tilde{\sigma}_i$ & \multicolumn{4}{c}{$\tilde{\mathcal{C}}$}  \\ \hline
1 & ${\TsV}3.30\%$ & $10.00\%$ & $12.06\%$ &      $100.00\%$ &                 &                 &            \\
2 &      $23.44\%$ & $15.00\%$ & $18.00\%$ & ${\TsV}43.87\%$ &      $100.00\%$ &                 &            \\
3 &      $43.21\%$ & $40.00\%$ & $20.59\%$ & ${\TsV}49.20\%$ & ${\TsV}51.79\%$ &      $100.00\%$ &            \\
4 &      $30.05\%$ & $35.00\%$ & $25.00\%$ & ${\TsV}61.43\%$ & ${\TsV}50.00\%$ & ${\TsV}41.18\%$ & $100.00\%$ \\ \hline
\end{tabular}
\end{table}

\begin{remark}
Constraints are inherent to Markowitz optimization. Indeed, the raw solution
given by the mean-variance optimization is generally not satisfied. This is why
Quants spend a lot of time adding and testing constraints. This is particular
true for strategic asset allocation, for which the annual exercises are very
time-consuming. However, adding constraints introduces the personal
views of the Quant in charge of the optimization. Moreover, this process
of trial and error must be repeated each time the allocation problem changes.
Therefore, Markowitz optimization is more a handmade solution, and not an
industrial solution. This is why it cannot be used \textquotedblleft as
is\textquotedblright\ by robo-advisors, whose mass production/customization approach is
incompatible with human intervention.
\end{remark}

\subsection{Adding a benchmark}

Let us now consider a benchmark which is represented by a portfolio $b$. The
tracking error between the portfolio $x$ and its benchmark $b$ is the
difference between the return of the portfolio and the return of the
benchmark:%
\begin{eqnarray*}
e_{t} &=&R_{t}\left( x\right) -R_{t}\left( b\right)  \\
&=&\left( x-b\right) ^{\top }R_{t}
\end{eqnarray*}%
where $R_{t}=\left( R_{t,1},\ldots ,R_{t,n}\right) $ is the vector of asset
returns. The expected excess return is:
\begin{equation*}
\mu \left( x\mid b\right) =\mathbb{E}\left[ e_{t}\right] =\left( x-b\right)
^{\top }\mu
\end{equation*}%
whereas the volatility of the tracking error is:
\begin{equation*}
\sigma \left( x\mid b\right) =\sigma \left( e_{t}\right) =\sqrt{\left(
x-b\right) ^{\top }\Sigma \left( x-b\right) }
\end{equation*}%
The investor's objective is to maximize the expected tracking error with a
constraint on the tracking error volatility. Like the Markowitz
problem, we transform this $\sigma $-problem into a $\gamma $-problem:%
\begin{eqnarray*}
x^{\star }\left( \gamma \right)  &=& \arg \min \frac{1}{2}\left( x-b\right) ^{\top
}\Sigma \left( x-b\right) -\gamma x^{\top }\mu \left( x\mid b\right)  \\
&\text{s.t.}&x\in \Omega
\end{eqnarray*}%
The objective function is then:
\begin{eqnarray*}
f\left( x\right)  &=&\frac{1}{2}\left( x-b\right) ^{\top }\Sigma \left(
x-b\right) -\gamma \left( x-b\right) ^{\top }\mu  \\
&=&\frac{1}{2}x^{\top }\Sigma x-x^{\top }\left( \gamma \mu +\Sigma b\right)
+\left( \frac{1}{2}b^{\top }\Sigma b+\gamma b^{\top }\mu \right)
\end{eqnarray*}%
We deduce that:%
\begin{eqnarray*}
x^{\star }\left( \gamma \right)  &=&\frac{1}{2}x^{\top }\Sigma x-\gamma
x^{\top }\tilde{\mu} \\
&\text{s.t.}&x\in \Omega
\end{eqnarray*}%
where $\tilde{\mu}=\mu +\dfrac{1}{\gamma }\Sigma b$. Let $\mu _{b}$ be the
vector of implied expected returns such that the benchmark $b$ is the optimal
portfolio. Since we have $b=\gamma \Sigma ^{-1}\mu _{b}$, the
optimization problem becomes:%
\begin{eqnarray*}
x^{\star }\left( \gamma \right)  & = & \arg \min \frac{1}{2}x^{\top }\Sigma x-\xi x^{\top
}\left( \frac{\mu +\mu _{b}}{2}\right)  \\
&\text{s.t.}&x\in \Omega
\end{eqnarray*}%
where $\xi =2\gamma $. Introducing a benchmark constraint is then equivalent to
regularizing the expected returns.

\subsection{Tikhonov and ridge regularization}

Previously, we have seen a method that regularizes the covariance matrix and an
approach that regularizes the vector of expected returns. We now turn to a
framework that regularizes the two input parameters of Markowitz optimization
problems, and not only the covariance matrix or the vector of expected returns.
While the two previous approaches are more specific to financial optimization,
the following methods have been developed in PDEs and later in statistics. This
is why we consider the following general optimization problem:
\begin{eqnarray}
x^{\star } & = &\arg \min \frac{1}{2}\left\Vert A_{1} x - b_1\right\Vert _{2}^{2} \label{eq:statistical-problem} \\
& \text{s.t.} & \left\{
  \begin{array}{l}
    A_{2} x = b_2   \\
    A_{3} x \geqslant b_3
  \end{array}
\right.
\notag
\end{eqnarray}%
We recognize a standard quadratic programming problem. Problems
(\ref{eq:sigma-problem}) -- (\ref{eq:qp3}) can easily be written as Problem
(\ref{eq:statistical-problem}). For instance, the $\gamma$-problem
(\ref{eq:qp1}) is obtained with $A_{1}^{\top} A_{1} = \Sigma$ and $A_{1}^{\top}
b_1 = \gamma \mu$, while we have $b_1 = \mathbf{0}$, $A_{3} = \mu^{\top}$ and
$b_3 = \mu^{\star}$ for the $\mu$-problem. If we prefer to use the empirical
model (\ref{eq:qp3}), we specify $A_{1} = W^{1/2}R = D_w^{1/2} C_T R$ and $b_1
= \gamma W^{-1/2}w = \gamma \left( C_T^{\top} D_w^{1/2}\right)^{-1} w$. We
notice that the $L_2$ norm is natural because of the specification of $A_1$.

\subsubsection{Formulation of the Tikhonov problem}

In order to regularize the Markowitz optimization problem, we can add a
penalty term. For instance, the most famous approach is the Tikhonov
regularization. The general problem can be written as follows:%
\begin{eqnarray}
x^{\star } &=&\arg \min \frac{1}{2}\left\Vert A_{1} x - b_{1}\right\Vert _{2}^{2}+
\frac{1}{2}\varrho_2 \left\Vert \Gamma_2 \left( x-x_{0}\right) \right\Vert _{2}^{2}
\label{eq:tikhonov1} \\
&\text{s.t.} & A_{2} x = b_2  \notag
\end{eqnarray}%
where $\varrho_2 > 0$ is a positive number, $\Gamma_2 \in \mathbb{R}^{n\times
n}$, $A_{2}\in \mathbb{R}^{m\times n}$ and $b_2\in \mathbb{R}^{m\times 1}$.
The vector $x_{0}$ is an initial solution. The Tikhonov regularization matrix
$\Gamma_2$ forces the solution to be close to $x_0$ with respect to the
semi-norm $x \mapsto \left\Vert \Gamma_2 x \right \Vert_2$ whereas the Tikhonov
regularization parameter $\varrho_2$ indicates the strength of the
regularization.

\begin{remark}
In portfolio optimization, $x_0$ can be seen as a reference portfolio. For
instance, it can be a benchmark, an heuristic portfolio%
\footnote{For instance, it can be the equally-weighted (EW) portfolio or the
equal risk contribution (ERC) portfolio (Roncalli, 2013).} or the investment
portfolio of the previous period. The $L_2$ penalty term may then be used to
control the deviation between the new portfolio and the reference portfolio,
the tracking error or the portfolio turnover.
\end{remark}

\begin{remark}
The previous approach was introduced in asset management by Jorion (1988,
1992), who considered the Bayes-Stein estimator based on the one-factor
model developed by Sharpe (1963). With the notations above, we have the
following correspondence: $\Gamma_2 = \mathbf{1} \mathbf{1}^{\top}$ and $x_0 =
\mathbf{0}$.
\end{remark}

In Appendix \ref{appendix:section-tikhonov} on page
\pageref{appendix:section-tikhonov}, we show that the optimal
solution is the $x$-coordinate of the linear system solution%
\footnote{We obtain a linear system of the form $Az = b$ where $A$ is a
symmetric $2 \times 2$ block matrix. The (1,1) block depends on the matrix
$A_{1}$ while the (2,1) block depends on the matrix $A_{2}$.}:
\begin{equation}
\left(
\begin{array}{cc}
A_{1}^{\top } A_{1}+\varrho_2 \Gamma_2 ^{\top }\Gamma_2  & A_{2}^{\top } \\
A_{2} & \mathbf{0}%
\end{array}%
\right) \left(
\begin{array}{c}
x \\
\lambda
\end{array}%
\right) =\left(
\begin{array}{c}
A_{1}^{\top }b_1+\varrho_2 \Gamma_2 ^{\top }\Gamma_2 x_{0} \\
b_2
\end{array}%
\right) \label{eq:tikhonov2}
\end{equation}%
where $\lambda $ is the vector of Lagrange coefficients associated with the
constraint $A_{2} x = b_2$. The OLS regression corresponds to $\Gamma_2
=\mathbf{0}$ whereas the ridge regression is obtained with $\Gamma_2 = I_{n}$.
For $\lambda = \mathbf{0}$ and $\varrho_2 = 0$, the OLS solution is simply
$x^{\star }=A_{1}^{\dagger}b_1$ where $A^{\dagger }_1=\left( A_1^{\top }A_1\right)
^{-1}A_1^{\top }$ is the Moore-Penrose pseudo-inverse matrix of $A_1$. For $\lambda
= \mathbf{0}$ and $\varrho_2 > 0$, the regularized solution becomes $x^{\star
}=A_{1}^{\#} b_1^{\#}$ where $A_{1}^{\# }$ may be interpreted as the
Tikhonov regularization of $A_{1}^{\dagger}$:
\begin{equation*}
A_{1}^{\#} = \left(A_{1}^{\top} A_{1} + \varrho_2 \Gamma_2^{\top} \Gamma_2\right)^{-1} A_{1}^{\top}
\end{equation*}
We also notice that $A_{1}^{\top} A_{1} + \varrho_2 \Gamma_2^{\top}
\Gamma_2$ is invertible if the matrix $\Gamma_2 $ is invertible. Indeed, if
$\left(A_{1}^{\top} A_{1} + \varrho_2 \Gamma_2^{\top} \Gamma_2
\right) x=\mathbf{0}$, we have:%
\begin{equation*}
0=x^{\top }\left( A_{1}^{\top} A_{1} + \varrho_2 \Gamma_2^{\top} \Gamma_2 \right)
x=\left\Vert A_{1}x\right\Vert _{2}^{2}+\varrho_2 \left\Vert \Gamma_2 x\right\Vert
_{2}^{2}\geqslant \varrho_2 \left\Vert \Gamma_2 x\right\Vert _{2}^{2}
\end{equation*}%
This ensures the property that the matrix $ A_{1}^{\top} A_{1} + \varrho_2
\Gamma_2^{\top} \Gamma_2$ is positive definite. This idea can be extended using
spectral decomposition of $A_1$, which naturally leads to defining the
regularization of the matrix $A_1$ through spectral filters.

\subsubsection{Relationship with covariance shrinkage methods}

Let us consider the regularized Markowitz problem:%
\begin{eqnarray*}
x^{\star } &=&\arg \min \frac{1}{2}x^{\top }\Sigma x-\gamma x^{\top }\mu +%
\mathcal{R}\left( x\right)  \\
&s.t.&\mathbf{1}^{\top }x=1
\end{eqnarray*}%
where $\mathcal{R}\left( x\right) $ is the regularization function. If we
consider the Tikhonov formulation (\ref{eq:tikhonov1}), we have the following
correspondence: $A_{1}^{\top }A_{1}=\Sigma $ and $A_{1}^{\top }b_1=\gamma
\mu $. We deduce that the regularization on the matrix $A_{1}^{\dagger }$ can
be written as a regularization on the covariance matrix $\Sigma $ when there is
no target portfolio ($x_0 = \mathbf{0}$):
\begin{equation*}
\Sigma\left(\varrho_2\right) = \Sigma +\varrho_2 \Gamma_2 ^{\top }\Gamma_2
\end{equation*}%
Therefore, there is a strong relationship between regularization and shrinkage.
Indeed, the empirical covariance matrix $\hat{\Sigma}$ is an unbiased estimator
of $\Sigma $, but its convergence is very slow in particular when $n$ is large.
We know also that the estimator $\hat{\Phi}$ based on factor models converges
more quickly, but it is biased. Ledoit and
Wolf (2003) propose combining the two estimators $\hat{\Sigma}$ and $%
\hat{\Phi}$ in order to obtain a more efficient estimator. Let
$\hat{\Sigma}\left(\alpha\right) = \alpha \hat{\Sigma}+\left( 1-\alpha \right)
\hat{\Phi}$ be this new estimator. Ledoit and Wolf estimate the optimal value
of $\alpha $ by
minimizing the expected value of the quadratic loss:%
\begin{equation*}
\alpha ^{\star }=\arg \min \mathbb{E}\left[ L\left( \alpha \right) \right]
\end{equation*}%
where the loss function is equal to:%
\begin{equation*}
L\left( \alpha \right) =\left\Vert \alpha \hat{\Sigma}+\left( 1-\alpha
\right) \hat{\Phi}-\Sigma \right\Vert _{2}^{2}
\end{equation*}%
We have, up to a scaling factor\footnote{%
This is not an issue since $\gamma $ is not a fixed parameter, but is
calibrated to solve a $\sigma $-problem or a $\mu $-problem.}, the following
correspondence:%
\begin{equation*}
\left\{
\begin{array}{l}
\varrho_2 =\dfrac{1-\alpha ^{\star }}{\alpha ^{\star }} \\
\Gamma_2 =\limfunc{chol} \hat{\Phi}
\end{array}%
\right.
\end{equation*}%
where $\limfunc{chol} M $ is the upper Cholesky factor of the matrix $M$.
Therefore, the Ledoit-Wolf shrinkage technique is a special case of Tikhonov
regularization. In a similar way, the double shrinkage method proposed by
Candelon \textsl{et al.} (2012) is obtained by setting $\Gamma_2 =I_{n}$ and $%
x_{0}\neq \mathbf{0}$.

\subsubsection{Ridge regularization}

The ridge regularization is defined by $\Gamma_2 =I_{n}$. We deduce that the
mean-variance objective function becomes:
\begin{eqnarray*}
f\left( x\right)  &=&\frac{1}{2}x^{\top }\Sigma x-\gamma x^{\top }\mu +
\frac{1}{2}\varrho_2 \left\Vert x-x_{0}\right\Vert _{2}^{2} \\
 & \propto &\frac{1}{2}x^{\top }\left( \Sigma +\varrho_2 I_{n}\right) x-x^{\top }\left(
\gamma \mu +\varrho_2 x_{0}\right)  \\
&=&\frac{1}{2}x^{\top }\Sigma\left(\varrho_2\right) x-\gamma x^{\top }\mu\left(\varrho_2\right)
\end{eqnarray*}
where $\Sigma\left(\varrho_2\right) = \Sigma +\varrho_2 I_{n}$ and
$\mu\left(\varrho_2\right) = \mu + \dfrac{\varrho_2 }{\gamma }x_{0}$. Let
$x^{\star }\left( \gamma ;\varrho_2 ,x_{0}\right) $ be the unconstrained
solution of the ridge optimization problem:
\begin{equation*}
x^{\star }\left( \gamma ;\varrho_2 ,x_{0}\right) = \arg \min \frac{1}{2}x^{\top }\Sigma
x-\gamma x^{\top }\mu +\frac{1}{2}\varrho_2 \left\Vert x-x_{0}\right\Vert
_{2}^{2}
\end{equation*}%
We have:
\begin{eqnarray*}
x^{\star }\left( \gamma ;\varrho_2 ,x_{0}\right)  &=&\gamma \Sigma\left(\varrho_2\right)^{-1}\mu\left(\varrho_2\right) \\
&=&\gamma \left( \Sigma +\varrho_2 I_{n}\right) ^{-1}\left( \mu +\frac{\varrho_2
}{\gamma }x_{0}\right)  \\
&=&\left( I_{n}+\varrho_2 \Sigma ^{-1}\right) ^{-1}\left( x^{\star }\left(
\gamma ;\mu \right) +x^{\star }\left( \varrho_2 ;x_{0}\right) \right)
\end{eqnarray*}
where $x^{\star }\left( \gamma ;\mu \right) =\gamma \Sigma ^{-1}\mu $ is the
Markowitz solution. We deduce that the regularized solution is the average of
two portfolios: the Markowitz portfolio $x^{\star }\left( \gamma ;\mu \right) $
and the optimal portfolio $x^{\star }\left( \varrho_2 ;x_{0}\right) $ when the
vector of expected returns is equal to $x_{0}$ and the risk/return trade-off
parameter is $\varrho_2 $. Bruder \textsl{et al.} (2013) also show that:
\begin{equation*}
x^{\star }\left( \gamma ;\varrho_2 ,x_{0}\right) =\omega\left(\varrho_2\right) x^{\star
}\left( \gamma ;\mu \right) +\left( I_{n}-\omega\left(\varrho_2\right)\right) x_{0}
\end{equation*}
where the matrix of weights $\omega\left(\varrho_2\right)$ is equal to $\left(
I_{n}+\varrho_2 \Sigma ^{-1}\right) ^{-1}$. We verify that:
\begin{equation*}
\lim_{\varrho_2\rightarrow \infty }\omega\left(\varrho_2\right) =\mathbf{0}
\end{equation*}
Without any constraints, the ridge regularization reduces the leverage of
Markowitz portfolio when there is no target portfolio. When we impose that the
portfolio is fully invested ($\mathbf{1}^{\top }x=1$), this is equivalent
imposing that the target portfolio is the equally-weighted portfolio.\smallskip

\begin{figure}[tbph]
\centering
\caption{Ridge regularization with a target portfolio}
\label{fig:ridge1}
\figureskip
\includegraphics[width = \figurewidth, height = \figureheight]{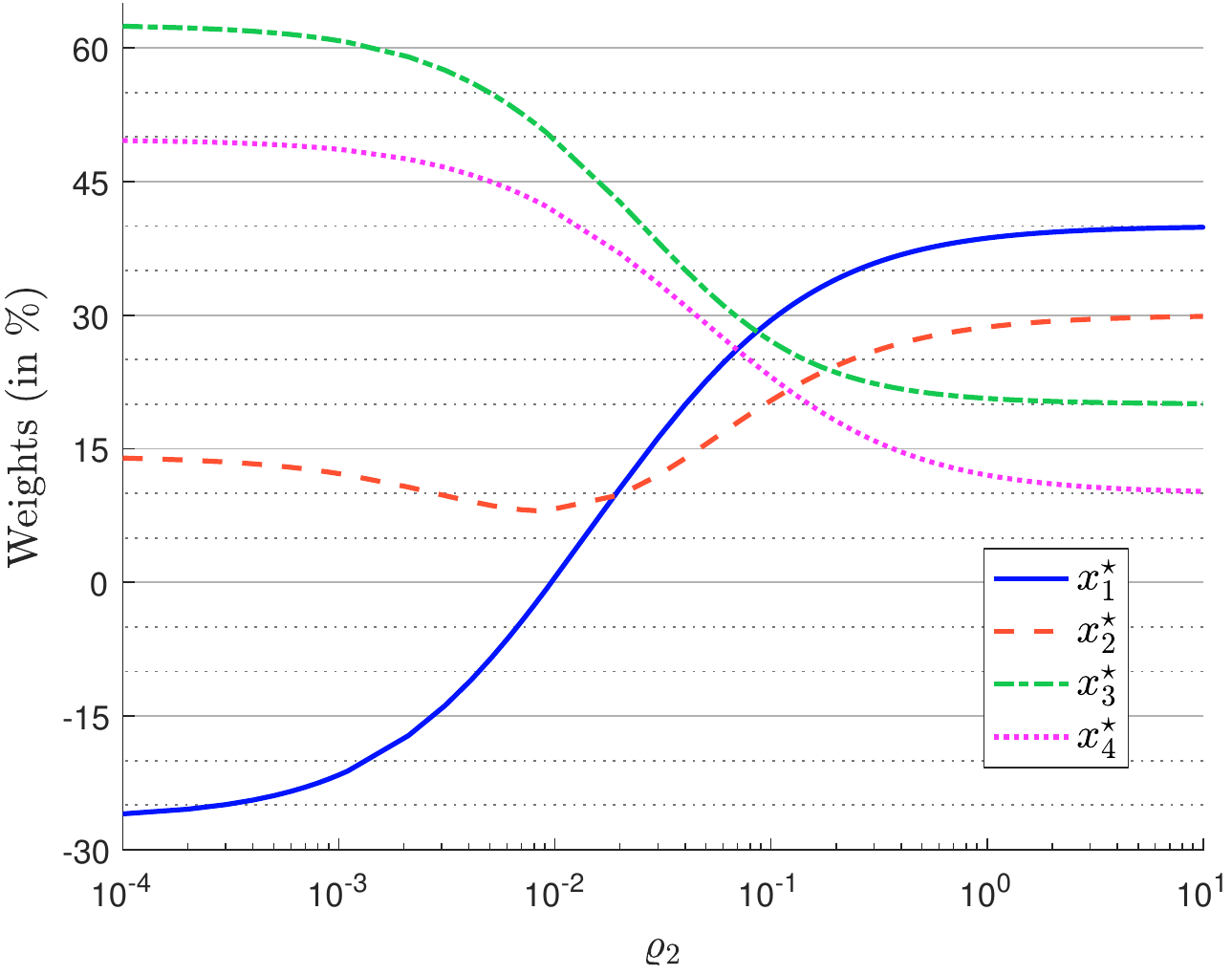}
\end{figure}

\begin{figure}[tbph]
\centering
\caption{Ridge regularization without a target portfolio}
\label{fig:ridge2}
\figureskip
\includegraphics[width = \figurewidth, height = \figureheight]{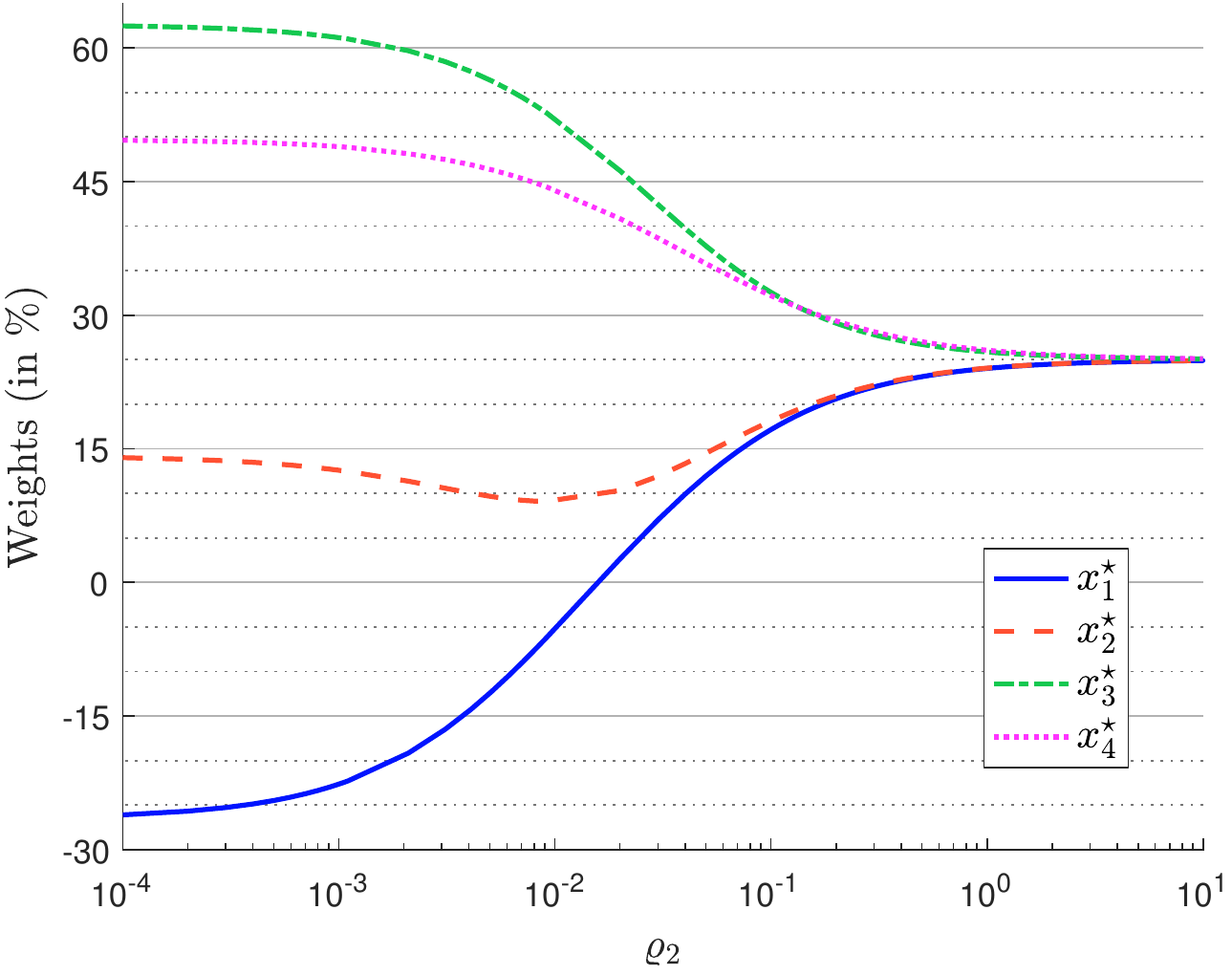}
\end{figure}

We consider an example\label{ex:example2} where the investment universe is
composed of 4 assets. The expected returns are equal to $\mu_1 = 4\%$, $\mu_2 =
5\%$, $\mu_3= 9\%$ and $\mu_4 = 10\%$ whereas the volatilities are equal to
$\sigma_1 = 15\%$, $\sigma_2 = 18\%$, $\sigma_3 = 20\%$ and $\sigma_4 = 25\%$.
The correlation matrix is the following:
\begin{equation*}
\mathcal{C} = \left(
\begin{array}{rrrr}
 1.00 &       &       &      \\
 0.70 &  1.00 &       &      \\
 0.10 &  0.10 &  1.00 &      \\
-0.20 & -0.20 & -0.70 & 1.00
\end{array}
\right)
\end{equation*}
We assume that $\gamma = 0.25$ and the portfolio is fully invested. We impose
that the target portfolio $x_0$ is equal to $\left(40\%, 30\%, 20\%,
10\%\right)$. Figure \ref{fig:ridge1} show the optimal weights with respect to
the penalization factor $\varrho_2$. We verify that the optimized portfolio
converges to the target portfolio when $\varrho_2$ increases. When there is no
target portfolio, it converges to the equally-weighted portfolio (see Figure
\ref{fig:ridge2}). This result is due to the capital budget constraint. Indeed,
if we do not impose the constraint $\sum_{i=1}^n x_i = 1$, the ridge portfolio
converges to the zero solution $x^{\star} = \mathbf{0}$. We also notice that
the paths of weights are not necessarily monotonous (increasing or decreasing).
For instance, the weight of the second asset decreases when $\varrho_2$ is
small and increases when $\varrho_2$ is large.\smallskip

We notice that the ridge regularization impacts entirely the covariance matrix.
Indeed, the shrinkage volatilities are equal to $\sqrt{\sigma
_{i}^{2}+\varrho_2 }$ whereas the shrinkage correlation matrix is defined by:%
\begin{equation*}
\left[ \mathcal{C}\left(\varrho_2\right) \right] _{i,j}=\rho _{i,j}\frac{\sigma
_{i}\sigma _{j}}{\sqrt{\left( \sigma _{i}^{2}+\varrho_2 \right) \left( \sigma
_{j}^{2}+\varrho_2 \right) }}
\end{equation*}%
It follows that $\lim_{\varrho_2 \rightarrow \infty
}\mathcal{C}\left(\varrho_2\right)=I_{n}$. Since the volatilities tend to
$\infty $, the ridge regularization can be viewed as a shrinkage covariance
method between the input covariance
matrix $\Sigma $ and the identity matrix:%
\begin{equation*}
\Sigma \left(\alpha\right) = \alpha \Sigma +\left( 1-\alpha \right) I_{n}
\end{equation*}

\begin{figure}[tbph]
\centering
\caption{Impact of the parameter $\varrho$ on the correlation (diagonal ridge regularization)}
\label{fig:ridge3}
\figureskip
\includegraphics[width = \figurewidth, height = \figureheight]{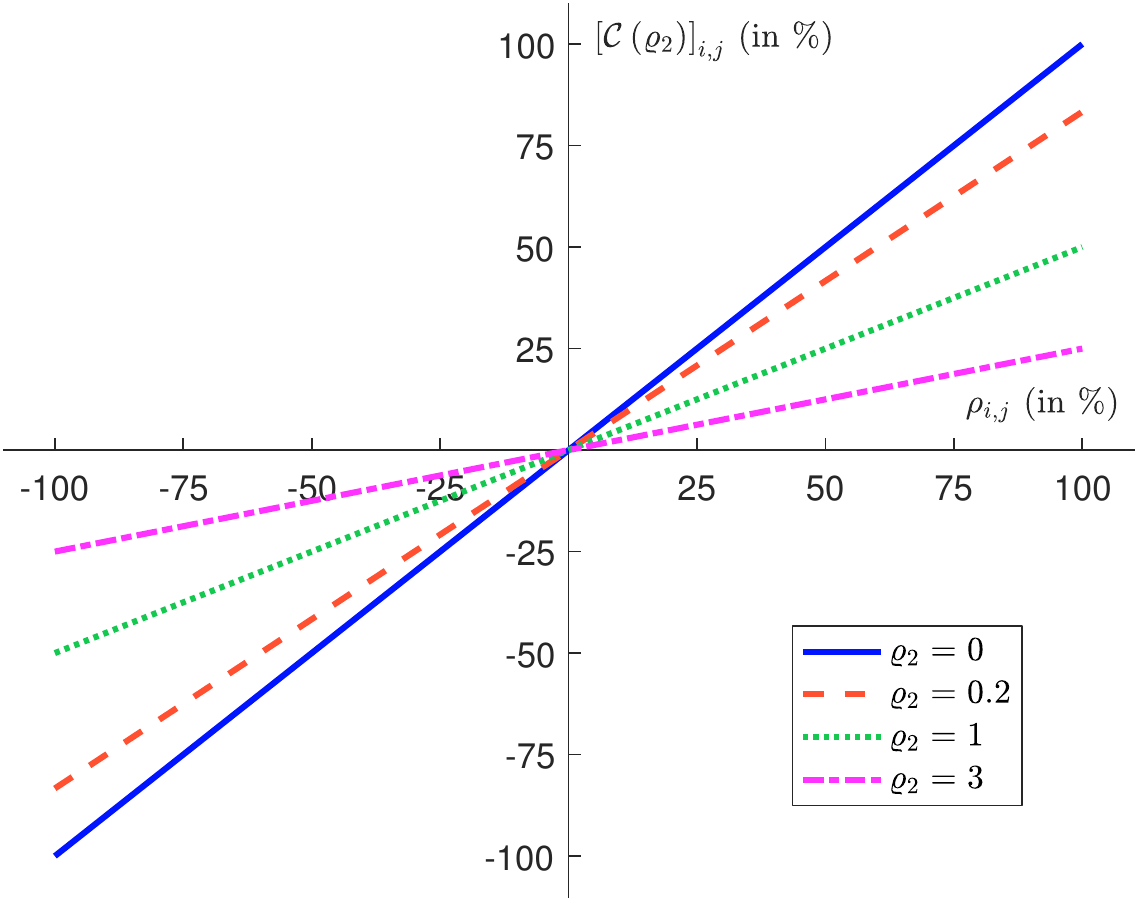}
\end{figure}

\begin{remark}
\label{remark:ridge} A variant of the ridge regularization is to
define $\Gamma_2 $ as a diagonal matrix. For instance, if $\Gamma_2
=\limfunc{diag}\Sigma $, the regularized
correlation matrix satisfies:%
\begin{equation*}
\left[ \mathcal{C}\left(\varrho_2\right)\right] _{i,j}=\frac{\rho _{i,j}}{1+\varrho_2 }
\end{equation*}
In Figure \ref{fig:ridge3}, we have reported the impact of the parameter
$\varrho_2$ on the correlation values.
\end{remark}

\subsection{Spectral filtering}

Spectral filtering is a general approach based on the singular value
decomposition (SVD) of the matrix $A_1$. Ridge regularization and denoising
techniques can be seen as special cases of the SVD method.

\subsubsection{General filters}

We consider the SVD decomposition of the matrix $A_{1}$ by assuming that
$\limfunc{rank}A_{1}=r$:
\begin{equation*}
A_{1}=USV^{\top }
\end{equation*}%
where the matrices\footnote{%
In the case of the empirical model, we have $U\in \mathbb{R}^{T\times r}$.}
$U\in \mathbb{R}^{n\times r}$, $V\in \mathbb{R}^{n\times r}$, $s=\left(
s_{1},\ldots ,s_{r}\right) \in \mathbb{R}^{r}$ and $S=\limfunc{diag}\left(
s\right) $ satisfy $U^{\top }U=V^{\top }V=I_{r}$ and $s_{k}\geqslant
s_{k+1}>0$. The Moore-Penrose pseudo-inverse of $A_{1}$ can be defined as:
\begin{equation*}
A_{1}^{\dagger }=VS^{-1}U^{\top }
\end{equation*}%
where $S^{-1}=\limfunc{diag}\left( s^{\dagger}\right) $. Let us denote $s_{\max
}\left( A_{1}\right) =s_{1}$ the largest singular value of
$A_{1}$.\smallskip

As instability is raised by small eigenvalues, filtering can be applied to keep
eigenvalues away from $0$. A filter $\mathcal{G}\left( s;\varrho \right)
=\left( G\left( s_{1};\varrho \right) ,\ldots ,G\left( s_{r};\varrho \right)
\right) $ is a vector-valued function, where the $k^{\mathrm{th}}$ entry
$G\left( s_{k};\varrho \right) :\left] 0,s_{\max }\left( A\right) \right]
\rightarrow \mathbb{R}$ satisfies:
\begin{equation*}
\lim_{\varrho \rightarrow 0}G\left( s_{k};\varrho \right) =\frac{1}{s_{k}}
\end{equation*}%
for all $\varrho \geqslant 0$ and $s_{k}\in \left] 0,s_{\max }\left( A\right)
\right] $. The parameter $\varrho $ controls the magnitude of the
regularization of $A_{1}^{\dagger }$:
\begin{equation*}
A_{1}^{\dagger }\left( \varrho \right) =V\limfunc{diag}\left( \mathcal{G}%
\left( s;\varrho \right) \right) U^{\top }
\end{equation*}%
As a consequence, we verify the property of convergence:%
\begin{equation*}
\lim_{\varrho \rightarrow 0}A_{1}^{\dagger }\left( \varrho \right)
=A_{1}^{\dagger }
\end{equation*}%
\smallskip

This method can be extended to regularize the matrix $A_{1}^{\top }A_{1}$. On
one hand, if $A_{1}$ has full rank, we can approximate $Q=A_{1}^{\top }A_{1}$
by $A_{1}^{\top }\left( A_{1}^{\dagger }\left( \varrho \right) \right) ^{-1}$.
On the other hand, a direct computation leads
to $Q=A_{1}^{\top }A_{1}=VS^{2}V^{\top }$. Therefore, we can regularize $%
Q=A_{1}^{\top }A_{1}$ by:
\begin{equation*}
Q\left( \varrho \right) =V\limfunc{diag}\left( s^{2}\left( \varrho \right)
\right) V^{\top }
\end{equation*}%
where $s^{2}\left( \varrho \right) $ is a vector that may be equal to $%
\mathcal{G}\left( s;\varrho \right) ^{\dagger }\, \odot\, \mathcal{G}\left(
s;\varrho \right) ^{\dagger }$, or $\mathcal{G}\left( s\odot s;\varrho \right)
^{\dagger }$ or $\mathcal{G}\left( s;\varrho \right) ^{\dagger
}\,\odot\, s$. Once again, we have the convergence property:%
\begin{equation*}
\lim_{\varrho \rightarrow 0}Q\left( \varrho \right) =A_{1}^{\top }A_{1}
\end{equation*}%
If we consider the problem:%
\begin{eqnarray*}
x^{\star } &=&\arg \min \frac{1}{2}\left\Vert A_{1}x-b_{1}\right\Vert
_{2}^{2} \\
&\text{s.t.}&A_{2}x=b_{2}
\end{eqnarray*}%
the normal equations are:%
\begin{equation}
\left(
\begin{array}{cc}
A_{1}^{\top }A_{1} & A_{2}^{\top } \\
A_{2} & \mathbf{0}%
\end{array}%
\right) \left(
\begin{array}{c}
x \\
\lambda
\end{array}%
\right) =\left(
\begin{array}{c}
A_{1}^{\top }b_{1} \\
b_{2}%
\end{array}%
\right)   \label{eq:spectral1}
\end{equation}%
Spectral filtering is then equivalent to replacing the linear system (\ref%
{eq:spectral1}) by the following set of normal equations:%
\begin{equation}
\left(
\begin{array}{cc}
V\limfunc{diag}\left( s^{2}\left( \varrho \right) \right) V^{\top } &
A_{2}^{\top } \\
A_{2} & \mathbf{0}%
\end{array}%
\right) \left(
\begin{array}{c}
x \\
\lambda
\end{array}%
\right) =\left(
\begin{array}{c}
U\limfunc{diag}\left( \mathcal{G}\left( s;\varrho \right) ^{\dagger }\right)
V^{\top }b_{1} \\
b_{2}%
\end{array}%
\right)   \label{eq:spectral2}
\end{equation}

\subsubsection{Application to Tikhonov regularization}

To define the spectral regularization of the Tikhonov problem, the matrices
$A_{1}$ and $\Gamma_2 $ have to be able to be factored in a coherent way:
\begin{equation*}
A_{1}=US_{1}V^{\top }
\end{equation*}%
and:%
\begin{equation*}
\Gamma _{2}=WS_{2}V^{\top }
\end{equation*}%
Direct computations gives:%
\begin{equation*}
A_{1}^{\top }A_{1}+\varrho _{2}\Gamma _{2}^{\top }\Gamma _{2}=V\left(
S_{1}^{2}+\varrho _{2}S_{2}^{2}\right) V^{\top }
\end{equation*}%
We deduce that the $k^{\mathrm{th}}$ entry of the spectral filter
$\mathcal{G}\left( s_{1};\varrho _{2}\right) $ is defined by:
\begin{equation*}
G\left( s_{1,k};\varrho _{2}\right) =\frac{s_{1,k}}{s_{1,k}^{2}+\varrho
_{2}s_{2,k}^{2}}
\end{equation*}%
Using the previous notations, we have:%
\begin{eqnarray*}
\varrho _{2}\Gamma _{2}^{\top }\Gamma _{2} &=&A_{1}^{\top }A_{1}+\varrho
_{2}\Gamma _{2}^{\top }\Gamma _{2}-A_{1}^{\top }A_{1} \\
&=&V\limfunc{diag}\left( s_{1}^{2}\left( \varrho _{2}\right) \right) V^{\top
}-V\limfunc{diag}\left( s_{1}\odot s_{1}\right) V^{\top } \\
&=&V\limfunc{diag}\left( s_{1}^{2}\left( \varrho _{2}\right)
-s_{1}^{2}\right) V^{\top }
\end{eqnarray*}%
where $s_{1}^{2}=s_{1}\odot s_{1}$. In this case, the optimal portfolio
$x^{\star }$ is the $x$-coordinate of the solution to the linear system:
\begin{equation}
\left(
\begin{array}{cc}
V\limfunc{diag}\left( s_{1}^{2}\left( \varrho _{2}\right) \right) V^{\top }
& A_{2}^{\top } \\
A_{2} & \mathbf{0}%
\end{array}%
\right) \left(
\begin{array}{c}
x \\
\lambda
\end{array}%
\right) =\left(
\begin{array}{c}
A_{1}^{\top }b_{1}+V\limfunc{diag}\left( s_{1}^{2}\left( \varrho
_{2}\right) -s_{1}^{2}\right) V^{\top }x_{0} \\
b_{2}%
\end{array}%
\right)   \label{eq:tikhonov3}
\end{equation}%
We notice that only the right singular vectors appear in Equation
(\ref{eq:tikhonov3}). Ridge regularization can be viewed as particular
filters\footnote{For $\Gamma _{2}=0$, we have $G\left( s_{1,k};\varrho
_{2}\right) =s_{1,k}^{\dagger }$. For $\Gamma _{2}=I_{n}$ (ridge
regularization), the $k^{\mathrm{th}}$ entry of the spectral filter
$\mathcal{G}\left( s_{1};\varrho _{2}\right) $ is defined by:
\begin{equation*}
G\left( s_{1,k};\varrho _{2}\right) =\frac{s_{1,k}}{s_{1,k}^{2}+\varrho _{2}}
\end{equation*}%
}. More generally, when $A_{1}$ and $\Gamma _{2}$ have the same right
singular vectors, Tikhonov regularization can be stated in terms of a
filter.\smallskip

\begin{figure}[t]
\centering
\caption{Spectral filtering (ridge regularization and denoising method)}
\label{fig:spectral1}
\figureskip
\includegraphics[width = \figurewidth, height = \figureheight]{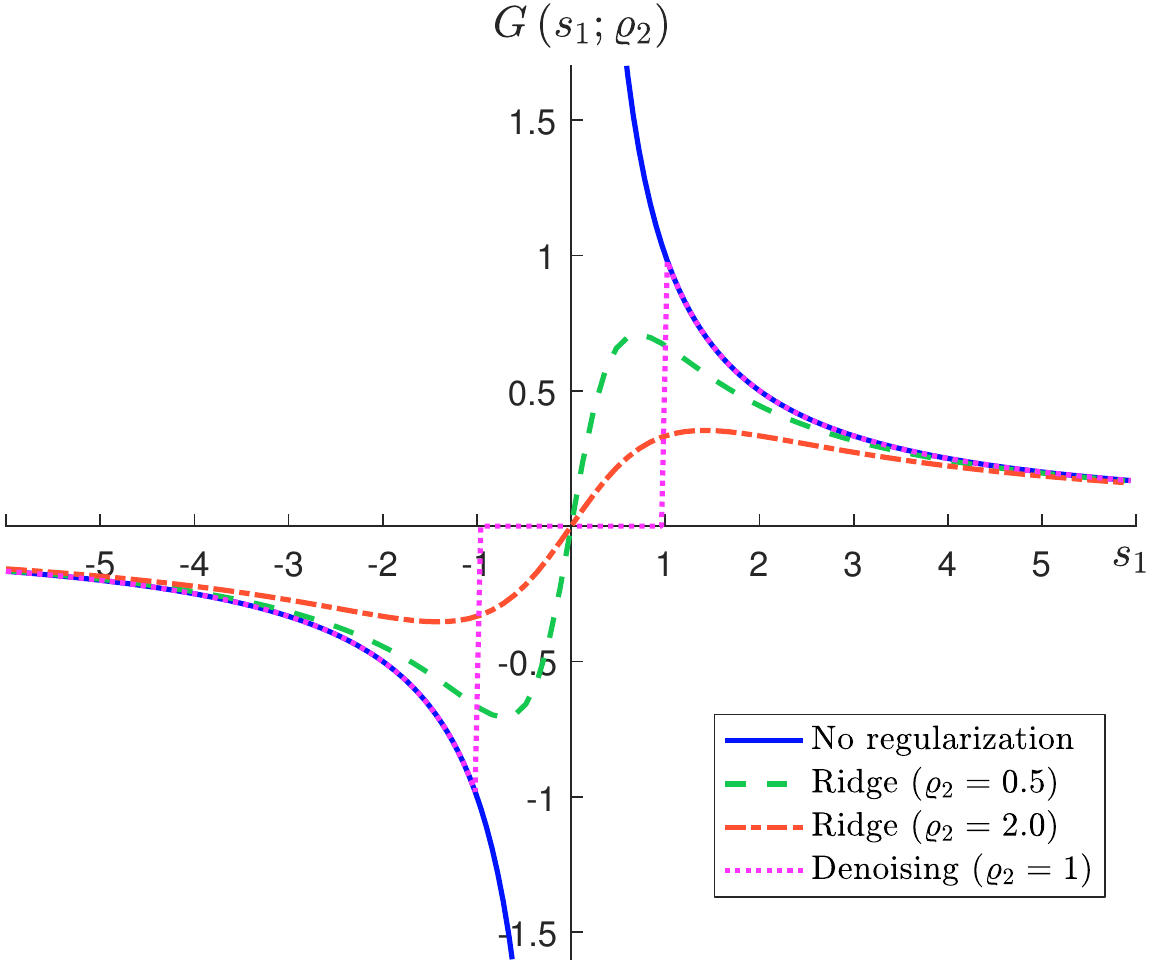}
\end{figure}

In Figure \ref{fig:spectral1}, we report the spectral filter of the ridge
regularization. The spectral filtering approach includes another popular
method, which is the denoising method (Laloux \textsl{et al.}, 1999):
\begin{equation*}
G\left( s_{1,k};\varrho _{2}\right) =\mathbf{1}\left\{ \left\vert
s_{1,k}\right\vert \geqslant \varrho _{2}\right\} \cdot s_{1,k}^{\dagger }
\end{equation*}%
We notice that deleting singular values is equivalent to applying a hard
thresholding method while ridge regularization is a smoothing approach.

\subsubsection{Improvement of the stability condition}

The condition number $\kappa \left( A\right) $ of the matrix $A$ summarizes the
level of difficulty when performing the optimization in a stable way. More
specifically, it measures how much an error on the vector $b$ changes the solution
of the linear equation $Ax=b$. We have:
\begin{equation*}
\kappa \left( A\right) =\left\Vert A^{\dagger }\right\Vert \cdot \left\Vert
A\right\Vert
\end{equation*}%
It follows that $\kappa \left( A^{\dagger }\right) =\kappa \left( A\right) $,
and we have the property $\kappa \left( A\right) \geqslant 1$. When $\kappa \left(
A\right) $ is low, the problem is numerically stable and easy to solve. The
closer to one, the better the stability.\smallskip

With the $L_{\infty}$ norm, we obtain:
\begin{equation}
\kappa \left( A\right) =\frac{\max_{k}\left\vert s_{k}\right\vert }{%
\min_{k}\left\vert s_{k}\right\vert }  \label{eq:cond1}
\end{equation}%
where the $s_{k}$'s are the singular values of $A$. Using the filter
$\mathcal{G}\left(s;\varrho\right)$, we obtain:
\begin{equation}
\kappa \left( A^{\dagger }\left(\varrho\right)\right) =\frac{\min_{k}\left\vert G\left( s_{k}; \varrho\right) \right\vert}%
{\max_{k}\left\vert G\left( s_{k}; \varrho \right) \right\vert}
\label{eq:cond2}
\end{equation}%
For a fixed value of $\varrho >0$, all previous filters satisfy the two
following properties:

\begin{enumerate}
\item $G\left( s_k; \varrho\right) \sim s_k^{-1}$ for $s_k\rightarrow \infty
    $;

\item $G\left( s_k; \varrho\right) $ is bounded from above on $\left[
    0,+\infty \right) $.
\end{enumerate}
As a consequence, if we compare Equations (\ref{eq:cond1}) and
(\ref{eq:cond2}), the denominator is essentially unchanged while the numerator
is decreased\footnote{From an unbounded function to a bounded function.}.
Therefore, spectral filtering decreases the condition number of $A$, because
these techniques reduce the dispersion of singular values.

\subsection{Mixed penalties}

The Euclidian regularization is natural because the $L_{2}$ norm
appears in Problem (\ref{eq:qp1}). Explicit formulas are obtained, and can be
implemented at once. Other regularization techniques have been introduced to
impose other constraints on the optimal solution $x^{\star }$. As the unit ball
for the $L_{1}$ norm is not uniformly convex, sparse solutions may be obtained
by penalizing with $L_{1}$ instead of $L_{2}$.

\subsubsection{$L_{p}$ regularization}

Instead of Tikhonov regularization, one may consider the $L_{p}$
regularization:
\begin{eqnarray}
x^{\star } &=&\arg \min \frac{1}{2}\left\Vert A_{1}x-b_{1}\right\Vert
_{2}^{2}+\frac{1}{p}\varrho _{p}\left\Vert \Gamma _{p}\left( x-x_{0}\right)
\right\Vert _{p}^{p}  \label{eq:mixed1} \\
&\text{s.t. }&A_{2}x=b_{2}  \notag
\end{eqnarray}%
where $x_{0}\in \mathbb{R}^{n}$ is a targeted portfolio and $p>0$%
.\smallskip

For $p>1$, the function $\Gamma _{p}\left( x\right) =\left\Vert \Gamma
_{p}\left( x-x_{0}\right) \right\Vert _{p}^{p}$ is strictly convex and its
gradient is Lipschitz continuous. Indeed, the gradient is equal $p\Gamma _{p}^{\top }%
\limfunc{sign}\left( \Gamma _{p}\left( x-x_{0}\right) \right) \odot \left\vert
\Gamma _{p}\left( x-x_{0}\right) \right\vert ^{p-1}$, where the functions
$\limfunc{sign}\left( x\right) $ and $\left\vert x\right\vert $ are taken
component wise. For $p=1$, the function $\Gamma _{1}\left(
x\right) $ is convex, lower semi-continuous but may not be differentiable at $%
x=x_{0}$. An explicit expression for its subgradient can be formulated in
terms of proximal operators. For $p\in \left] 0,1\right[ $, the function $%
\Gamma _{p}\left( x\right) $ is not convex, and Problem (\ref{eq:mixed1}) is
not convex.\smallskip

\begin{figure}[tbph]
\centering
\caption{Lasso regularization with a target portfolio}
\label{fig:lasso1}
\figureskip
\includegraphics[width = \figurewidth, height = \figureheight]{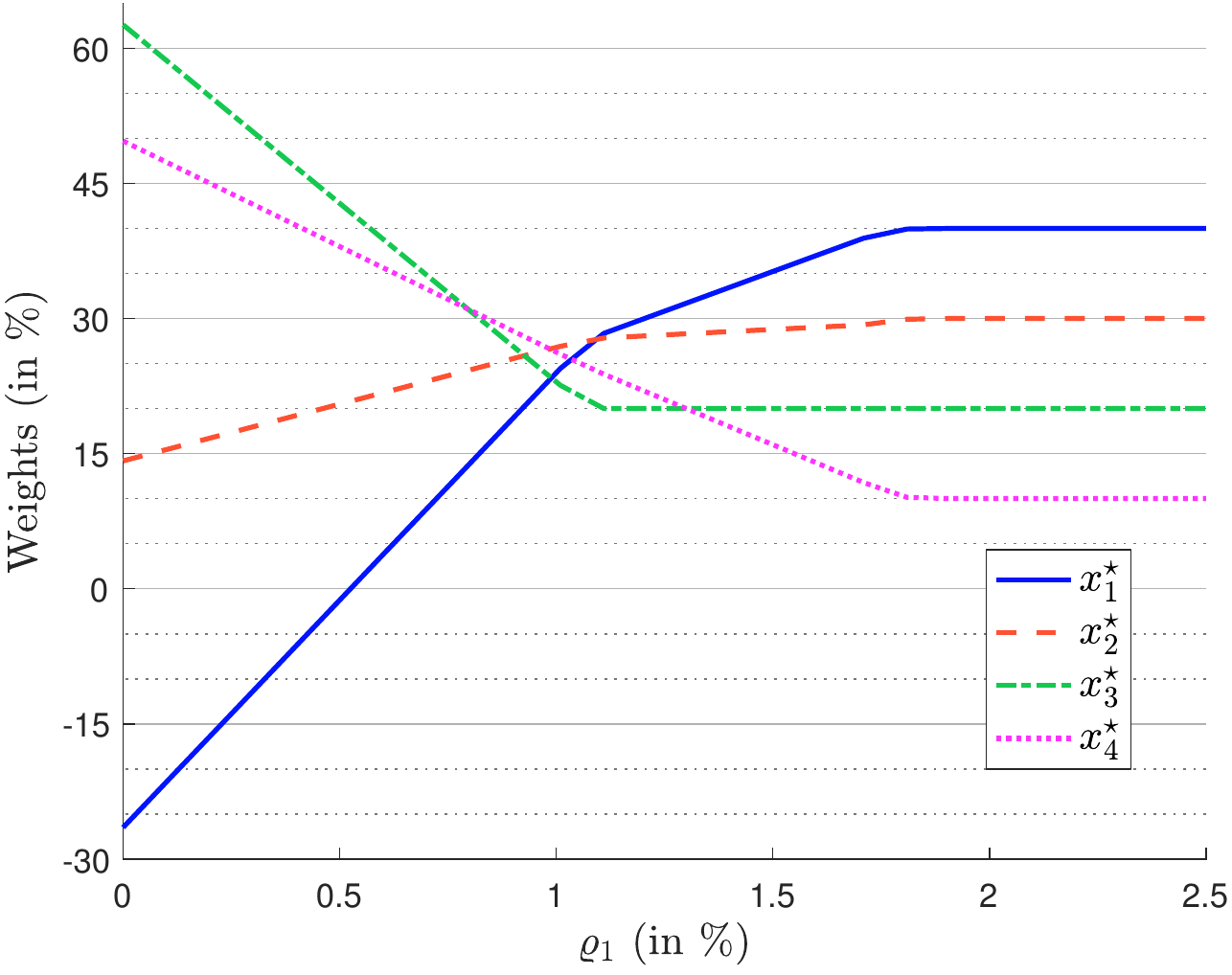}
\end{figure}

\begin{figure}[tbph]
\centering
\caption{Lasso regularization without a target portfolio}
\label{fig:lasso2}
\figureskip
\includegraphics[width = \figurewidth, height = \figureheight]{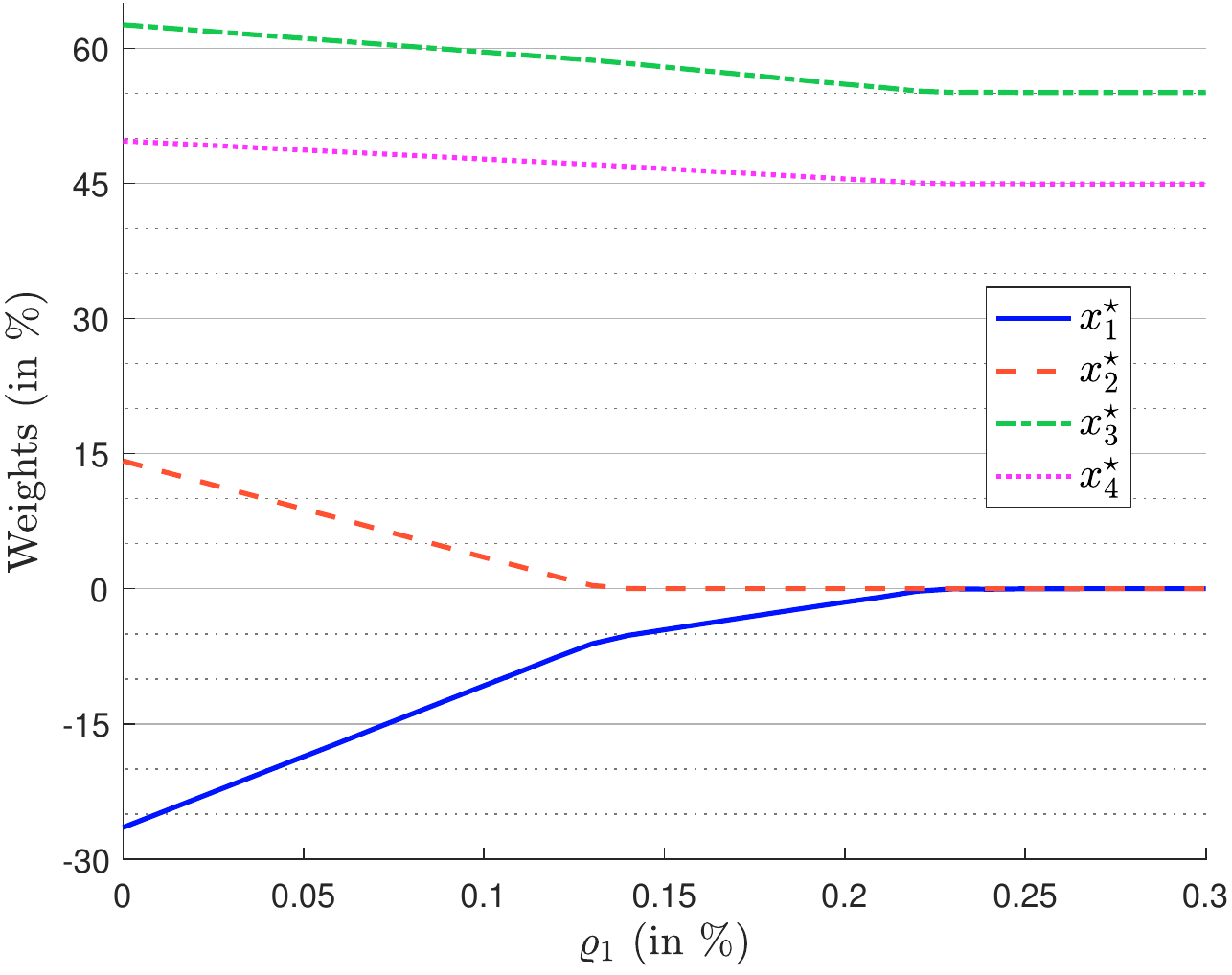}
\end{figure}

The penalties $L_{p}$ for $p\geqslant 1$ are used for regularization, while the
penalties $L_{p}$ for $p\leqslant 1$ are used for sparsity.
The case $p=1$ is the most interesting since it corresponds to the lasso
regression (Tibshirani, 1996). In this case, a large value of $\varrho _{1}$
associated with the constraint $\mathbf{1}^{\top }x=1$ forces the optimal
portfolio to have long-only positions (Brodie \textsl{et al.}, 2009).\smallskip

We consider the example given on page \pageref{ex:example2}. We use a $L_1$ (or
lasso) penalty with $\Gamma_1 = I_n$. Figure \ref{fig:lasso1} show the optimal
weights with respect to the penalization factor $\varrho_1$. Like in the ridge
approach, the optimized portfolio converges to the optimal portfolio when the
parameter $\varrho$ increases. When there is no target portfolio, we observe a
divergence of the limit portfolio between ridge and lasso approaches. While the
ridge portfolio converges to the equally-weighted portfolio, the lasso
portfolio converges to the long-only mean-variance optimized portfolio (Figure
\ref{fig:lasso2}). If we compare Figures \ref{fig:ridge1} and \ref{fig:lasso1},
we notice that the magnitude of the regularization factor is not the same. We
also observe that the paths are different. The path is smoothed and continuous
for the ridge approach, while it is more a piecewise linear function for the
lasso approach. We verify that the $L_1$ penalty produces a sparse optimized
portfolio. This is obvious for the case where there is no target portfolio
since weights may be equal to zero. When there is a target portfolio, the
sparsity concerns the bets between the optimized portfolio $x^{\star}$ and the
target portfolio $x_0$. In this case, relative (and not absolute) weights are
equal to zero. Another difference between the two approaches is that the lasso
method produces a monotonic path (decreasing or increasing) contrary to the
ridge method.

\subsubsection{$L_{1}-L_{2}$ regularization}

We can also consider a mixed penalty:
\begin{eqnarray}
x^{\star } &=&\arg \min \frac{1}{2}\left\Vert A_1 x - b_1\right\Vert
_{2}^{2}+\varrho _{p}\left\Vert \Gamma _{p}\left( x-x_{0}\right) \right\Vert
_{p}^{p}+\frac{1}{2}\varrho _{2}\left\Vert \Gamma _{2}\left( x-x_{0}\right)
\right\Vert _{2}^{2}  \label{eq:mixed2} \\
&\text{s.t. }&A_{2}x=b_{2}  \notag
\end{eqnarray}%
where $p\neq 2$. In the case $p=1$, we obtain:%
\begin{eqnarray}
x^{\star } &=&\arg \min \frac{1}{2}\left\Vert A_1 x - b_1\right\Vert
_{2}^{2}+\varrho _{1}\left\Vert \Gamma _{1}\left( x-x_{0}\right) \right\Vert
_{1}+\frac{1}{2}\varrho _{2}\left\Vert \Gamma _{2}\left( x-x_{0}\right)
\right\Vert _{2}^{2}  \label{eq:mixed3} \\
&\text{s.t. }&A_{2}x=b_{2}  \notag
\end{eqnarray}%
This regularization is called elastic net (Hastie \textsl{et al.}, 2009). This
is the most common mixed penalty used in portfolio optimization (Roncalli,
2013).\smallskip

\begin{figure}[tbph]
\centering
\caption{Mixed regularization with a target portfolio}
\label{fig:mixed1}
\figureskip
\includegraphics[width = \figurewidth, height = \figureheight]{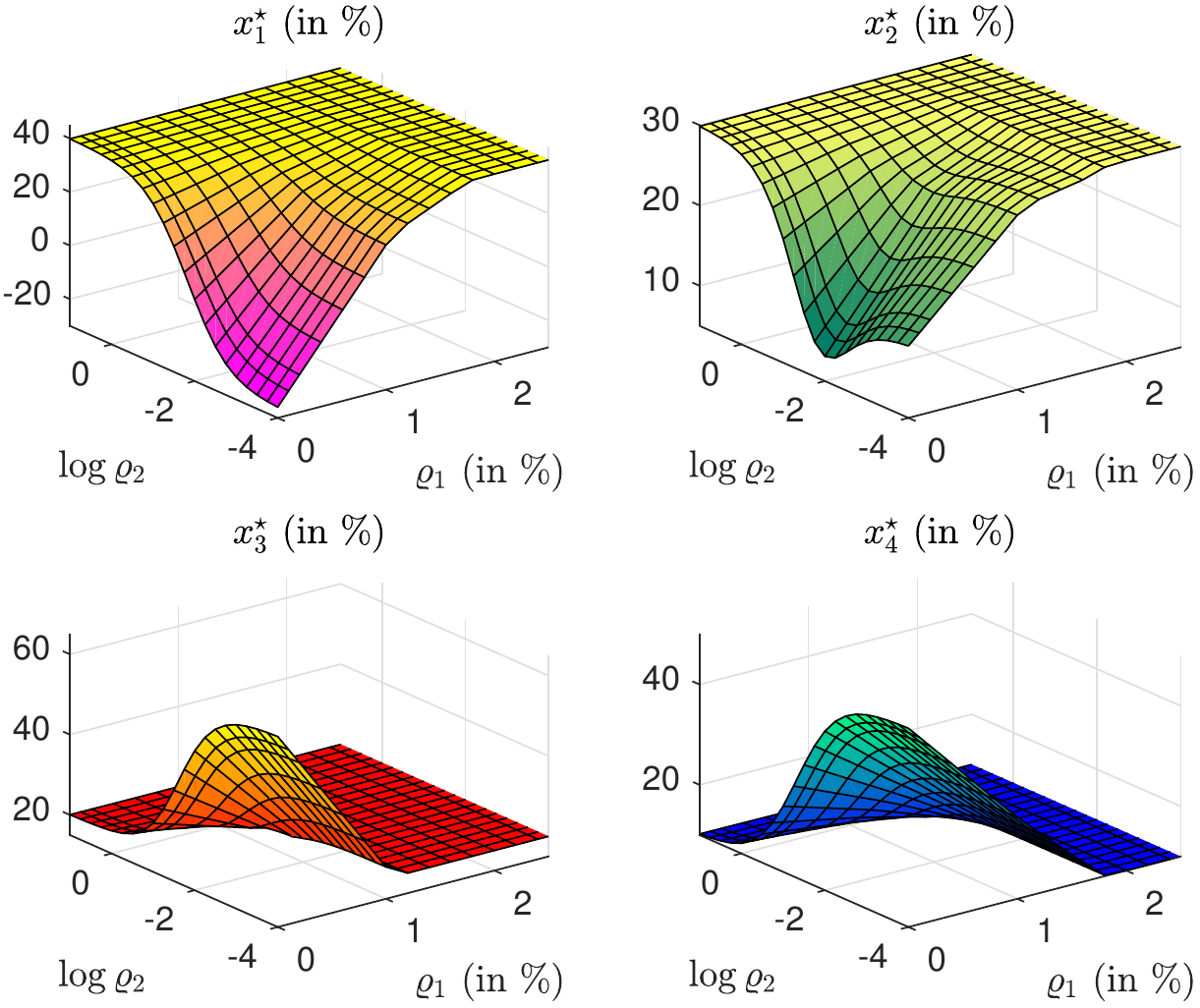}
\end{figure}

\begin{figure}[tbph]
\centering
\caption{Mixed regularization without a target portfolio}
\label{fig:mixed2}
\figureskip
\includegraphics[width = \figurewidth, height = \figureheight]{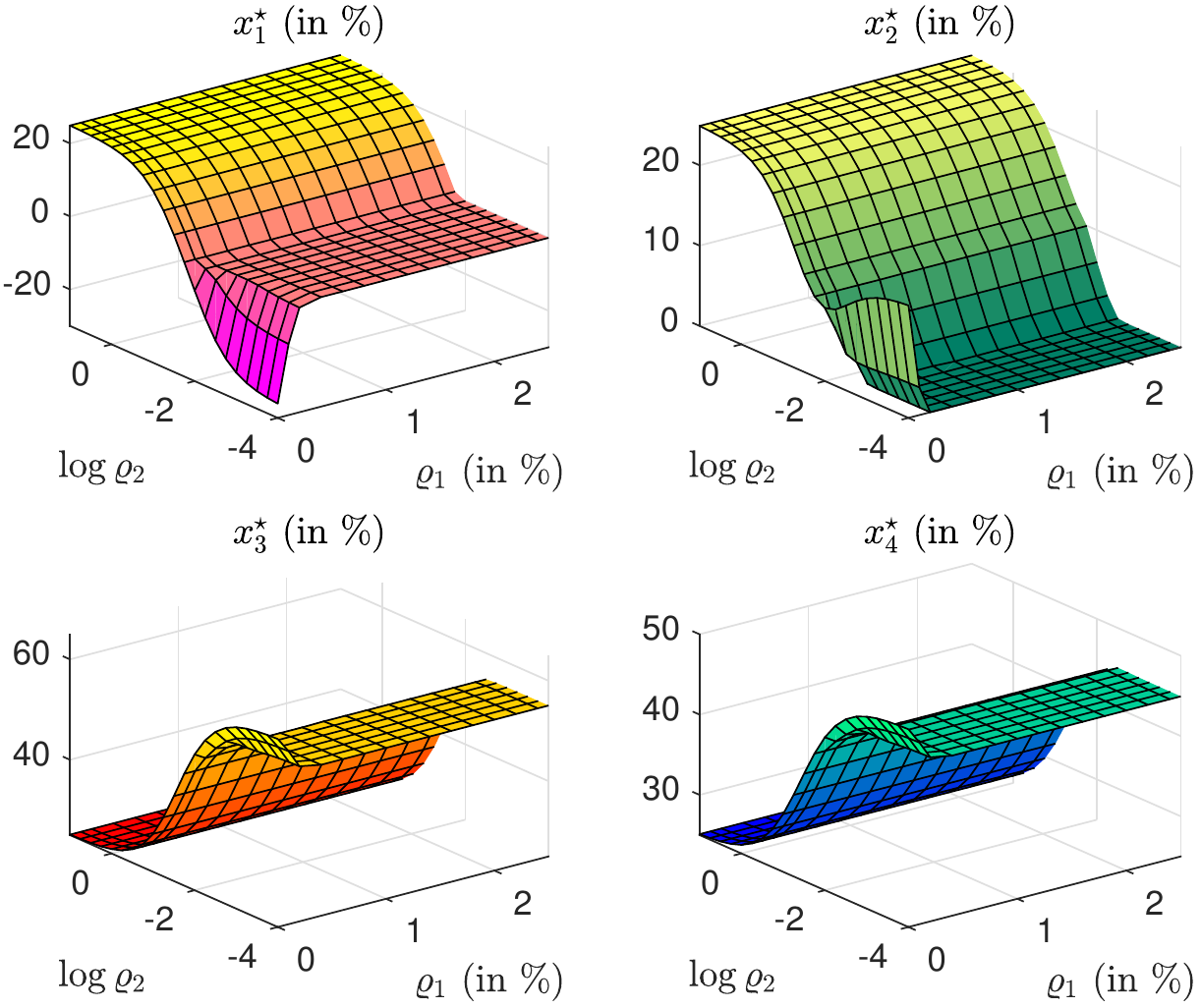}
\end{figure}

We consider again the example given on page \pageref{ex:example2}. We use a
lasso-ridge penalty with $\Gamma_1 = \Gamma_2 = I_n$. Results are reported in
Figures \ref{fig:mixed1} and \ref{fig:mixed2}. We notice a large difference
concerning the convergence. Indeed, we recall that the lasso and ridge
approaches converge to the same portfolio when we impose a target portfolio,
but to two different portfolios when there is no target portfolio. When mixing
the two norms, the limit portfolio is generally the ridge portfolio, because of the magnitude
of $\varrho_1$ and $\varrho_2$ in portfolio management (see Appendix \ref{appendix:section-limit-solution}
on page \pageref{appendix:section-limit-solution}). This result
is true because we have imposed $\Gamma_1 = \Gamma_2 = I_n$.

\subsubsection{Solving the mixed penalty problem}

Problems (\ref{eq:mixed2}) and (\ref{eq:mixed3}) are more complex to solve
than a traditional quadratic programming problem. In the case of the $%
L_{1}-L_{2}$ regularization problem and if we assume that $\Gamma
_{1}$ is a matrix with non-negative entries\footnote{Which is generally the case
(Bruder \textsl{et al.}, 2013; Roncalli, 2013).}, we can use a
modified QP solver. The underlying idea is to write $\Gamma
_{1}\left( x\right) $ in the following way:%
\begin{equation*}
\Gamma _{1}\left( x\right) =\mathbf{1}^{\top }\Gamma _{1}\delta ^{-}+\mathbf{%
1}^{\top }\Gamma _{1}\delta ^{+}
\end{equation*}%
where $\delta ^{-}=\max \left( \mathbf{0},x_{0}-x\right) $ and $\delta
^{+}=\max \left( \mathbf{0},x-x_{0}\right) $. Therefore we obtain a standard
QP problem by augmenting the vector of unknown variables\footnote{%
See Appendix \ref{appendix:section-qprog} on page
\pageref{appendix:section-qprog} for a comprehensive presentation.}.
Thus, the optimization is performed with respect to $y=\left(
x,\delta ^{-},\delta ^{+}\right) $ and no longer with respect to
$x$. In the other cases, when we consider an $L_{p}$ penalty with
$p\neq 2$ or when $\Gamma _{1}$ is a matrix with some negative entries, the
general approach is to use the ADMM algorithm, which is
described in Appendix \ref{appendix:section-admm} on page \pageref%
{appendix:section-admm}. For instance, Problem (\ref{eq:mixed3}) can be
written as:
\begin{eqnarray*}
\left\{ x^{\star },z^{\star }\right\} &=&\arg \min f\left( x\right) +g\left(
z\right) \\
&\text{s.t.}&x-z=\mathbf{0}
\end{eqnarray*}%
where:
\begin{equation*}
f\left( x\right) =\frac{1}{2}\left\Vert A_{1}x-b_{1}\right\Vert _{2}^{2}+%
\frac{1}{2}\varrho _{2}\left\Vert \Gamma _{2}\left( x-x_{0}\right)
\right\Vert _{2}^{2}+\mathds{1}_{\Omega }\left( x\right)
\end{equation*}%
and:%
\begin{eqnarray*}
g\left( z\right) &=&\varrho _{1}\Gamma _{1}\left( x\right) \\
&=&\varrho _{1}\left\Vert \Gamma _{1}\left( x-x_{0}\right) \right\Vert _{1}
\end{eqnarray*}%
where $\Omega =\left\{ x\in \mathbb{R}^{n}:A_{2}x=b_{2}\right\} $. The
interest of this choice is that the $x$-step includes the constraint
and can be explicitly computed\footnote{%
We have%
\begin{equation*}
x^{\left( k+1\right) }=\arg \min \left\{ f\left( x\right) +\frac{\varphi }{2}%
\left\Vert x+z^{\left( k\right) }u^{\left( k\right) }\right\Vert
_{2}^{2}\right\}
\end{equation*}%
}, while the $z$-step requires to compute the proximal operator of the
function $\Gamma _{1}\left( x\right) $:%
\begin{equation*}
z^{\left( k+1\right) }=\arg \min \left\{ g\left( z\right) +\frac{\varphi }{2}%
\left\Vert x^{\left( k+1\right) }-z+u^{\left( k\right) }\right\Vert
_{2}^{2}\right\}
\end{equation*}%
The update of the scaled dual variable is:%
\begin{equation*}
u^{\left( k+1\right) }=u^{\left( k\right) }+\left( x^{\left( k+1\right)
}-z^{\left( k+1\right) }\right)
\end{equation*}%
The previous results can be extended when $p\neq 1$ and $\Omega $ is a set
of more complex constraints.

Appendices \ref{appendix:section-admm} and \ref{appendix:section-proximal} on
pages \pageref{appendix:section-admm}--\pageref%
{appendix:section-proximal-end} contain all the information for solving the
following optimization problem:
\begin{eqnarray*}
x^{\star } &=&\arg \min \frac{1}{2}\left\Vert A_{1}x-b_{1}\right\Vert
_{2}^{2}+\varrho _{p}\left\Vert \Gamma _{p}\left( x-x_{0}\right) \right\Vert
_{p}^{p}+\frac{1}{2}\varrho _{2}\left\Vert \Gamma _{2}\left( x-x_{0}\right)
\right\Vert _{2}^{2} \\
&\text{s.t. }&x\in \Omega
\end{eqnarray*}%
where $\Omega $ may be equality, inequality, bound and $L_{q}$ norm
constraints.

\subsection{Optimal choice of the regularization factor}
\label{section:cross-validation}

To choose the optimal regularization parameter, we first have to define an
optimization criterion. For instance, the optimal value of $\varrho _{1}$ or
$\varrho _{2}$ is generally obtained by cross-validation techniques. Exhaustive
methods such as leave-$p$-out cross-validation (LpOCV) or leave-one-out
cross-validation (LOOCV) are computationally intensive. This is why it may be
better to use non-exhaustive methods such as $k$-fold cross-validation or
out-of-sample testing. However, in the case of the Tikhonov regularization, an
explicit formula is known. Indeed, the generalized cross-validation procedure for
choosing $\varrho _{2}$ does not depend on the dual variable or the
constraints. In the case of the $L_{1}$ penalty, no explicit formula is known and
the brute force algorithm must be used for finding the optimal value of $\varrho _{1}$.

\subsubsection{Cross-validation and the PRESS statistic}

Let us consider the data matrix $X=\left( x_{1}^{\top },\ldots ,x_{T}^{\top
}\right) \in \mathbb{R}^{T\times K}$ where $x_{t}\in \mathbb{R}^{K}$, and a
response vector $Y=\left( y_{1},\ldots ,y_{T}\right) \in \mathbb{R}^{T}$ where
$y_{t}\in \mathbb{R}$. Since the Tikhonov regularization problem is
defined as follows:%
\begin{equation*}
\hat{\beta}=\arg \min \frac{1}{2}\left\Vert Y-X\beta \right\Vert _{2}^{2}+%
\frac{1}{2}\varrho _{2}\left\Vert \Gamma _{2}\beta \right\Vert _{2}^{2}
\end{equation*}%
we have:%
\begin{equation*}
\hat{\beta}=S\left( \varrho _{2}\right) X^{\top }Y
\end{equation*}%
where:%
\begin{equation*}
S\left( \varrho _{2}\right) =\left( X^{\top }X+\varrho _{2}\Gamma _{2}\Gamma
_{2}^{\top }\right) ^{-1}
\end{equation*}
It follows that $\hat{\beta}$ is a function of $\varrho _{2}$. Therefore,
the underlying idea is to find the optimal value $\hat{\varrho}_{2}$%
.\smallskip

In order to accurately estimate the hyperparameters of the model and to avoid
overfitting problems, the cross-validation (CV) method comprises several steps:

\begin{enumerate}
\item the sample of data is partitioned into two sets, the training set and
    the test (or validation) set;

\item the model is fitted on the training set;

\item the model is tested on the validation set.
\end{enumerate}

\noindent In order to reduce variability, steps 2 and 3 are performed using
different partitions of the data sample (step 1). The validation
results are combined, according to a measure of fit, to give an estimate of the
model predictive performance. The hyperparameters are then chosen in order to
maximize this goodness-of-fit measure. Two types of CV may be performed: exhaustive
and non-exhaustive cross-validation. For the first type, the model is estimated
and tested on all possible ways to divide the original sample into
training/test sets. This type of CV consists of the leave-$p$-out cross
validation (LpOCV). In this approach, $p$ observations are used in the test set
and the remaining observations are used in the
training set\footnote{%
The leave-one-out cross validation (LOOCV) procedure corresponds to the
special case $p=1$.}. This requires training and validating the model $%
\binom{T}{p}$ times, which can be extremely expensive if $T$ is large, even for
$p=1$. Nevertheless, an explicit expression for the sum of squares of the
errors is known in the case of Tikhonov regression. This formula may lead to
$O\left( T\right) $ operations. For this reason, non-exhaustive
cross-validation may be preferred in practice, such as $k$-fold CV, holdout
method, repeated random sub-sampling, jackknife, etc. Performing $k$-fold CV is
the most popular tool for model selection (Stone, 1974; Wahba, 1977; Stone,
1978).\smallskip

In $k$-fold CV, the sample of data is randomly shuffled and split into $k$
(almost) equally sized groups, the model is fitted using all but the $j^{%
\mathrm{th}}$ group of data, and the $j^{\mathrm{th}}$ group of data is used
for the test set. We repeat the procedure $k$ times, in such a way that each
group is tested exactly once. The $k$-fold cross validated error is
generally computed as:
\begin{equation*}
\mathcal{E}_{\mathrm{cv}}=\frac{1}{T}\sum_{j=1}^{k}\sum_{t\in \mathcal{G}%
_{k}}\left( y_{t}-x_{t}^{\top }\beta \left( k\right) \right) ^{2}
\end{equation*}%
where $t\in \mathcal{G}_{k}$ denotes the observations of the $k^{\mathrm{th}} $
group and $\beta \left( k\right) $ the estimation of $\beta $ obtained by
leaving out the $k^{\mathrm{th}}$ group. Even in simple cases, it cannot be
guaranteed that the function $\mathcal{E}_{\mathrm{cv}}$ has a unique minimum.
The simple grid search approach is probably the best approach. The
exhaustive Leave-one-out cross validation (LOOCV) is a particular case when $%
k$ is equal to the size of the dataset. The LOOCV is asymptotically equivalent
to Akaike Information Criterion (AIC), which is commonly used in statistics
(Stone, 1977). Interestingly, For Tikhonov regression, the cross validated
error $\mathcal{E}_{\mathrm{cv}}$ has an explicit expression known as the
Predicted Sum of Squares (or PRESS) statistic (Allen, 1971 \& 1974).\smallskip

We note $Y_{-t}$ and $X_{-t}$ the $\left( T-1\right) $ vector and $\left(
T-1\right) \times K$ matrix by leaving out the $t^{th}$ observation to the
vector $Y$ and the matrix $X$. We have:%
\begin{equation*}
\hat{\beta}_{-t}=\left( X_{-t}^{\top }X_{t}+\varrho _{2}\Gamma _{2}\Gamma
_{2}^{\top }\right) ^{-1}X_{-t}^{\top }Y_{-t}
\end{equation*}%
The explicit expression for the LOOCV procedure is\footnote{%
Proof is given in Appendix \ref{appendix:press} on page
\pageref{appendix:press}.}:
\begin{eqnarray*}
\mathcal{P}\mathrm{ress}\left( \varrho _{2}\right)  &=&\sum_{t=1}^{T}\left(
y_{t}-x_{t}^{\top }\hat{\beta}_{-t}\right) ^{2} \\
&=&\sum_{t=1}^{T}\left( 1-x_{t}^{\top }S\left( \varrho _{2}\right)
x_{t}\right) ^{-2}\left( y_{t}-x_{t}^{\top }\hat{\beta}\right) ^{2} \\
&=&\sum_{t=1}^{T}\left( \frac{\left[ \mathbf{L}\left( \varrho _{2}\right) Y%
\right] _{t}}{\left[ \mathbf{L}\left( \varrho _{2}\right) \right] _{t,t}}%
\right) ^{2}
\end{eqnarray*}%
where $\mathbf{L}\left( \varrho _{2}\right) $ is the projection matrix
defined as:%
\begin{eqnarray*}
\mathbf{L}\left( \varrho _{2}\right)  &=&I_{T}-XS\left( \varrho _{2}\right)
X^{\top } \\
&=&I_{T}-X\left( X^{\top }X+\varrho _{2}\Gamma _{2}\Gamma _{2}^{\top
}\right) ^{-1}X^{\top }
\end{eqnarray*}%
If $S\left( \varrho _{2}\right) $ is a band matrix, which is the case for
spline models, the coefficients $\left[ \mathbf{L}\left( \varrho _{2}\right) %
\right] _{t,t}$ and $\left[ \mathbf{L}\left( \varrho _{2}\right) Y\right] _{t}$
can be computed in $O\left( T\right) $ operations thanks to the Hutchinson-De Hoog
algorithm (Hutchinson and De Hoog, 1985).

\subsubsection{GCV for centered data as the selection criterion}

The generalized cross-validation (GCV) method is a rotation-invariant version of
LOOCV (Craven and Wahba, 1978). Even if it is not its main purpose, this
approach replaces the factor $\left[ \mathbf{L}\left( \varrho _{2}\right) %
\right] _{t,t}$ by the average value $T^{-1}\limfunc{trace}\mathbf{L}\left(
\varrho _{2}\right) $:%
\begin{equation}
GCV\left( \varrho _{2}\right) =\frac{T^{2}}{\limfunc{trace}^{2}\mathbf{L}%
\left( \varrho _{2}\right) }\sum_{t=1}^{T}\left( y_{t}-x_{t}^{\top }\hat{%
\beta}\right) ^{2} \label{eq:gcv-ridge}
\end{equation}%
We deduce that the GCV criterion depends on $\mathbf{L}\left( \varrho
_{2}\right) $ and the residual sum of squares $\sum_{t=1}^{T}\left( y_{t}-x_{t}^{\top }%
\hat{\beta}\right) ^{2}$. We recall that $\mathbf{H}\left( \varrho _{2}\right)
=I_{T}-\mathbf{L}\left( \varrho _{2}\right) $ is the hat matrix. The value
$\left[ \mathbf{H}\left( \varrho _{2}\right) \right] _{t,t} $ is called the
leverage value (Craven and Wahba, 1978) and determines the amount by which the
predicted value $\hat{y}_{t}=x_{t}^{\top }\hat{\beta}$ is influenced by
$y_{t}$. We also know that $\limfunc{trace}\mathbf{L}\left(
\varrho _{2}\right) =T-K$. From the Woodbury formula, we have\footnote{%
The Woodbury matrix identity is:%
\begin{equation*}
\left( A+BCD\right) ^{-1}=A^{-1}-A^{-1}B\left( C^{-1}+DA^{-1}B\right)
^{-1}DA^{-1}
\end{equation*}%
}:%
\begin{eqnarray*}
\mathbf{L}\left( \varrho _{2}\right)  &=&I_{T}-X\left( X^{\top }X+\varrho
_{2}\Gamma _{2}\Gamma _{2}^{\top }\right) ^{-1}X^{\top } \\
&=&\left( I_{T}+X\left( \varrho _{2}\Gamma _{2}\Gamma _{2}^{\top }\right)
^{-1}X^{\top }\right) ^{-1}
\end{eqnarray*}%
Let $\lambda _{t}$ be the eigenvalues\footnote{%
Computing the eigenvalues of $X\left( \Gamma _{2}\Gamma _{2}^{\top }\right)
^{-1}X^{\top }$ can be done in $O\left( T^{3}\right) $ operations.} of the
symmetric real matrix $X\left( \Gamma _{2}\Gamma _{2}^{\top }\right)
^{-1}X^{\top }$. We have:%
\begin{equation*}
\limfunc{trace}\mathbf{L}\left( \varrho _{2}\right) =\sum_{t=1}^{T}\left(
1+\frac{\lambda _{t}}{\varrho _{2}}\right) ^{-1}
\end{equation*}%
This formula allows the value of $\limfunc{trace}^{-2}\mathbf{L}%
\left( \varrho _{2}\right) $ to be computed for every value of $\varrho _{2}$.
Like the PRESS statistic, the optimal value of $\varrho _{2}$ is obtained by
minimizing the GCV function given by Equation (\ref{eq:gcv-ridge}).


\section{Application to robo-advisory}

The previous techniques are of particular interest for portfolio optimization
when building a strategic asset allocation (SAA), a trend-following strategy or
more generally a mean-variance diversified portfolio. Depending on the
approach, they can diversify or concentrate the portfolio. By mixing the
different approaches, we can also obtain a diversified allocation on some
selected stocks. In this case, portfolio regularization and portfolio sparsity
are combined. The previous techniques can also be used when implementing
tactical asset allocation (TAA). In this case, regularization and sparsity are
imposed in a relative way with respect to a benchmark or a current investment
portfolio. In this section, we show why these techniques are necessary when
building a robo-advisor based on an automated allocation engine.

\subsection{Robo-advisory and the secret sauce of portfolio optimization}

The idea that portfolio optimization is a simple mathematical problem is mistaken.
It is a process that requires manual interventions and may
take considerable time before a solution is found. And this human intervention
has little in common with numerical algorithms. Indeed, Quants know that
the secret sauce of portfolio optimization lies in the alchemy of defining the
right constraints in order to obtain an acceptable solution that makes sense.
Let us consider the traditional strategic asset allocation exercise that is
performed by institutional investors almost every year. We assume that the SAA team
has already produced the two inputs: the vector $\mu $ of expected returns and
the covariance $\Sigma$ of asset returns. We could think that the hard work
has therefore been done, and that computing the SAA portfolio will take
a matter of seconds since we just
have to run a Markowitz optimization. In reality, solving one Markowitz
optimization generally produces a bad solution and is not sufficient. This is
why Quants will use an iterative process based on this optimization
program:
\begin{eqnarray}
x_{\left( k\right) }^{\star } &=&\arg \min \frac{1}{2}x^{\top }\Sigma
x-\gamma x^{\top }\mu   \label{eq:saa1} \\
\text{{}} &\text{s.t.}&\left\{
\begin{array}{l}
\mathbf{1}^{\top }x=1 \\
\mathbf{0}\leqslant x\leqslant \mathbf{1} \\
x\in \Omega _{\left( k\right) }%
\end{array}%
\right.   \notag
\end{eqnarray}%
where $\Omega _{\left( 0\right) }=\mathbb{R}^{n}$ and $k$ is the step. They
will begin by solving the traditional Markowitz problem with long-only
constraints and will find an initial solution $x_{\left( 0\right) }^{\star }$. Then,
they will analyze this solution and define a new set of constraints $\Omega
_{\left( 1\right) }$ that might produce a more acceptable solution. The concept
of \textquotedblleft acceptable solution\textquotedblright\ remains unclear,
but it means one that can be accepted by the chief investment officer. Once
$\Omega _{\left( 1\right) }$ is defined, Quants will run the optimization
problem (\ref{eq:saa1}) and obtain a new solution $x_{\left( 1\right) }^{\star
}$. Next, they will analyze this new solution and define a new set of
constraints $\Omega _{\left( 2\right) }$ that might produce an even more
acceptable solution. They will iterate this process a number of times.
Therefore, this iterative process can be represented by the sequence
$\boldsymbol{P}$ defined as follows:
\begin{equation*}
\boldsymbol{P}=\left\{ x_{\left( 0\right) }^{\star },\Omega _{\left(
1\right) },x_{\left( 1\right) }^{\star },\Omega _{\left( 2\right)
},x_{\left( 2\right) }^{\star },\Omega _{\left( 3\right) },x_{\left(
3\right) }^{\star },\ldots \right\}
\end{equation*}%
Using this tool, we can evaluate Quants and draw some conclusions:
\begin{itemize}
\item A good Quant is a person that is able to \textquotedblleft close\textquotedblright\ this
sequence in a limited number of steps.
\item A bad Quant is a person that produces an infinite sequence and is not able to end the process.
\item Quant $Q_1$ is more efficient than Quant $Q_2$ if:
\begin{equation*}
\func{card}\boldsymbol{P}\left( Q_{1}\right) < \func{card}\boldsymbol{P}\left( Q_{2}\right)
\end{equation*}
\end{itemize}
\smallskip

Let us illustrate the previous process with an example%
\footnote{This example is taken from Roncalli (2013) on page 287.}.
We consider a universe of nine asset classes: (1) US 10Y Bonds, (2)
Euro 10Y Bonds, (3) Investment Grade Bonds, (4) High Yield Bonds,
(5) US Equities, (6) Euro Equities, (7) Japan Equities, (8) EM Equities
and (9) Commodities. In Tables \ref{tab:saa1-1} and
\ref{tab:saa1-2}, we indicate the statistics used to compute the
optimal allocation. The objective is to find the optimal allocation
for an ex-ante volatility of around $7\%$.\smallskip

\begin{table}[tbph]
\centering
\caption{Expected returns and risks (in \%)}
\label{tab:saa1-1}
\tableskip
\begin{tabular}{c|cccc:cccc:c}
\hline
           & (1)        & (2)        & (3)        & (4)    & (5)        & (6)        & (7)        & (8)    & (9)        \\
\hline
$\mu_i$    & ${\TsV}4.2$ & ${\TsV}3.8$ & ${\TsV}5.3$ & $10.4$ & ${\TsV}9.2$ & ${\TsV}8.6$ & ${\TsV}5.3$ & $11.0$ & ${\TsV}8.8$ \\
$\sigma_i$ & ${\TsV}5.0$ & ${\TsV}5.0$ & ${\TsV}7.0$ & $10.0$ &     $15.0$ &     $15.0$ &     $15.0$ & $18.0$ &     $30.0$ \\
\hline
\end{tabular}
\end{table}

\begin{table}[tbph]
\caption{Correlation matrix of asset returns (in \%)}
\centering
\label{tab:saa1-2}
\tableskip
\begin{tabular}{c|cccc:cccc:c}
\hline
    & (1)           & (2)           & (3)        & (4)        & (5)        & (6)        & (7)        & (8)        & (9)    \\ \hline
(1) & ${\TsIII}100$ &               &            &            &            &            &            &            &        \\
(2) & ${\TsVIII}80$ & ${\TsIII}100$ &            &            &            &            &            &            &        \\
(3) & ${\TsVIII}60$ & ${\TsVIII}40$ & $100$      &            &            &            &            &            &        \\
(4) & $-20$         & $-20$         & ${\TsV}50$ & $100$      &            &            &            &            &        \\ \hdashline
(5) & $-10$         & $-20$         & ${\TsV}30$ & ${\TsV}60$ & $100$      &            &            &            &        \\
(6) & $-20$         & $-10$         & ${\TsV}20$ & ${\TsV}60$ & ${\TsV}90$ & $100$      &            &            &        \\
(7) & $-20$         & $-20$         & ${\TsV}20$ & ${\TsV}50$ & ${\TsV}70$ & ${\TsV}60$ & $100$      &            &        \\
(8) & $-20$         & $-20$         & ${\TsV}30$ & ${\TsV}60$ & ${\TsV}70$ & ${\TsV}70$ & ${\TsV}70$ & $100$      &        \\ \hdashline
(9) & ${\TsXIII}0$  & ${\TsXIII}0$  & ${\TsV}10$ & ${\TsV}20$ & ${\TsV}20$ & ${\TsV}20$ & ${\TsV}30$ & ${\TsV}30$ & $100$  \\ \hline
\end{tabular}
\end{table}

Using these figures, we obtain an initial allocation
$x^{\star}_{\left(0\right)}$ that is reported in Table%
\footnote{The weights and the risk/return statistics are given in
$\%$.} \ref{tab:saa1-3}. The optimal portfolio is invested in only
four asset classes. The allocation in US 10Y Bonds is $28\%$, while
the allocation in High Yield Bonds is $70\%$. It is obvious that
this portfolio cannot be a SAA policy. This is why the Quant will
add some constraints in order to obtain a better solution. We can
impose that the weight of one asset class cannot exceed $25\%$.
Using this new set of constraints $\Omega_{\left(1\right)}$, we
obtain Portfolio $x^{\star}_{\left(1\right)}$ that is less
concentrated than Portfolio $x^{\star}_{\left(0\right)}$. The
allocation in US 10Y Bonds and High Yield Bonds reaches the cap of
$25\%$. The portfolio is now invested in Euro 10Y Bonds ($15.90\%$),
US Equities ($10.70\%$) and EM Equities ($21.27\%$). The drawback of
this solution could be the allocation in equities, which is too small.
This is why the Quant will add another constraint in order to obtain
an equity allocation that is larger than $40\%$. At the
third iteration, we then obtain Portfolio
$x^{\star}_{\left(3\right)}$. If we assume that the SAA exercise is
complete for a European institutional investor, this solution is not
acceptable because it contains many US assets and too few
European assets. This is why the Quant will add two new constraints.
He can require that the allocation in Euro 10Y Bonds is larger than
the allocation in US 10Y Bonds, and that the allocation in Euro Equities
is larger than the allocation in US Equities. By using this new set
of constraints $\Omega_{\left(4\right)}$, we obtain the following
solution: the weight of US 10Y Bonds is $12.13\%$, the weight of
Euro 10Y Bonds is $22.13\%$, the weight of IG Bonds is $15.00\%$,
etc. Again, this solution may not be acceptable, because there is no
allocation in Japanese equities. Therefore, the Quant may impose
that there is at least $5\%$ invested in this asset class. After few
additional iterations, the solution is given by the last column in
Table \ref{tab:saa1-3}.\smallskip

\begin{table}[tbph]
\caption{The iterative trial-and-error solutions}
\centering
\label{tab:saa1-3}
\tableskip
\begin{tabular}{cc|ccccccc}
\hline
\multicolumn{2}{c}{Step $k$}                          & \#0          & \#1          & \#2          & \#3          & \#4          & $\cdots$ & \#K          \\ \hline
US 10Y Bonds    & (1)                                 &      $28.39$ &      $25.00$ &      $24.99$ &      $25.00$ &      $12.13$ &          &      $10.00$ \\
Euro 10Y Bonds  & (2)                                 & ${\TsV}0.00$ &      $15.90$ &      $18.60$ &      $16.50$ &      $22.13$ &          &      $30.00$ \\
IG Bonds        & (3)                                 & ${\TsV}0.00$ & ${\TsV}0.00$ & ${\TsV}0.00$ & ${\TsV}4.86$ &      $15.00$ &          &      $10.00$ \\
HY Bonds        & (4)                                 &      $69.64$ &      $25.00$ &      $16.41$ &      $10.00$ &      $10.00$ &          & ${\TsV}5.00$ \\ \hdashline
US Equities     & (5)                                 & ${\TsV}0.00$ &      $10.70$ &      $20.86$ &      $25.00$ &      $10.00$ &          &      $10.00$ \\
Euro Equities   & (6)                                 & ${\TsV}0.00$ & ${\TsV}0.00$ & ${\TsV}3.16$ & ${\TsV}5.00$ &      $20.00$ &          &      $20.00$ \\
Japan Equities  & (7)                                 & ${\TsV}0.00$ & ${\TsV}0.00$ & ${\TsV}0.00$ & ${\TsV}0.00$ & ${\TsV}0.00$ &          & ${\TsV}5.00$ \\
EM Equities     & (8)                                 & ${\TsV}1.17$ &      $21.27$ &      $15.98$ &      $10.00$ &      $10.00$ &          & ${\TsV}8.00$ \\ \hdashline
Commodities     & (9)                                 & ${\TsV}0.79$ & ${\TsV}2.13$ & ${\TsV}0.00$ & ${\TsV}3.64$ & ${\TsV}0.73$ &          & ${\TsV}2.00$ \\ \hline
\multicolumn{2}{c|}{$\mu\left(x\right)$}              & ${\TsV}8.63$ & ${\TsV}7.77$ & ${\TsV}7.41$ & ${\TsV}7.12$ & ${\TsV}6.99$ &          & ${\TsV}6.57$ \\
\multicolumn{2}{c|}{$\sigma\left(x\right)$}           & ${\TsV}7.00$ & ${\TsV}7.00$ & ${\TsV}7.00$ & ${\TsV}7.00$ & ${\TsV}7.00$ &          & ${\TsV}6.84$ \\
\multicolumn{2}{c|}{$\func{SR}\left(x \mid r\right)$} &      $80.49$ &      $68.08$ &      $63.03$ &      $58.93$ &      $57.00$ &          &      $52.17$ \\ \hline
\end{tabular}
\end{table}

\begin{figure}[tbph]
\centering
\caption{How does the secret sauce of portfolio optimization work?}
\label{fig:saa1}
\figureskip
\includegraphics[width = \figurewidth, height = \figureheight]{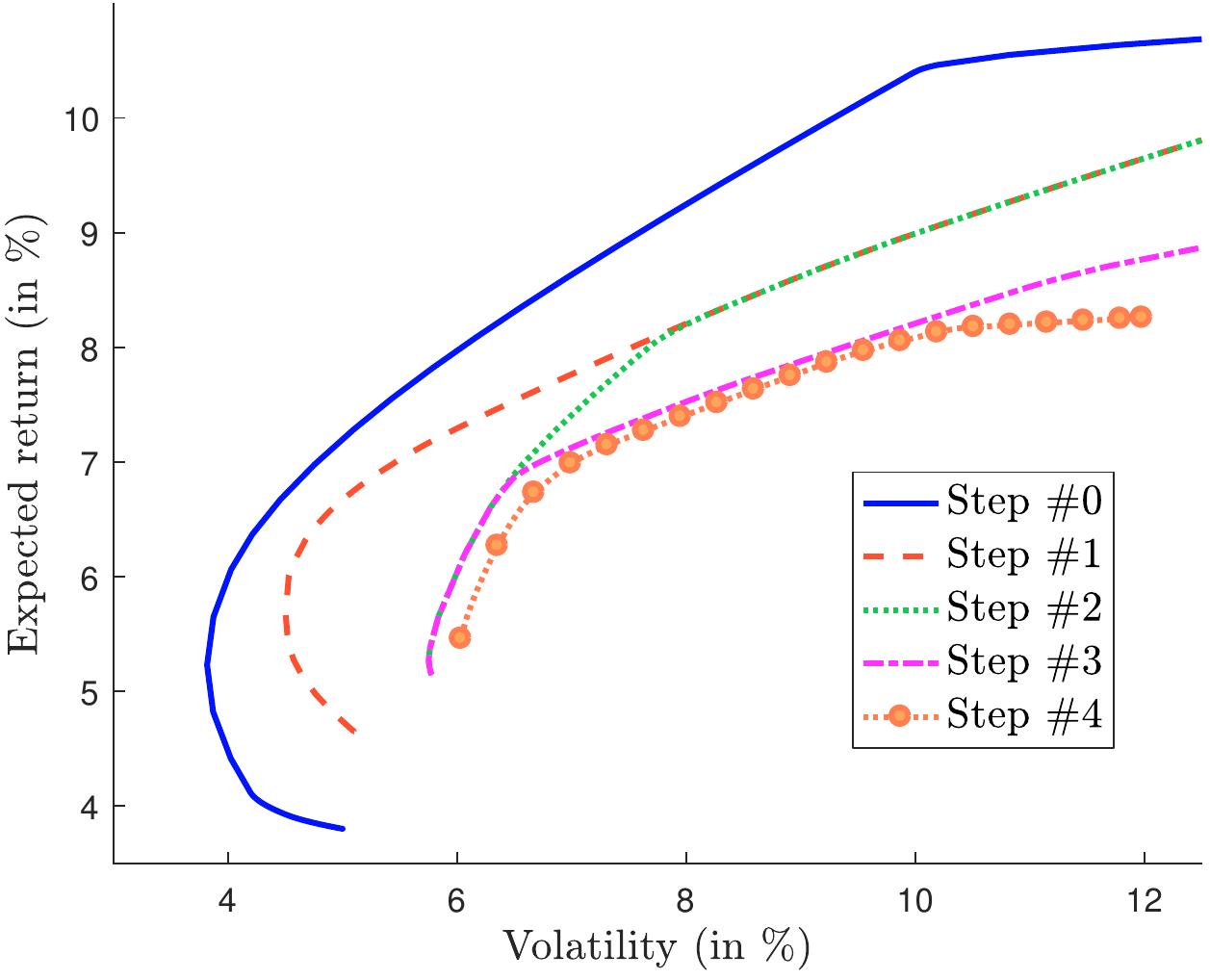}
\end{figure}

We notice that the previous iterative process $\boldsymbol{P}$ satisfies:
\begin{equation*}
\Omega _{\left( k+1\right) }\subset \Omega _{\left( k\right) }\subset \cdots
\subset \Omega _{\left( 2\right) }\subset \Omega _{\left( 1\right) }
\end{equation*}%
The underlying idea is to define an increasingly constrained investment
universe. For instance, we verify that the efficient frontiers are ordered and
that they are more and more constrained (see Figure \ref{fig:saa1}).

\begin{remark}
Quants may use variants of Problem (\ref{eq:saa1}). When they are also in
charge of producing $\mu $ and $\Sigma $, they may also consider the iterative
process with the following optimization problem:
\begin{equation*}
x_{\left( k\right) }^{\star }=\arg \min \frac{1}{2}x^{\top }\Sigma _{\left(
k\right) }x-\gamma x^{\top }\mu _{\left( k\right) }
\end{equation*}%
In this case, the sequence $\boldsymbol{P}$ is defined as follows:
\begin{equation*}
\boldsymbol{P}=\left\{ x_{\left( 0\right) }^{\star },\Omega _{\left(
1\right) },\Sigma _{\left( 1\right) },\mu _{\left( 1\right) },x_{\left(
1\right) }^{\star },\Omega _{\left( 2\right) },\Sigma _{\left( 2\right)
},\mu _{\left( 2\right) },x_{\left( 2\right) }^{\star },\ldots \right\}
\end{equation*}
\end{remark}

It is obvious that the iterative process for defining the optimal
portfolio conflicts with an automated and algorithm-driven
robo-advisor. First, this is not the intent of a robo-advisor,
unless we reduce the concept of robo-advisory to a digital
application or a data-visualization tool, meaning that allocation
decisions are made outside the robo-advisor. Second, a
robo-advisor should be able to manage many portfolios on an
industrial scale. If we consider the traditional lifestyle approach
based on three portfolios (defensive, balanced and dynamic), which
are rebalanced at the end of each month, it is obvious that the
robo-advisor can be manually loaded every month. Again, this
approach does not correspond to the robo-advisory concept. Indeed,
robo-advisors claim that they better meet the expectations of
investors by taking into account their constraints and by being more
granular. This is particularly true with the emergence of goal-based
investing in wealth management:
\begin{quote}
\textquotedblleft \textsl{While mass production has happened a long time ago
in investment management through the introduction of mutual funds and more
recently exchange traded funds, a new industrial revolution is currently
under way, which involves mass customization, a production and distribution
technique that will allow individual investors to gain access to scalable
and cost-efficient forms of goal-based investing solutions}%
\textquotedblright\ (Martellini, 2016, page 5).
\end{quote}
Lastly, the iterative process does not help improve the portfolio management
in a scientific manner. Indeed, it is a blind-eye approach, because it is
difficult to explain the performance of the portfolio. We don't know if it
comes from the expected returns step (or the active bets) or the portfolio
optimization step. In robo-advisory, these two steps must be easily identified
and distinguished. Indeed, the portfolio optimization engine is part of the
robo-advisor while expected returns may be designed outside the robo-advisor.
This is generally the case because they can be imposed by the final investor
himself, they can change from one third-party distributor to another, some
investors will want to introduce trend-following patterns, etc. Contrary to the
optimization method, the engine of expected returns is therefore not necessarily
decided by the fintech that produces the robo-advisor. This is why the two
steps must be perfectly differentiated.

\subsection{Formulation of the optimization problem}

We note $\tilde{x}$ the reference portfolio\footnote{which is also called the
strategic or the benchmark portfolio.} and $x_{t}$ the current portfolio. The
optimized portfolio for the next period is the solution of this comprehensive
optimization program:
\begin{eqnarray}
x_{t+1}^{\star } &=&\arg \min f\left( x\right) +\tilde{\varrho}%
_{1}\left\Vert \tilde{\Gamma}_{1}\left( x-\tilde{x}\right) \right\Vert _{1}+%
\frac{1}{2}\tilde{\varrho}_{2}\left\Vert \tilde{\Gamma}_{2}\left( x-\tilde{x}%
\right) \right\Vert _{2}^{2}+  \label{eq:robo1} \\
&&+\varrho _{1}\left\Vert \Gamma _{1}\left( x-x_{t}\right) \right\Vert _{1}+%
\frac{1}{2}\varrho _{2}\left\Vert \Gamma _{2}\left( x-x_{t}\right)
\right\Vert _{2}^{2}  \notag \\
&\text{s.t.}&\left\{
\begin{array}{l}
\mathbf{1}^{\top }x=1 \\
\mathbf{0}\leqslant x\leqslant \mathbf{1} \\
x\in \Omega
\end{array}%
\right.   \notag
\end{eqnarray}
where $\Omega $ is a set of predetermined constraints. This problem
considers both $L_{1}$ and $L_{2}$ penalty functions with respect to
the reference portfolio and the current portfolio. Concerning
$f\left( x\right) $, we can use the Markowitz function:
\begin{equation*}
f\left( x\right) =\frac{1}{2}x^{\top }\Sigma x-\gamma x^{\top }\mu
\end{equation*}%
However, it is certainly better to consider the tracking-error
function with respect to the reference portfolio:
\begin{eqnarray*}
f\left( x\right)  &=&\frac{1}{2}\left( x-\tilde{x}\right) ^{\top }\Sigma
\left( x-\tilde{x}\right) -\gamma \left( x-\tilde{x}\right) ^{\top }\mu  \\
&=&\frac{1}{2}x^{\top }\Sigma x-\gamma x^{\top }\left( \mu +\frac{1}{\gamma }%
\Sigma \tilde{x}\right) +C
\end{eqnarray*}%
where $C$ is a constant that does not depend on the variable $x$.
\smallskip

The aims of Problem (\ref{eq:robo1}) are multiple:
\begin{enumerate}
\item The first objective is naturally to optimize the traditional
    risk/return trade-off.

\item The second objective is to control the active bets between the
reference portfolio $\tilde{x}$ and the new optimized portfolio $%
x_{t+1}^{\star }$ at various levels:

\begin{enumerate}
\item The first layer is to target a tracking error by using the TE
    objective function in place of the MVO objective function;

\item The second layer is the $L_{2}$ penalty
    $\tilde{\varrho}_{2}\left\Vert \tilde{\Gamma}_{2}\left(
    x-\tilde{x}\right) \right\Vert _{2}^{2}$ that helps to smooth the
    tactical allocation with respect to the strategic allocation. This
    layer implies shrinking the covariance matrix $\Sigma $;

\item The third layer is the $L_{1}$ penalty $\tilde{\varrho}_{1}\left\Vert
    \tilde{\Gamma}_{1}\left( x-\tilde{x}\right) \right\Vert _{1}$ that
    helps to sparsify the relative bets with respect to Portfolio
    $\tilde{x}$;
\end{enumerate}

\item The third objective is to control the turnover ($L_{1}$ penalty) and
the quadratic costs ($L_{2}$ penalty) with respect to the current portfolio $%
x_{t}$.
\end{enumerate}

With all these safeguards, we are equipped to perform stable and
robust dynamic allocation for robo-advisors. However, three issues
remain unsolved: the specification of expected returns, the choice of
the tracking error level and the calibration of the regularization
parameters. The idea of the next section is not to give a solution
or to publish our know-how on these topics (Malongo \textsl{et al.},
2016). However, we will indicate the shortcomings to be avoided.

\subsection{Practical considerations}

\subsubsection{Incorporating active management views}

In some cases, robo-advisors are closed systems, but most of the
time, they are open systems. Often, the fintech that developed
the robo-advisor technology enters into bilateral agreements with
third-party distributors (asset managers, private banks, wealth
managers, insurance companies, retail distributors, etc.). In this
case, the robo-advisor platform is adapted to take into account the
distributor's specific requirements, constraints and objectives.
For instance, the robo-advisor platform may be plugged with the distributor's
risk/return profiling system. The number of funds
and the investment universe changes from one distributor to another
one. One of the big specific features is the engine that produces expected
returns. It is rare that the distributor uses the default engine
provided by the fintech. For instance, some investors will want to
incorporate momentum patterns, others prefer to use expected returns
produced by their economic experts, etc.\smallskip

In practice, it is extremely difficult to express bets in terms of absolute
returns. Portfolio managers prefer to use a rating scale $\mathcal{S}$ with
different grades. The typical rating scale contains $7$ grades:
\begin{equation*}
\begin{tabular}{cc}
\hline
Grade & Definition     \\
\hline
$---$ & Strong bearish \\
$--$  & Bearish        \\
$-$   & Weak bearish   \\ \hdashline
$0$   & Neutral        \\ \hdashline
$+$   & Weak bullish   \\
$++$  & Bullish        \\
$+++$ & Strong bullish \\
\hline
\end{tabular}%
\end{equation*}%
The challenge is then to transform these grades into expected returns. The most
frequent empirical approach is based on the Black-Litterman model, which is
described in Appendix \ref{appendix:section-bl} on page
\pageref{appendix:section-bl}.\smallskip

Given a strategic portfolio $\tilde{x}$, we compute the implied expected
returns $\tilde{\mu}_{i}$ of Asset $i$ thanks to the CAPM equation:
\begin{equation}
\tilde{\mu}_{i}=r+\limfunc{SR}\left( \tilde{x}\mid r\right) \frac{\left(
\Sigma \tilde{x}\right) _{i}}{\sqrt{\tilde{x}^{\top }\Sigma \tilde{x}}}
\label{eq:robo2}
\end{equation}%
We assume that the signal $s_{i}$ on Asset $i$ is homogeneous to a Sharpe
ratio. In particular, we have:
\begin{equation*}
\Delta \func{SR}{}_{i}=\delta \frac{s_{i}}{n_{s}}
\end{equation*}%
where $n_{s}$ is the range index of the rating scale\footnote{%
It is equal to:%
\begin{equation*}
n_{s}=\frac{-1+\limfunc{card}\mathcal{S}}{2}
\end{equation*}} and $\delta $ is a scalar that indicates the flexibility of active
tactical management\footnote{Typically, $\delta $ is set to one.}.
Then, we deduce that the expected return of the portfolio manager is
equal to:
\begin{eqnarray*}
\breve{\mu}_{i} &=&\left( \func{SR}{}_{i}+\Delta \func{SR}{}_{i}\right) \cdot
\sigma _{i} \\
&=&\tilde{\mu}_{i}+\delta \frac{s_{i}}{n_{s}}\sigma_{i}
\end{eqnarray*}%
where $\func{SR}_{i}=\left( \tilde{\mu}_{i}-r\right) /\sigma _{i}$ is the
implied Sharpe ratio of Asset $i$ relative to the strategic portfolio $%
\tilde{x}$ and $\sigma _{i}$ is the estimated volatility of Asset $i$. The
final step is to combine $\tilde{\mu}_{i}$ and $\breve{\mu}_{i}$ using the
Black-Litterman framework:%
\begin{equation*}
\mu _{i}=\frac{\tau }{\tau +1}\tilde{\mu}_{i}+\left( 1-\frac{\tau }{\tau +1}%
\right) \breve{\mu}_{i}
\end{equation*}%
where $\tau $ is a parameter that measures the confidence into active bets. For
instance, when $\tau \rightarrow \infty $, the manager's views are not taken
into account, while the conditional expected returns tends to manager's views
when $\tau \rightarrow 0$.\smallskip

\begin{table}[tbph]
\caption{Covariance matrix of asset classes (Jan. 2016 -- Dec. 2016)}
\centering
\label{tab:robo1-1}
\tableskip
\centering
\begin{tabular}{c|rrrrrr:rrrr}
\hline \hline
\multicolumn{11}{c}{Volatility (in \%)} \\
\multicolumn{1}{c}{}
         & \multicolumn{1}{c}{(1)} & \multicolumn{1}{c}{(2)} & \multicolumn{1}{c}{(3)} & \multicolumn{1}{c}{(4)} &
           \multicolumn{1}{c}{(5)} & \multicolumn{1}{c}{(6)} & \multicolumn{1}{c}{(7)} & \multicolumn{1}{c}{(8)} &
           \multicolumn{1}{c}{(9)} & \multicolumn{1}{c}{(10)} \\
\multicolumn{1}{c}{}
         &  9.2 &   7.0 &   9.4 &   7.6 &  10.1 & \multicolumn{1}{r}{7.6} & 16.1  &  20.5 &  24.3 &  17.8 \\ \hline \hline
\multicolumn{11}{c}{Correlation matrix (in \%)} \\
\multicolumn{1}{c}{}
         & \multicolumn{1}{c}{(1)} & \multicolumn{1}{c}{(2)} & \multicolumn{1}{c}{(3)} & \multicolumn{1}{c}{(4)} &
           \multicolumn{1}{c}{(5)} & \multicolumn{1}{c}{(6)} & \multicolumn{1}{c}{(7)} & \multicolumn{1}{c}{(8)} &
           \multicolumn{1}{c}{(9)} & \multicolumn{1}{c}{(10)} \\ \hdashline
(1)  &   100.0 &       &       &       &       &       &       &       &       &       \\
(2)  &    17.7 & 100.0 &       &       &       &       &       &       &       &       \\
(3)  &    98.1 &  19.4 & 100.0 &       &       &       &       &       &       &       \\
(4)  &    16.5 &  99.5 &  18.1 & 100.0 &       &       &       &       &       &       \\
(5)  &    71.1 &   2.4 &  76.3 &   2.1 & 100.0 &       &       &       &       &       \\
(6)  &    85.9 &  12.7 &  87.6 &  11.8 &  89.1 & 100.0 &       &       &       &       \\ \hdashline
(7)  &    34.5 &   0.7 &  38.1 &   1.3 &  68.8 &  57.8 & 100.0 &       &       &       \\
(8)  &   -13.2 &   2.8 &  -4.0 &   3.6 &  41.0 &  18.2 &  59.5 & 100.0 &       &       \\
(9)  &    20.3 &   2.0 &  27.6 &   0.8 &  21.6 &  25.3 &   8.0 &  15.6 & 100.0 &       \\
(10) &    16.6 &  10.2 &  26.0 &  10.5 &  57.2 &  44.6 &  54.3 &  67.7 &  42.9 & 100.0 \\
\hline \hline
\end{tabular}
\end{table}

We consider an example with 10 asset classes: (1) US Sovereign Bonds, (2)
Euro Sovereign Bonds, (3) US Investment Grade Bonds, (4) EMU Investment Grade Bonds, (5) US High Yield Bonds,
(6) EM Bonds, (7) US Equities, (8) Europe Equities, (9) Japan Equities
and (10) EM Equities. In Table \ref{tab:robo1-1}, we report
the estimated covariance matrix for the period January 2016 --
December 2016. We consider an equally-weighted portfolio
$\tilde{x}$, which corresponds to a 40/60 strategic allocation. By
assuming that $r=0$ and $\limfunc{SR}\left( \tilde{x}\mid r\right)
=0.5$, we calculate the vector of implied expected returns using
Equation (\ref{eq:robo2}). The results are given in the second column in
Table \ref{tab:robo1-2}. For instance, the implied expected return
of US Sovereign bonds is equal to $2.57\%$. We now consider a set of
manager's views. The first scenario \#1 corresponds to a weak
bearish scenario on equity markets. Therefore, the grades are set to
$-$ for the four equity asset classes and $+$ for the two
sovereign bond asset classes. In Table \ref{tab:robo1-2}, we calculate%
\footnote{We assume that $\delta = 1$ and $\tau = 1$.} the expected
returns $\breve{\mu}$ implied by these views, and the final expected
returns $\mu$. For instance, $\breve{\mu}_i$ and $\mu_i$ are equal
to $5.46\%$ and $4.10\%$ for US Sovereign bonds. We verify that
expected returns are increased for sovereign bonds, decreased for
equities and neutral for the other asset classes.\smallskip

\begin{table}
\caption{Expected returns in \% (scenario \#1)}
\centering
\label{tab:robo1-2}
\tableskip
\centering
\begin{tabular}{l|cccc}
\hline
Asset class     & $\tilde{\mu}_i$ & $s_i$ & $\breve{\mu}_i$ & $\mu_i$ \\ \hline
US Sov. Bonds   & $2.57$ & $+$ & ${\TsVIII}5.64$ & $4.10$  \\
Euro Sov. Bonds & $0.96$ & $+$ & ${\TsVIII}3.29$ & $2.12$  \\
US IG Bonds     & $3.02$ & $0$ & ${\TsVIII}3.02$ & $3.02$  \\
EMU IG Bonds    & $1.02$ & $0$ & ${\TsVIII}1.02$ & $1.02$  \\
US HY Bonds     & $4.09$ & $0$ & ${\TsVIII}4.09$ & $4.09$  \\
EM Bonds        & $2.88$ & $0$ & ${\TsVIII}2.88$ & $2.88$  \\ \hdashline
US Equities     & $5.76$ & $-$ & ${\TsVIII}0.40$ & $3.08$  \\
Europe Equities & $6.35$ & $-$ &         $-0.48$ & $2.94$  \\
Japan Equities  & $6.76$ & $-$ &         $-1.34$ & $2.71$  \\
EM Equities     & $7.18$ & $-$ & ${\TsVIII}1.24$ & $4.21$  \\
\hline
\end{tabular}
\end{table}

\begin{table}
\caption{Scenario \#2}
\centering
\label{tab:robo1-3}
\tableskip
\centering
\begin{tabular}{l|cccc}
\hline
Asset class     & $\tilde{\mu}_i$ & $s_i$ & $\breve{\mu}_i$ & $\mu_i$ \\ \hline
US Sov. Bonds   & $2.57$ & $0$ & ${\TsV}2.57$ & ${\TsV}2.57$  \\
Euro Sov. Bonds & $0.96$ & $0$ & ${\TsV}0.96$ & ${\TsV}0.96$  \\
US IG Bonds     & $3.02$ & $0$ & ${\TsV}3.02$ & ${\TsV}3.02$  \\
EMU IG Bonds    & $1.02$ & $0$ & ${\TsV}1.02$ & ${\TsV}1.02$  \\
US HY Bonds     & $4.09$ & $0$ & ${\TsV}4.09$ & ${\TsV}4.09$  \\
EM Bonds        & $2.88$ & $0$ & ${\TsV}2.88$ & ${\TsV}2.88$  \\ \hdashline
US Equities     & $5.76$ & $+$ &      $11.13$ & ${\TsV}8.45$  \\
Europe Equities & $6.35$ & $+++$ &    $26.85$ &      $16.60$  \\
Japan Equities  & $6.76$ & $+$ &      $14.86$ &      $10.81$  \\
EM Equities     & $7.18$ & $+$ &      $13.11$ &      $10.14$  \\
\hline
\end{tabular}
\end{table}

\begin{table}
\caption{Scenario \#3}
\centering
\label{tab:robo1-4}
\tableskip
\centering
\begin{tabular}{l|cccc}
\hline
Asset class     & $\tilde{\mu}_i$ & $s_i$ & $\breve{\mu}_i$ & $\mu_i$ \\ \hline
US Sov. Bonds   & $2.57$ &   $0$ & ${\TsXIII}2.57$ & ${\TsVIII}2.57$ \\
Euro Sov. Bonds & $0.96$ &   $0$ & ${\TsXIII}0.96$ & ${\TsVIII}0.96$ \\
US IG Bonds     & $3.02$ &   $0$ & ${\TsXIII}3.02$ & ${\TsVIII}3.02$ \\
EMU IG Bonds    & $1.02$ &   $0$ & ${\TsXIII}1.02$ & ${\TsVIII}1.02$ \\
US HY Bonds     & $4.09$ &   $0$ & ${\TsXIII}4.09$ & ${\TsVIII}4.09$ \\
EM Bonds        & $2.88$ & $---$ &   ${\TsV}$$-4.72$ &       $-2.18$ \\ \hdashline
US Equities     & $5.76$ &   $0$ & ${\TsXIII}5.76$ & ${\TsVIII}5.76$ \\
Europe Equities & $6.35$ &   $0$ & ${\TsXIII}6.35$ & ${\TsVIII}6.35$ \\
Japan Equities  & $6.76$ &   $0$ & ${\TsXIII}6.76$ & ${\TsVIII}6.76$ \\
EM Equities     & $7.18$ & $---$ &        $-10.62$ &         $-4.69$ \\
\hline
\end{tabular}
\end{table}

We consider a second scenario that is more favorable to stock
markets, in particular European stocks (see Table
\ref{tab:robo1-3}). By construction, the implied expected returns do
not change because we consider the same strategic allocation.
However, the expected returns $\breve{\mu}_i$ and $\mu_i$ are
different because we have changed the scenario. Finally, we consider
a third scenario in Table \ref{tab:robo1-4}, which is an adverse
scenario on
emerging markets%
\footnote{$\tau$ is set to $0.5$ in order to reflect stronger confidence in
this scenario.}.

\subsubsection{Choosing the right tracking error level}

Volatility target strategies are very popular among Quants
(Hallerbach, 2012; Hocquard \textsl{et al.}, 2013). This explains
why many robo-advisors are based on volatility or tracking error
targeting. As said previously, we prefer TE objective function to
MVO objective function. In this case, there is no constraint on
the portfolio volatility, which is related to the volatility $\sigma
\left( \tilde{x}\right) $ of the reference portfolio. However, the
question of the TE level remains open. We provide some methods to
set the right level of tracking error.\smallskip

\begin{figure}[tbph]
\centering
\caption{Relationship between volatility and tracking error levels}
\label{fig:robo3}
\figureskip
\includegraphics[width = \figurewidth, height = \figureheight]{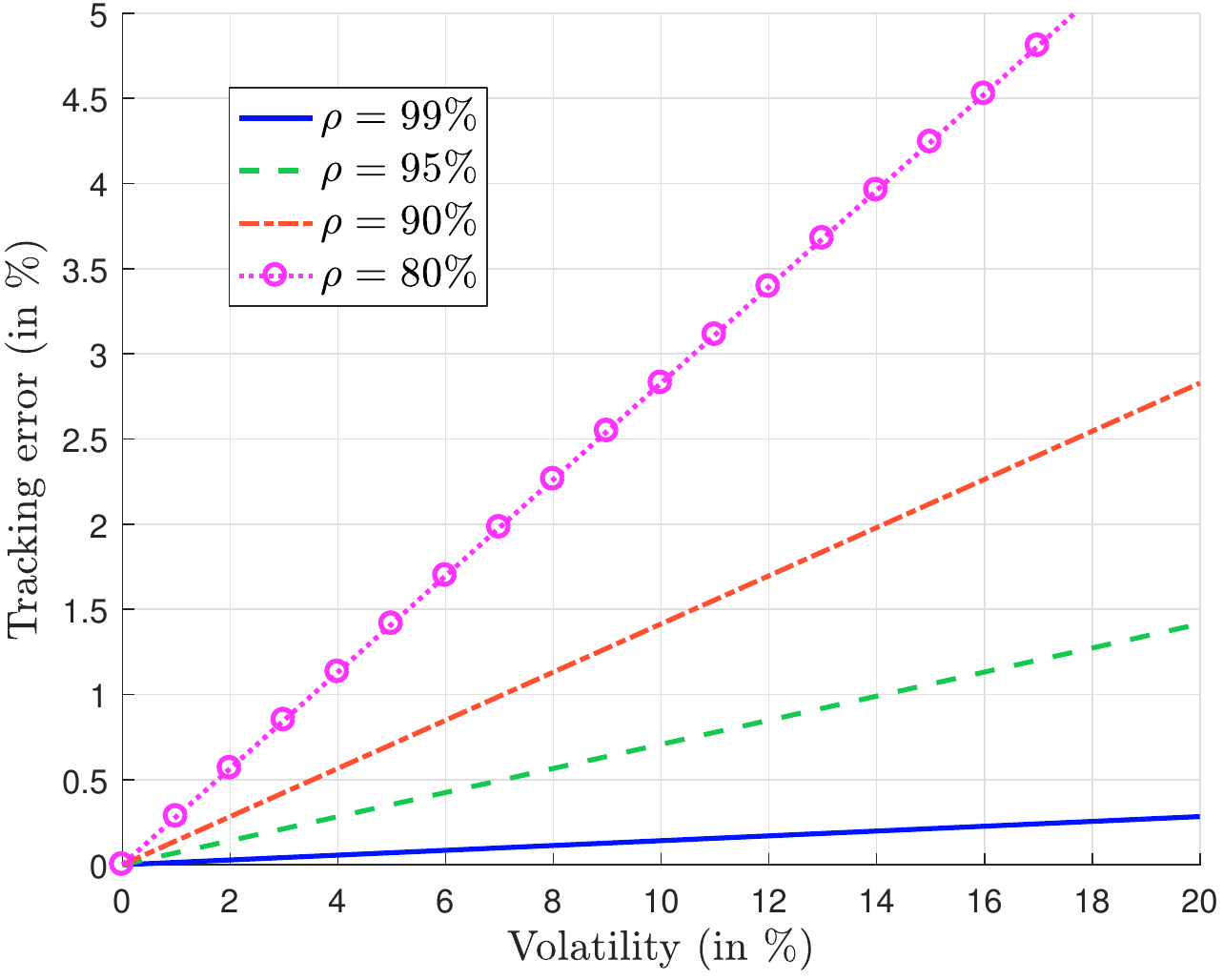}
\end{figure}

Let $x$ and $\tilde{x}$ be the tactical and strategic portfolios. We have:
\begin{eqnarray*}
\sigma ^{2}\left( x\mid \tilde{x}\right)  &=&\sigma ^{2}\left( R_{t}\left(
x\right) -R_{t}\left( \tilde{x}\right) \right)  \\
&=&\sigma ^{2}\left( x\right) +\sigma ^{2}\left( \tilde{x}\right) -2\rho
\left( x,\tilde{x}\right) \sigma \left( x\right) \sigma \left( \tilde{x}%
\right)
\end{eqnarray*}%
where $\rho \left( x,\tilde{x}\right) $ is the correlation between the
portfolio $x$ and the benchmark $\tilde{x}$. Generally, we have $\sigma \left(
x\right) \approx \sigma \left( \tilde{x}\right) $, implying that:
\begin{equation}
\sigma \left( x\mid \tilde{x}\right) =\sqrt{2\left( 1-\rho \left( x,\tilde{x}%
\right) \right) }\cdot \sigma \left( \tilde{x}\right) \label{eq:robo3}
\end{equation}%
In Figure \ref{fig:robo3}, we have reported the relationship between the
volatility of the strategic portfolio and the tracking error of the portfolio.
We notice that it depends on the correlation level. It follows that if the
strategic portfolio's volatility is low (less than $5\%$), we cannot
target a high level of tracking error volatility. A level of $1\%$ is certainly
the maximum. When the volatility is moderate between $5\%$ and $10\%$, we can
target a value between $1\%$ and $2\%$. We can achieve a higher tracking error
only if the portfolio's volatility is high.\smallskip

The previous result is of major importance, because it states that the tracking
error level of the tactical portfolio must be related to the volatility of the
strategic portfolio. In practice, the volatility is time-varying, implying that
using a constant tracking error strategy is not optimal.\smallskip

There is a second reason to consider a time-varying tracking error
level, because another issue concerns the relationship between the
tracking error and the active bets. We can show that (Grinold,
1994):
\begin{equation*}
\mu \left( x\mid \tilde{x}\right) =\sigma \left( x\mid \tilde{x}\right)
\cdot \func{TC}\cdot \func{IC}\cdot \sqrt{n}
\end{equation*}%
where $\func{TC}$ is the transfer coefficient, $\func{IC}$ is the information
coefficient and $n$ is the number of assets. This relationship is known as
\textquotedblleft the fundamental law of active management\textquotedblright.
If we assume that $\func{TC}$ and $\func{IC}$ are constant for a given active
manager and a given portfolio, it follows that the excess return is
proportional to the tracking error volatility:
\begin{equation*}
\mu \left( x\mid \tilde{x}\right) \propto \sigma \left( x\mid \tilde{x}%
\right)
\end{equation*}%
However, alpha generation is also linked to the number and strength of active
bets:
\begin{equation*}
\mu \left( x\mid \tilde{x}\right) =g_{\mu }\left( s_{1},\ldots ,s_{n}\right)
\end{equation*}%
We deduce that the tracking error must be a function of the scores $s_{i}$:%
\begin{equation}
\sigma \left( x\mid \tilde{x}\right) =g_{\sigma }\left( s_{1},\ldots
,s_{n}\right)   \label{eq:robo4}
\end{equation}%
This relationship is essential when considering tactical allocation.
Indeed, if all the scores are equal to zero, there is no active bet,
implying that we must target a zero tracking error level. If all the
scores are equal, we are in the same situation. Indeed, since we are
bullish in all the asset classes, there is no reason to deviate from
the strategic portfolio. In order to take a high tracking error
risk, we need the bets to present a high dispersion:
\begin{equation*}
\begin{tabular}{ccccc}
\hline
$s_{i}$ & \#1 & \#2 & \#3 & \#4 \\ \hline
$s_{1}$ & $0$ & $++$ & $+$ & $+++$ \\
$s_{2}$ & $0$ & $++$ & $-$ & $+++$ \\
$s_{3}$ & $0$ & $++$ & $+$ & $---$ \\
$s_{4}$ & $0$ & $++$ & $+$ & $---$ \\ \hline
$\sigma \left( x\mid \tilde{x}\right) $ & zero & zero & moderate & high \\
\hline
\end{tabular}%
\end{equation*}
Since the function $g_{\mu }$ is unknown and difficult to estimate, the
function $g_{\sigma }$ is also unknown. However, we may use the following rule
of thumb:
\begin{equation}
\sigma \left( x\mid \tilde{x}\right) \approx c\cdot \left( \frac{\sigma
\left( s\right) +\func{mad}\left( s\right) }{2}\right) \cdot \sigma ^{+}
\label{eq:robo5}
\end{equation}%
where $\sigma \left( s\right) $ is the standard deviation of scores,
$\func{mad}\left( s\right) $ is the mean absolute difference of scores, and
$\sigma ^{+}$ is the maximum tracking error. The value of $\sigma ^{+}$ may be
deduced from the relationship (\ref{eq:robo3}). By construction, we have:
\begin{equation*}
0\leq \left( \frac{\sigma \left( s\right) +\func{mad}\left( s\right) }{2}\right) \leq 3.6213
\end{equation*}%
and:
\begin{equation*}
0\leq \lim_{n\rightarrow \infty }\left( \frac{\sigma \left( s\right) +
\func{mad}\left( s\right) }{2}\right) \leq 3
\end{equation*}%
where $n$ is the number of assets. It follows that the scaling factor $c$ is
approximatively equal to $\dfrac{1}{3}$.\smallskip

\begin{figure}[tbph]
\centering
\caption{Relationship between active bets and tracking errors}
\label{fig:robo2}
\figureskip
\includegraphics[width = \figurewidth, height = \figureheight]{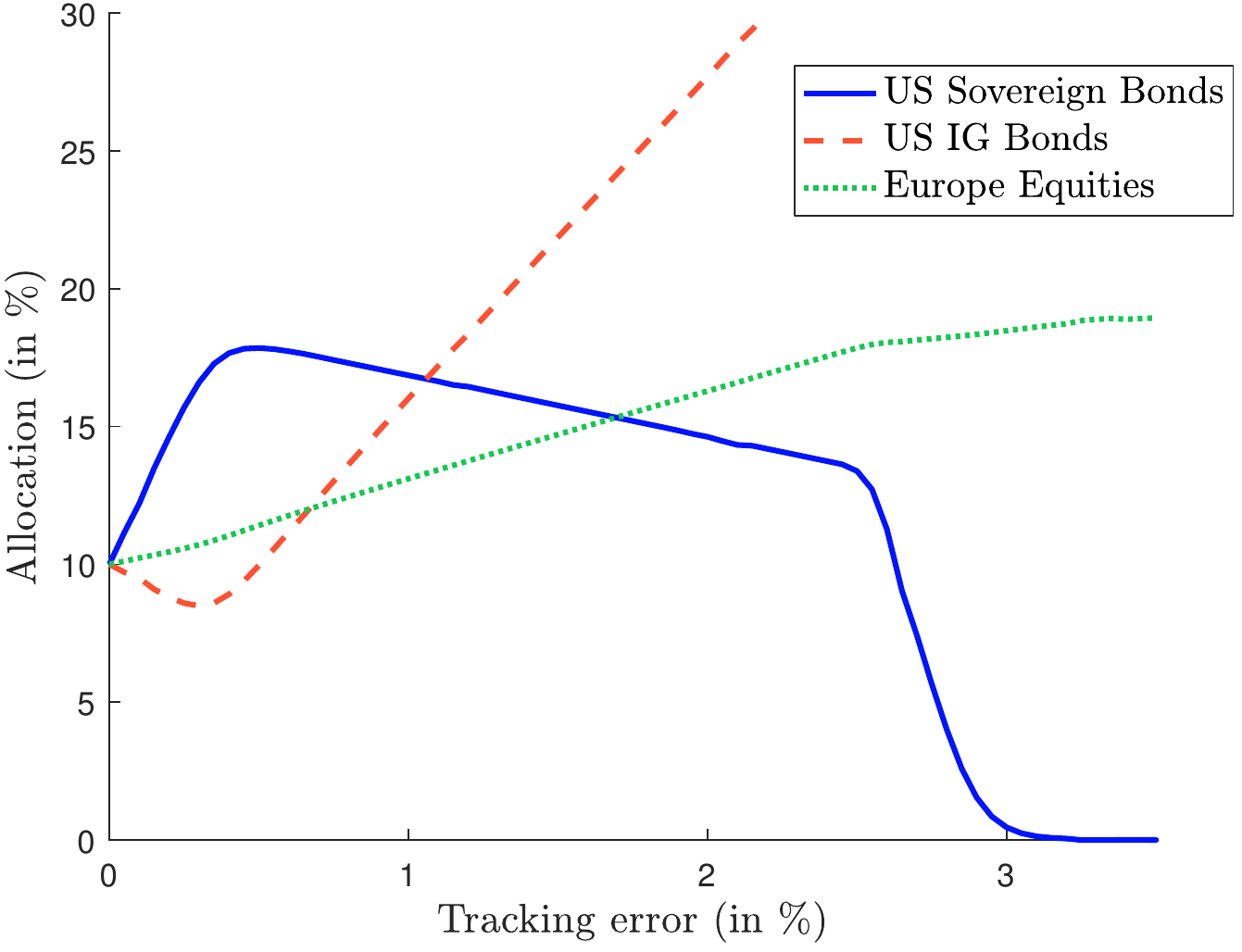}
\end{figure}

Equation (\ref{eq:robo5}) is a preliminary approach to set the level
of tracking error. Nevertheless, this rule of thumb has a major
drawback. It does not depend on the asset classes and their scores.
Let us consider the previous example described on page
\pageref{tab:robo1-1}. We assume that the signals are
respectively $+$, $0$, $+$, $0$, $+$, $0$, $0$, $+$, $0$ and $+$. In Figure %
\ref{fig:robo2}, we report the tactical allocation when we target a tracking
error level. For Europe Equities, we have a signal equal to $+$, and we verify
that the allocation is increasing with respect to the tracking error. For
US Sovereign and IG Bonds, we also have a signal equal to $+$, but the relationship between
the allocation and the tracking error is not monotonically increasing. The case of
US IG Bonds will be easily solved once we consider Problem
(\ref{eq:robo1}) instead of a simple tracking error optimization. The case of
US Sovereign Bonds is more problematic. Indeed, in an initial period when the tracking
error is low, the relationship is increasing. However, when the tracking error
increases too much, we obtain the opposite result. The reason is that the
volatility of US Sovereign Bonds is low compared to the other asset classes
(equities, investment grade and high yield). If we increase the tracking error,
there is a threshold beyond which it is better to play only active bets on
the most risky assets. Indeed, playing active bets on low risk assets does
not give rise to a high tracking error budget. This is why the optimizer
switches from low-risk assets to high-risk assets. This means that the choice
of a tracking error level depends on the set of parameters: the maximum
tracking error that depends on the strategic portfolio, the scores or active
bets and the volatility of the assets that compose the tactical portfolio.

\begin{figure}[tbph]
\centering
\caption{Impact of the ridge parameter on the shrinkage correlation}
\label{fig:robo4}
\figureskip
\includegraphics[width = \figurewidth, height = \figureheight]{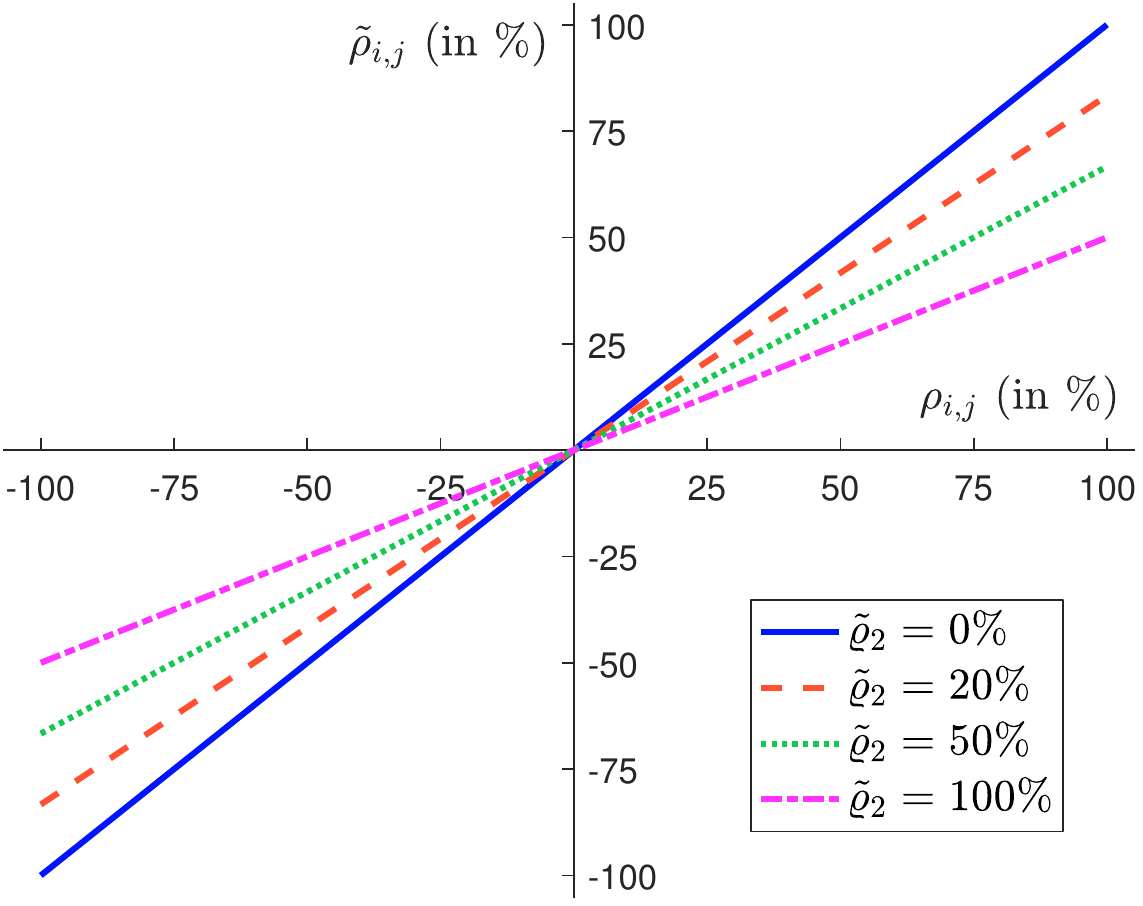}
\end{figure}

\subsubsection{Calibrating the regularization parameters}

As said previously, the choice of the regularization parameters is
not straightforward and requires a solid expertise and experience.
However, we will provide some tips that can help to calibrate the
model%
\footnote{We can also implement cross-validation methods presented in Section
\ref{section:cross-validation} on page \pageref{section:cross-validation}}.
The first thing to notice concerns the magnitude of $\varrho
_{1}$ and $\varrho _{2}$. On page \pageref{remark:ridge}, we have
seen that if $\tilde{\Gamma}_{2}=\limfunc{diag}\Sigma $, the
regularized correlations are:%
\begin{equation*}
\tilde{\rho}_{i,j}=\frac{\rho _{i,j}}{1+\tilde{\varrho}_{2}}
\end{equation*}%
In Figure \ref{fig:robo4}, we have reported the relationship between
the initial correlation $\rho _{i,j}$ and the shrinkage correlation
$\tilde{\rho}_{i,j}$. When $\tilde{\varrho}_{2}$ is equal to zero,
$\tilde{\rho}_{i,j}=\rho _{i,j}$. When
$\tilde{\varrho}_{2}\rightarrow \infty $, the shrinkage correlation
tends to zero. We then obtain a diagonal matrix with equal
volatilities. Therefore, there is a trade-off between considering
the initial covariance matrix and ignoring the dependence between
assets. A good way to choose $\tilde{\varrho}_{2}$ is to reduce the
impact of arbitrage factors while keeping the significance of common
risk factors. If we now consider the $L_{1}$ penalty $\varrho
_{1}\left\Vert \Gamma _{1}\left( x-x_{t}\right) \right\Vert _{1}$
and if we set $\Gamma _{1}=I_{n}$, the $L_{1}$ norm measures the portfolio's
two-way turnover:
\begin{equation*}
\left\Vert \left( x-x_{t}\right) \right\Vert _{1}=\sum_{i=1}^{n}\left\vert
x_{i}-x_{i,t}\right\vert
\end{equation*}%
The parameter $\varrho _{1}$ may then be used to control the
turnover. If $\Gamma _{1}$ is a matrix with non-negative entries that contains the
unit transaction costs, the $L_{1}$ norm measures the portfolio's transaction
cost (Scherer, 2007). This means that $\varrho
_{1}$ is the average transaction cost if $\Gamma _{1}$ is the
identity matrix. It follows that the order of magnitude of
$\tilde{\varrho}_{2}$ is not comparable to the order of magnitude of
$\varrho _{1}$. In the first case, it is expressed as a percentage (for
instance, $\tilde{\varrho}_{2}=25\%$) whereas
in the lasso problem it is expressed in
basis points (for instance, $\varrho _{1}=5$
bps). This is in line with the practice that shows that optimal
values of $L_{2}$ regularization are higher than those of $L_{1}$
regularization. The second thing to notice concerns the
specification of regularization matrices $\Gamma _{1}$,
$\tilde{\Gamma}_{1}$, $\Gamma _{2}$ and $\tilde{\Gamma}_{2}$. Most
of the time, they correspond to diagonal matrices, because it is
not easy to consider the cross effects of regularization. The simplest
way is to consider identity matrices, meaning that the
regularization patterns reduce to ridge and lasso approaches. If we
use the same parameters $\varrho _{1}=\tilde{\varrho}_{1} $ and $\varrho
_{2}=\tilde{\varrho}_{2}$, it is equivalent to considering that the two
portfolios plays a symmetric role. However, this is not the case.
Portfolio $x_{t}$ is used in order to limit the turnover and to
smooth the dynamic allocation. Portfolio $\tilde{x}$ is used in
order to control the relative active bets. This is why $\tilde{x}$
is more important than $x_{t}$ for implementing the active
management. Last but not least, the calibration of the parameters
highly depends on the investment profile. If the fund is composed of
equities, we need to use more aggressive parameters in order to
be more active than with a multi-asset fund. This means that
there is no magic formula, and the calibration stage requires much
empirical research and many tests in order to understand the
interconnectedness between the different terms of the portfolio
optimization problem.

\section{Conclusion}

According to Fisch \textsl{et al.} (2017), robo-advisors are \textquotedblleft
\textsl{computer algorithms that provide advice on investment portfolios and
then manage those portfolios}\textquotedblright . Since they are digital-based
tools that are generally implemented as web online services, fintechs compete
in order to offer better customization, data visualization, analytics, process
automation, etc. And the concepts of artificial intelligence, big data and
machine learning are never far away when we see the presentation of a
robo-advisor. Most of the time, fintechs prefer to insist on the application's
ergonomics and functionalities, and give little insight into the robo-advisor's \textit{raison
d'\^etre}: an automated portfolio allocation engine.\smallskip

One of the reasons may be that portfolio allocation is more
human-based than computer-based. It is true that automation in
portfolio optimization is a big issue. Indeed, portfolio
optimization is a hard task and does not always produce the desired results.
This is because the mathematical problem is not necessarily
well defined when we would like to obtain a smooth, sparse, active
and dynamic allocation.\smallskip

In this article, we come back to the traditional mean-variance
optimization, and identify the reason for the issues. We have shown
that it primarily corresponds to an alpha optimizer, and not to a beta
optimizer. Then we have presented the theory of regularization and
sparsity, and have demonstrated how it improves portfolio
optimization. Finally, this approach is applied for building
automated robo-advisory.

\clearpage

\appendix

\section{Mathematical results}

\subsection{Notations}
\label{appendix:notations}

We use the following notations:
\begin{itemize}

\item $\mathds{1}_{\Omega}\left( x\right) $ is the convex indicator function
    of $\Omega$: $ \mathds{1}_{\Omega}\left( x\right) =0$ for $x\in \Omega$
    and $\mathds{1}_{\Omega}\left( x\right) = +\infty $ for $x\notin \Omega$.

\item $A^{\dagger}$ is the Moore-Penrose pseudo-inverse matrix of $A$; in the
    scalar case, we have $0^{\dagger }=0$ and $a^{\dagger }=a^{-1}$ if $a\neq
    0$.

\item $\mathcal{C}=\left(\rho_{i,j}\right)$ denotes the correlation
    matrix with entries $\rho_{i,j}$.

\item $\mathcal{C}_n\left(\rho\right)$ is the constant correlation matrix of
    dimension $n$, whose uniform correlation is $\rho$.

\item $\mu$ is the vector of expected return.

\item $\Sigma$ is the covariance matrix.

\item $\left\Vert x\right\Vert _{p}=\left( \sum_{i=1}^{n}\left\vert
    x_{i}\right\vert ^{p}\right) ^{1/p}$ is the $L_{p}$ norm.

\item $\left\Vert x\right\Vert _{A}=\left( x^{\top }Ax\right) ^{1/2}$ is the
    weighted $L_{2}$ norm.

\item $\left[M\right]_{i,j}$ is the $\left(i,j\right)$ entry of the matrix
    $M$.

\item $x\odot y$ is the Hadamard element-wise product: $\left[ x\odot y%
\right] _{i,j}=\left[ x\right] _{i,j}\left[ y\right] _{i,j}$.

\item $\mathcal{P}_{\Omega }\left( x\right) $ is the projection of $x$ on
the set $\Omega $:%
\begin{equation*}
\mathcal{P}_{\Omega }\left( x\right) =\arg \min\nolimits_{y\in \Omega }\frac{1}{2}%
\left\Vert y-x\right\Vert _{2}^{2}
\end{equation*}

\item $\mathbf{prox}_{f}\left( v\right) $ is the proximal operator of $%
f\left( x\right) $:%
\begin{equation*}
\mathbf{prox}_{f}\left( v\right) =\arg \min\nolimits_{x}\left\{ f\left(
x\right) +\frac{1}{2}\left\Vert x-v\right\Vert _{2}^{2}\right\}
\end{equation*}

\end{itemize}

\subsection{Matrix form of the estimators $\hat{\mu}$ and $\hat{\Sigma}$}
\label{appendix:matrix-form}

Since we have $\hat{\mu}=\sum_{t=1}^{T}w_{t}R_{t}$, it follows that $\hat{\mu%
}=R^{\top }w$ where $w=\left( w_{1},\ldots ,w_{T}\right) \in \mathbb{R}^{T}$
and $R=\left( R_{1},\ldots ,R_{T}\right) \in \mathbb{R}^{T\times n}$. By noting
$D_{w}=\limfunc{diag}\left( w\right) $, the expression of the covariance
matrix becomes:%
\begin{eqnarray*}
\hat{\Sigma} &=&\sum_{t=1}^{T}w_{t}R_{t}R_{t}^{\top }-\hat{\mu}\hat{\mu}%
^{\top } \\
&=&\sum_{t=1}^{T}w_{t}R_{t}R_{t}^{\top }-\left(
\sum_{t=1}^{T}w_{t}R_{t}\right) \left( \sum_{t=1}^{T}w_{t}R_{t}\right)
^{\top } \\
&=&R^{\top }D_{w}R-R^{\top }w\left( R^{\top }w\right) ^{\top } \\
&=&R^{\top }\left( D_{w}-ww^{\top }\right) R
\end{eqnarray*}%
An alternative form is:%
\begin{eqnarray*}
\hat{\Sigma} &=&\sum_{t=1}^{T}w_{t}\left( R_{t}-\hat{\mu}\right) \left(
R_{t}-\hat{\mu}\right) ^{\top } \\
&=&\left( R-\mathbf{1}\hat{\mu}^{\top }\right) ^{\top }D_{w}\left( R-\mathbf{1}%
\hat{\mu}^{\top }\right)  \\
&=&\left( R-\mathbf{1}w^{\top }R\right) ^{\top }D_{w}\left( R-\mathbf{1}w^{\top
}R\right)  \\
&=&R^{\top }\left( C_{T}^{\top }D_{w}C_{T}\right) R
\end{eqnarray*}%
where $C_{T}=I_{T}-\mathbf{1}w^{\top }$ is the (weighted) centering matrix%
\footnote{We verify that:%
\begin{eqnarray*}
C_{T}^{\top }D_{w}C_{T} &=&\left( I_{T}-\mathbf{1}w^{\top }\right) ^{\top
}D_{w}\left( I_{T}-\mathbf{1}w^{\top }\right)  \\
&=&D_{w}-w\mathbf{1}^{\top }D_{w}-D_{w}\mathbf{1}w^{\top }+w\mathbf{1}^{\top }D_{w}\mathbf{1}%
w^{\top } \\
&=&D_{w}-ww^{\top }
\end{eqnarray*}%
because $D_{w}\mathbf{1}=w$ and $\mathbf{1}^{\top }D_{w}\mathbf{1}=1$.}. In the
case
of uniform weights $w_{t}=1/T$, $C_{T}$ is equal to $I_{T}-\dfrac{1}{T}%
\mathbf{11}^{\top }$. We observe that it is symmetric and
idempotent. We deduce that $\hat{\Sigma}=\dfrac{1}{T}R^{\top
}C_{T}R$.

\subsection{Relationship between the conditional normal distribution and the
linear regression} \label{appendix:section-conditional-expectation}

Let us consider a Gaussian random vector defined as follows:%
\begin{equation*}
\left(
\begin{array}{c}
X \\
Y%
\end{array}%
\right) \sim \mathcal{N}\left( \left(
\begin{array}{c}
\mu _{x} \\
\mu _{y}%
\end{array}%
\right) ,\left(
\begin{array}{cc}
\Sigma _{xx} & \Sigma _{xy} \\
\Sigma _{yx} & \Sigma _{yy}%
\end{array}%
\right) \right)
\end{equation*}
The conditional distribution of $Y$\ given $X=x$ is a multivariate normal
distribution:
\begin{equation*}
Y\mid X=x\sim \mathcal{N}\left( \mu _{y\mid x},\Sigma _{yy\mid x}\right)
\end{equation*}%
where:%
\begin{equation*}
\mu _{y\mid x}=\mathbb{E}\left[ Y\mid X=x\right] =\mu _{y}+\Sigma
_{yx}\Sigma _{xx}^{-1}\left( x-\mu _{x}\right)
\end{equation*}%
and:%
\begin{equation*}
\Sigma _{yy\mid x}=\sigma ^{2}\left[ Y\mid X=x\right] =\Sigma _{yy}-\Sigma
_{yx}\Sigma _{xx}^{-1}\Sigma _{xy}
\end{equation*}%
It follows that $Y=\mu _{y\mid x}+U$ where $U$ is a centered Gaussian random
variable with variance $s^{2}=\Sigma _{yy\mid x}$.We recognize the linear
regression of $Y$ on $X$:%
\begin{eqnarray*}
Y &=&\mu _{y}+\Sigma _{yx}\Sigma _{xx}^{-1}\left( x-\mu _{x}\right) +U \\
&=&\left( \mu _{y}-\Sigma _{yx}\Sigma _{xx}^{-1}\mu _{x}\right) +\Sigma
_{yx}\Sigma _{xx}^{-1}x+U \\
&=&\alpha + \beta^{\top }x +U
\end{eqnarray*}%
where $\alpha = \mu _{y}-\Sigma _{yx}\Sigma _{xx}^{-1}\mu _{x}$ and $\beta
=\Sigma _{yx}\Sigma _{xx}^{-1}$. Moreover, we have:%
\begin{eqnarray*}
\mathfrak{R}^{2} &=&1-\frac{\limfunc{var}\left( U\right) }{\limfunc{var}%
\left( Y\right) } \\
&=&1-\frac{s^{2}}{\Sigma _{yy}} \\
&=&\frac{\Sigma _{yx}\Sigma _{xx}^{-1}\Sigma _{xy}}{\Sigma _{yy}}
\end{eqnarray*}

\begin{remark} In the case where the correlation matrix of the random vector
$\left( X,Y\right) $ is constant --
$\mathcal{C}=\mathcal{C}_{n+1}\left( \rho \right) $, Maillard
\textsl{et al.} (2010) proved that:
\begin{equation*}
\mathcal{C}_{xx}^{-1}=\frac{\rho \mathbf{11}^{\top }-\left( \left(
n-1\right) \rho +1\right) I_{n}}{\left( n-1\right) \rho ^{2}-\left(
n-2\right) \rho -1}
\end{equation*}%
We deduce that:%
\begin{eqnarray*}
\beta  &=&\Sigma _{xx}^{-1}\Sigma _{xy} \\
&=&\left( \frac{\sigma _{x}}{\sigma _{y}}\right) \odot \mathcal{C}_{xx}^{-1}\mathcal{C}_{x,y} \\
&=&\left( \frac{\sigma _{x}}{\sigma _{y}}\right) \odot \left(
\frac{\rho \mathbf{11}^{\top }-\left( \left( n-1\right) \rho
+1\right) I_{n}}{\left( n-1\right) \rho ^{2}-\left( n-2\right) \rho
-1}\right) \rho \mathbf{1}
\end{eqnarray*}%
and:%
\begin{equation*}
\beta _{i}=\frac{\rho \left( \rho -1\right) }{\left( n-1\right) \rho
^{2}-\left( n-2\right) \rho -1}\cdot \frac{\sigma _{y}}{\sigma _{x_{i}}}
\end{equation*}%
where $\sigma _{y}$ and $\sigma _{x}$ are the standard deviation of random
vectors $Y$ and $X$. The coefficient of determination becomes:
\begin{eqnarray*}
\mathfrak{R}^{2} &=&\frac{\Sigma _{yx}\Sigma _{xx}^{-1}\Sigma _{xy}}{\Sigma
_{yy}} \\
&=&\frac{n\rho ^{2}}{n\rho -\left( \rho -1\right) }
\end{eqnarray*}%
In the two-asset case, we obtain the famous result: $\mathfrak{R}^{2}=\rho
^{2}$. When the number of assets is very large, the coefficient of
determination is equal to the uniform correlation:
\begin{equation*}
\lim_{n\rightarrow \infty }\mathfrak{R}^{2}=\left\{
\begin{array}{ll}
1 & \text{if }\rho <0 \\
\rho  & \text{if }\rho \geqslant 0%
\end{array}%
\right.
\end{equation*}
\end{remark}

\subsection{Tikhonov regularization}
\label{appendix:section-tikhonov}

We consider the following optimization problem:%
\begin{eqnarray}
x^{\star } &=&\arg \min \frac{1}{2}\left\Vert A_1 x - b_1\right\Vert _{2}^{2}+\frac{%
1}{2}\varrho_2 \left\Vert \Gamma_2 \left( x-x_{0}\right) \right\Vert _{2}^{2}
\label{eq:app-tikhonov1} \\
&\text{s.t. }& A_2 x = b_2  \notag
\end{eqnarray}%
where $A_1\in \mathbb{R}^{T\times n}$, $b_1\in \mathbb{R}^{T\times 1}$, $\varrho_2
>0$, $\Gamma \in \mathbb{R}^{n\times n}$, $A_2 \in \mathbb{R}^{m\times n}$, $%
b_2\in \mathbb{R}^{m\times 1}$ and $d\in \mathbb{R}^{m\times 1}$. We
assume that $A_1$ has full rank. The Tikhonov matrix $\Gamma_2 $
forces desirable properties of the solution whereas $\varrho_2 $
indicates the strength of the regularization. $x_{0}$ is an initial
solution. In the case of portfolio optimization, it could be an
heuristic portfolio (like the EW portfolio) or the current
allocation in order to control the turnover (Scherer, 2007). The
Lagrange function is equal to:
\begin{equation*}
\mathcal{L}\left( x,\lambda \right) =\frac{1}{2}\left\Vert A_1 x - b_1\right\Vert
_{2}^{2}+\frac{1}{2}\varrho_2 \left\Vert \Gamma_2 \left( x-x_{0}\right)
\right\Vert _{2}^{2}+\lambda ^{\top }\left( A_2 x - b_2\right)
\end{equation*}%
Computation of the gradient leads to:%
\begin{equation*}
\partial _{x}\,\mathcal{L}\left( x,\lambda \right) =A_1^{\top }\left(
A_1 x - b_1\right) +\varrho_2 \Gamma_2 ^{\top }\Gamma_2 \left( x-x_{0}\right) + A_2^{\top
}\lambda
\end{equation*}%
Since we have $\partial _{x}\,\mathcal{L}\left( x,\lambda \right) =\mathbf{0} $
and $A_2 x = b_2$, the optimal portfolio $x^{\star }$ is the $x$-coordinate
solution of the linear system:%
\begin{equation}
\left(
\begin{array}{cc}
A_1^{\top }A_1+\varrho_2 \Gamma_2 ^{\top }\Gamma_2  & A_2^{\top } \\
A_2 & \mathbf{0}%
\end{array}%
\right) \left(
\begin{array}{c}
x \\
\lambda
\end{array}%
\right) =\left(
\begin{array}{c}
A_1^{\top }b_1+\varrho_2 \Gamma_2 ^{\top }\Gamma_2 x_{0} \\
b_2%
\end{array}%
\right)   \label{eq:app-tikhonov2}
\end{equation}%
This linear system gives the primal and dual variables.

\subsection{Limit solutions of $L_p$ - $L_2$ regularization}
\label{appendix:section-limit-solution}

As $\varrho _{p}$ is fixed and $\varrho_2$ tends to $+\infty$, the formal limit
to Problem \eqref{eq:mixed2} is given by:
\begin{equation*}
x^{\star } = \arg \min \left\Vert \Gamma _{2}\left( x-x_{0}\right) \right\Vert_{2}^{2}
\quad \text{s.t.} \quad A_{2}x=b_{2}
\end{equation*}
As $\varrho_2$ is fixed and $\varrho_p$ tends to $+\infty$, the formal limit to
Problem \eqref{eq:mixed2} is given by:
\begin{equation*}
x^{\star } = \arg \min \left\Vert \Gamma _{p}\left( x-x_{0}\right) \right\Vert_{p}^{p}
\quad \text{s.t.} \quad A_{2}x=b_{2}
\end{equation*}
If $\varrho_p$ and $\varrho_2$ both tend to $+\infty$, the formal limit to
Problem \eqref{eq:mixed2} depends on the regime $\varrho_p/\varrho_2$.

\subsection{Augmented QP algorithm}
\label{appendix:section-qprog}

A quadratic programming (QP) problem is an optimization problem with a
quadratic objective function and linear constraints:
\begin{eqnarray}
x^{\star } &=&\arg \min \frac{1}{2}x^{\top }A_{1}x-x^{\top }b_{1}
\label{eq:qprog1} \\
&\text{s.t.}&A_{3}x\geqslant b_{3}  \notag
\end{eqnarray}%
With the inequality constraints, we can easily manage equality constraints
and bounds\footnote{%
An equality constraint $A_{2}x=b_{2}$ is equivalent to two inequality
constraints $A_{2}x\geqslant b_{2}$ and $A_{2}x\leqslant b_{2}$. The same
result applies to bounds $x^{-}\leqslant x\leqslant x^{+}$, which can be
written as $x\geqslant x^{-}$ and $-x\geqslant -x^{+}$.}. If we introduce a $%
L_{2}$ penalization, the optimization program becomes:
\begin{eqnarray*}
\left( \ast \right)  &=&\frac{1}{2}x^{\top }A_{1}x-x^{\top }b_{1}+\frac{1}{2}%
\varrho _{2}\left\Vert \Gamma _{2}\left( x-x_{0}\right) \right\Vert _{2}^{2}
\\
&=&\frac{1}{2}x^{\top }A_{1}x-x^{\top }b_{1}+\frac{1}{2}\varrho _{2}x^{\top
}\Gamma _{2}x -\varrho _{2}x^{\top }\Gamma _{2}x_{0}+\frac{1}{2}\varrho
_{2}x_{0}^{\top }\Gamma _{2}x_{0}
\end{eqnarray*}%
We deduce that the regularization program can be cast into a QP problem:
\begin{eqnarray}
x^{\star } &=&\arg \min \frac{1}{2}x^{\top }A_{1}\left( \varrho _{2}\right)
x-x^{\top }b_{1}\left( \varrho _{2}\right)   \label{eq:qprog2} \\
&\text{s.t.}&A_{3}x\geqslant b_{3}  \notag
\end{eqnarray}%
where $A_{1}\left( \varrho _{2}\right) =A_{1}+\varrho _{2}\Gamma _{2}$ and $%
b_{1}\left( \varrho _{2}\right) =b_1 + \varrho _{2}\Gamma _{2}x_{0}$.\smallskip

Let us now introduce an $L_{1}$ penalization. We have:%
\begin{eqnarray}
x^{\star } &=&\arg \min f\left( x\right)   \notag \\
&\text{s.t.}&A_{3}x\geqslant b_{3}  \notag
\end{eqnarray}%
where:%
\begin{equation*}
f\left( x\right) =\frac{1}{2}x^{\top }A_{1}x-x^{\top }b_{1}+\varrho
_{1}\left\Vert \Gamma _{1}\left( x-x_{0}\right) \right\Vert _{1}
\end{equation*}%
and $\Gamma _{1}$ is a matrix with non-negative entries. If we use a decomposition
of the following form:
\begin{equation}
x=x_{0}+\delta ^{+}-\delta ^{-}  \label{eq:qprog3}
\end{equation}%
with $\delta ^{-}=\left( \delta _{1}^{-},\ldots ,\delta _{n}^{-}\right) $, $%
\delta ^{+}=\left( \delta _{1}^{+},\ldots ,\delta _{n}^{+}\right) $, $\delta
_{i}^{-}\geqslant 0$ and $\delta _{i}^{+}\geqslant 0$, we deduce that:%
\begin{equation*}
\left\Vert \Gamma _{1}\left( x-x_{0}\right) \right\Vert _{1}=\left\Vert
\Gamma _{1}\left( \delta ^{+}-\delta ^{-}\right) \right\Vert _{1}=\mathbf{1}%
^{\top }\left( \Gamma _{1}\left( \delta ^{+}+\delta ^{-}\right) \right)
\end{equation*}%
The objective function becomes:%
\begin{equation*}
f\left( x\right) =\frac{1}{2}x^{\top }A_{1}x-x^{\top }b_{1}+\mathbf{1}^{\top
}\Gamma _{1}\delta ^{+}+\mathbf{1}^{\top }\Gamma _{1}\delta ^{-}
\end{equation*}%
Let $y=\left( x,\delta ^{-},\delta ^{+}\right) $ be the vector of unknown
variables. We obtain an augmented QP problem of dimension $3\times n$:
\begin{eqnarray}
y^{\star } &=&\arg \min \frac{1}{2}y^{\top }\tilde{A}_{1}y-y^{\top }\tilde{b}%
_{1}  \label{eq:qprog4} \\
&\text{s.t.}&\tilde{A}_{3}y\geqslant \tilde{b}_{3}  \notag
\end{eqnarray}%
where:%
\begin{equation*}
\tilde{A}_{1}=\left(
\begin{array}{ccc}
A_{1} & \mathbf{0} & \mathbf{0} \\
\mathbf{0} & \mathbf{0} & \mathbf{0} \\
\mathbf{0} & \mathbf{0} & \mathbf{0}%
\end{array}%
\right)
\end{equation*}%
and:%
\begin{equation*}
\tilde{b}_{1}=\left(
\begin{array}{c}
b_{1} \\
-\Gamma _{1}^{\top} \mathbf{1} \\
-\Gamma _{1}^{\top} \mathbf{1}%
\end{array}%
\right)
\end{equation*}%
We can write Equation (\ref{eq:qprog3}) as follows:%
\begin{equation*}
I_{n}x+I_{n}\delta ^{-}-I_{n}\delta ^{+}=x_{0}
\end{equation*}%
Since we have $\delta ^{+}\geqslant \mathbf{0}$ and $\delta ^{-}\geqslant
\mathbf{0}$, we deduce that:%
\begin{equation*}
\tilde{A}_{3}=\left(
\begin{array}{rrr}
A_{3} & \mathbf{0} & \mathbf{0} \\
I_{n} & I_{n} & -I_{n} \\
-I_{n} & -I_{n} & I_{n} \\
\mathbf{0} & I_{n} & \mathbf{0} \\
\mathbf{0} & \mathbf{0} & I_{n}%
\end{array}%
\right)
\end{equation*}%
and:
\begin{equation*}
\tilde{b}_{3}=\left(
\begin{array}{r}
b_{3} \\
x_{0} \\
-x_{0} \\
\mathbf{0} \\
\mathbf{0}%
\end{array}%
\right)
\end{equation*}

\subsection{ADMM algorithm}
\label{appendix:section-admm}

\subsubsection{Dual ascent principle and method of multipliers}

The alternating direction method of multipliers (ADMM) is an algorithm
introduced by Gabay and Mercier (1976) to solve problems which can be expressed
as\footnote{We follow the standard presentation of Boyd \textsl{et al.} (2011)
on ADMM.}:
\begin{eqnarray}
\left\{ x^{\star },z^{\star }\right\} &=&\arg \min f\left( x\right) +g\left(
z\right)  \label{eq:appendix-admm1} \\
&\text{s.t.}&Ax+Bz=c  \notag
\end{eqnarray}%
where $A\in \mathbb{R}^{p\times n}$, $B\in \mathbb{R}^{p\times m}$, $c\in
\mathbb{R}^{p}$, and the functions $f:\mathbb{R}^{n}\rightarrow \mathbb{R}%
\cup \{+\infty \}$ and $g:\mathbb{R}^{m}\rightarrow \mathbb{R}\cup \{+\infty
\}$ are proper closed convex functions. The expression of the augmented
Lagrange function is:
\begin{equation*}
\mathcal{L}_{\varphi }\left( x,z,\lambda \right) =f\left( x\right) +g\left(
z\right) +\lambda ^{\top }\left( Ax+Bz-c\right) +\frac{\varphi }{2}%
\left\Vert Ax+Bz-c\right\Vert _{2}^{2}
\end{equation*}%
where $\varphi >0$. The ADMM algorithm uses the property that the objective
function is separable, and consists of the following iterations:
\begin{eqnarray*}
x^{\left( k+1\right) } &=&\arg \min \mathcal{L}_{\varphi }\left( x,z^{\left(
k\right) },\lambda ^{\left( k\right) }\right) \\
&=&\arg \min \left\{ f\left( x\right) +\lambda ^{\left( k\right) ^{\top
}}\left( Ax+Bz^{\left( k\right) }-c\right) +\frac{\varphi }{2}\left\Vert
Ax+Bz^{\left( k\right) }-c\right\Vert _{2}^{2}\right\}
\end{eqnarray*}%
and:%
\begin{eqnarray*}
z^{\left( k+1\right) } &=&\arg \min \mathcal{L}_{\varphi }\left( x^{\left(
k+1\right) },z,\lambda ^{\left( k\right) }\right) \\
&=&\arg \min \left\{ g\left( z\right) +\lambda ^{\left( k\right) ^{\top
}}\left( Ax^{\left( k+1\right) }+Bz-c\right) +\frac{\varphi }{2}\left\Vert
Ax^{\left( k+1\right) }+Bz-c\right\Vert _{2}^{2}\right\}
\end{eqnarray*}%
The update for the dual variable $\lambda $ is then:
\begin{equation*}
\lambda ^{\left( k+1\right) }=\lambda ^{\left( k\right) }+\varphi \left(
Ax^{\left( k+1\right) }+Bz^{\left( k+1\right) }-c\right)
\end{equation*}%
We repeat the iterations until convergence.\smallskip

Boyd \textsl{et al.} (2011) notice that the previous algorithm can be
simplified. Let $r=Ax+Bz-c$ be the (primal) residual. By combining
linear and quadratic terms, we have:
\begin{equation*}
\lambda ^{\top }r+\frac{\varphi }{2}r^{2}=\frac{\varphi }{2}\left\Vert
r+u\right\Vert ^{2}-\frac{\varphi }{2}\left\Vert u\right\Vert ^{2}
\end{equation*}%
where $u=\varphi ^{-1}\lambda $ is the \textit{scaled} dual variable. We can
then write the Lagrange function (\ref{eq:appendix-admm1}) as follows:
\begin{equation}
\mathcal{L}_{\varphi }\left( x,z,u\right) =f\left( x\right) +g\left(
z\right) +\frac{\varphi }{2}\left\Vert Ax+Bz-c+u\right\Vert _{2}^{2}-\frac{1%
}{2\varphi }\left\Vert \lambda \right\Vert ^{2}  \label{eq:appendix-admm2}
\end{equation}%
Since the last term is a constant, we deduce that the $x$- and $z$-updates
become:
\begin{eqnarray}
x^{\left( k+1\right) } &=&\arg \min \mathcal{L}_{\varphi }\left( x,z^{\left(
k\right) },u^{\left( k\right) }\right)   \notag \\
&=&\arg \min \left\{ f\left( x\right) +\frac{\varphi }{2}\left\Vert
Ax+Bz^{\left( k\right) }-c+u^{\left( k\right) }\right\Vert _{2}^{2}\right\}
\label{eq:appendix-admm3a}
\end{eqnarray}%
and:%
\begin{eqnarray}
z^{\left( k+1\right) } &=&\arg \min \mathcal{L}_{\varphi }\left( x^{\left(
k+1\right) },z,u^{\left( k\right) }\right)   \notag \\
&=&\arg \min \left\{ g\left( z\right) +\frac{\varphi }{2}\left\Vert
Ax^{\left( k+1\right) }+Bz-c+u^{\left( k\right) }\right\Vert
_{2}^{2}\right\}   \label{eq:appendix-admm3b}
\end{eqnarray}%
For the scaled dual variable $u^{\left( k\right) }$, we have:
\begin{eqnarray}
u^{\left( k+1\right) } &=&u^{\left( k\right) }+r^{\left( k+1\right) }  \notag
\\
&=&u^{\left( k\right) }+\left( Ax^{\left( k+1\right) }+Bz^{\left( k+1\right)
}-c\right)   \label{eq:appendix-admm3c}
\end{eqnarray}%
where $r^{\left( k+1\right) }=Ax^{\left( k+1\right) }+Bz^{\left( k+1\right)
}-c$ is the primal residual at iteration $k+1$. Boyd \textsl{et al.} (2011)
also defined the variable $s^{\left( k+1\right) }=\varphi A^{\top }B\left(
z^{\left( k+1\right) }-z^{\left( k\right) }\right) $ and refer to $s^{\left(
k+1\right) }$ as the dual residual\footnote{%
We can interpret $s^{\left( k+1\right) }$ as the residual of the dual
feasibility conditions: $0\in \partial f\left( x^{\star }\right) +A^{\top
}\lambda ^{\star }$ and $0\in \partial g\left( z^{\star }\right) +B^{\top
}\lambda ^{\star }$ (Boyd \textsl{et al.}, 2011).} at iteration
$k+1$.\smallskip

This algorithm benefits from the dual ascent principle and the method of
multipliers. The difference with the latter is that the $x$ and $z$-updates are
performed in an alternating way. Therefore, it is more flexible because the
updates are equivalent to compute proximal operators for $f$ and $g$,
independently.

\subsubsection{Convergence and stopping criteria}

Under the assumption that the traditional Lagrange function $\mathcal{L}_{0}$
has a saddle point, one can prove that the residual $r^{\left( k\right) }$
converges to zero, the objective function $f\left( x^{\left( k\right)
}\right) +g\left( z^{\left( k\right) }\right) $ to the optimal value $%
f\left( x^{\star }\right) +g\left( z^{\star }\right) ,$ and the dual variable
$\lambda ^{\left( k\right) }$ to a dual optimal point. However, the rate of
convergence is not known and the primal variables $x^{\left( k\right) }$ and
$z^{\left( k\right) }$ do not necessarily converge to the optimal values
$x^{\star }$ and $z^{\star }$. Nevertheless, in the context of Markowitz
optimization with bound constraints, the results found by Raghunathan and Di
Cairano (2014) may be applied to obtain linear convergence for the primal
variables.\smallskip

In general, the stopping criterion is defined with respect to the
residuals:%
\begin{equation*}
\left\{
\begin{array}{l}
\left\Vert r^{\left( k\right) }\right\Vert _{2}\leqslant \varepsilon  \\
\left\Vert s^{\left( k\right) }\right\Vert _{2}\leqslant \varepsilon
^{\prime }%
\end{array}%
\right.
\end{equation*}%
where $r^{\left( k\right) }=Ax^{\left( k\right) }+Bz^{\left( k\right) }-c$ and
$s^{\left( k\right) }=\varphi A^{\top }B\left( z^{\left( k\right) }-z^{\left(
k-1\right) }\right) $. Typical values when implementing this stopping criterion
are $\varepsilon =\varepsilon ^{\prime }=10^{-18}$.

\subsubsection{Penalization parameter and initialization}

The convergence result holds regardless of the choice of the penalization
parameter $\varphi >0$. But the choice of $\varphi $ affects the speed of
convergence (Ghadimi \textsl{et al.}, 2015; Giselsson and Boyd, 2017). In
practice, the penalization parameter $\varphi $ may be changed at each
iteration, implying that $\varphi $ is replaced by $\varphi ^{\left( k\right)
}$ and the scaled dual variable $u^{k}$ is equal to $\lambda ^{\left( k\right)
}/\varphi ^{\left( k\right) }$. This may improve the
convergence and make the performance independent of the initial choice $%
\varphi ^{\left( 0\right) }$. To update $\varphi ^{\left( k\right)
}$ in practice, He \textsl{et al.} (2000) and Wang and Liao (2001)
provide a simple and efficient scheme. On the one hand, the $x$ and
$z$-updates in ADMM essentially comes from placing a penalty on
$\left\Vert r^{\left( k\right) }\right\Vert _{2}^{2}$. As a
consequence, if $\varphi ^{\left( k\right) }$ is large, $\left\Vert
r^{\left( k\right) }\right\Vert _{2}^{2}$ tends to be
small. On the other hand, $s^{\left( k\right) }$ depends linearly on $%
\varphi $. As a consequence, if $\varphi ^{\left( k\right) }$ is small, $%
\left\Vert s^{\left( k\right) }\right\Vert _{2}^{2}$ is small (and $%
\left\Vert r^{\left( k\right) }\right\Vert _{2}^{2}$ may be large). To keep $%
\left\Vert r^{\left( k\right) }\right\Vert _{2}^{2}$ and $\left\Vert s^{\left(
k\right) }\right\Vert _{2}^{2}$ within a factor $\mu $, one may consider:
\begin{equation*}
\varphi ^{\left( k+1\right) }=\left\{
\begin{array}{ll}
\tau \varphi ^{\left( k\right) } & \text{if }\left\Vert r^{\left( k\right)
}\right\Vert _{2}^{2}>\mu \left\Vert s^{\left( k\right) }\right\Vert _{2}^{2}
\\
\varphi ^{\left( k\right) }/\tau ^{\prime } & \text{if }\left\Vert s^{\left(
k\right) }\right\Vert _{2}^{2}>\mu \left\Vert r^{\left( k\right)
}\right\Vert _{2}^{2} \\
\varphi ^{\left( k\right) } & \text{otherwise}%
\end{array}%
\right.
\end{equation*}%
where $\mu $, $\tau $ and $\tau ^{\prime }$ are parameters that are greater
than one. In practice, we use $\varphi ^{\left( 0\right) }=1$, $u^{\left(
0\right) }=0$, $\mu =10^{3}$ and $\tau =\tau ^{\prime }=2$.

\subsubsection{Tikhonov regularization}

Let us consider the Tikhonov problem:%
\begin{eqnarray}
x^{\star } &=&\arg \min \frac{1}{2}\left\Vert A_{1}x-b_{1}\right\Vert
_{2}^{2}+\frac{1}{2}\varrho _{2}\left\Vert \Gamma _{2}\left( x-x_{0}\right)
\right\Vert _{2}^{2} \label{eq:appendix-admm-tikhonov} \\
&s.t.&\left\{
\begin{array}{l}
\left\Vert x\right\Vert _{q}\leqslant c_{q} \\
A_{2}x=b_{2} \\
A_{3}x\leqslant b_{3} \\
x^{-}\leqslant x\leqslant x^{+}%
\end{array}%
\right. \notag
\end{eqnarray}
where $q\in \lbrack 1,\infty )$. We note:%
\begin{eqnarray*}
\Omega _{1} &=&\left\{ x\in \mathbb{R}^{n}:\left\Vert x\right\Vert
_{q}\leqslant c_{q}\right\}  \\
\Omega _{2} &=&\left\{ x\in \mathbb{R}^{n}:A_{2}x=b_{2}\right\}  \\
\Omega _{3} &=&\left\{ x\in \mathbb{R}^{n}:A_{3}x\geqslant b_{3}\right\}
\\
\Omega _{4} &=&\left\{ x\in \mathbb{R}^{n}:x^{-}\leqslant x\leqslant
x^{+}\right\}
\end{eqnarray*}
We define:%
\begin{equation*}
f\left( x\right) =\frac{1}{2}\left\Vert A_{1}x-b_{1}\right\Vert _{2}^{2}+%
\frac{1}{2}\varrho _{2}\left\Vert \Gamma _{2}\left( x-x_{0}\right)
\right\Vert _{2}^{2}+\mathbf{1}_{\Omega _{2}}\left( x\right)
\end{equation*}%
and:%
\begin{equation*}
g\left( x\right) =\mathbf{1}_{\Omega _{1}}\left( x\right) +\mathbf{1}%
_{\Omega _{3}}\left( x\right) +\mathbf{1}_{\Omega _{4}}\left( x\right)
\end{equation*}%
The Tikhonov problem becomes:%
\begin{eqnarray*}
\left\{ x^{\star },z^{\star }\right\}  &=&\arg \min f\left( x\right)
+g\left( z\right)  \\
& \text{s.t.} & x-z=\mathbf{0}
\end{eqnarray*}%
Therefore, the ADMM algorithm is:
\begin{eqnarray*}
x^{\left( k+1\right) } &=&\arg \min \left\{ f\left( x\right) +\frac{\varphi
^{\left( k\right) }}{2}\left\Vert x-z^{\left( k\right) }+u^{\left( k\right)
}\right\Vert _{2}^{2}\right\}  \\
z^{\left( k+1\right) } &=&\arg \min \left\{ g\left( z\right) +\frac{\varphi
^{\left( k\right) }}{2}\left\Vert x^{\left( k+1\right) }-z+u^{\left(
k\right) }\right\Vert _{2}^{2}\right\}  \\
u^{\left( k+1\right) } &=&u^{\left( k\right) }+\left( x^{\left( k+1\right)
}-z^{\left( k+1\right) }\right)
\end{eqnarray*}%
We notice that we can replace the second step by:%
\begin{equation*}
z^{\left( k+1\right) }=\mathcal{P}_{\left\{ g\left( z\right) <\infty
\right\} }\left( x^{\left( k+1\right) }+u^{\left( k\right) }\right)
\end{equation*}%
where $\mathcal{P}_{\left\{ g\left( z\right) <\infty \right\} }\left( x^{\left(
k+1\right) }+u^{\left( k\right) }\right) $ is the orthogonal projection of
$x^{\left( k+1\right) }+u^{\left( k\right) }$ onto the convex set $\left\{ z\in
\mathbb{R}^{n}:g\left( z\right) <\infty \right\} $.With
this formulation, the $x$-step is explicit%
\footnote{%
The $x$-step is also given by:%
\begin{equation*}
\left(
\begin{array}{cc}
A_{1}^{\top }A_{1}+\varrho _{2}\Gamma _{2}^{\top }\Gamma _{2}+\varphi
^{\left( k\right) }I_{n} & A_{2}^{\top } \\
A_{2} & 0%
\end{array}%
\right) \left(
\begin{array}{c}
x^{\left( k+1\right) } \\
\lambda
\end{array}%
\right) =\left(
\begin{array}{c}
A_{1}^{\top }b_{1}+\varrho _{2}\Gamma _{2}^{\top }\Gamma _{2}x_{0}+\varphi
^{\left( k\right) }\left( z^{\left( k\right) }-u^{\left( k\right) }\right)
\\
b_{2}%
\end{array}%
\right)
\end{equation*}%
}, while the $z$-step consists in computing orthogonal projections onto a
convex set. Explicit formulas for orthogonal projections are presented in
Appendix \ref{appendix:section-proximal} on page
\pageref{appendix:section-proximal}.

\subsubsection{Mixed regularization}

We now replace the objective function of the Tikhonov problem by:
\begin{equation}
x^{\star }=\arg \min \frac{1}{2}\left\Vert A_{1}x-b_{1}\right\Vert
_{2}^{2}+\frac{1}{2}\varrho _{2}\left\Vert \Gamma _{2}\left( x-x_{0}\right)
\right\Vert _{2}^{2}+\frac{1}{p}\varrho _{p}\left\Vert \Gamma _{p}\left(
x-x_{0}\right) \right\Vert _{p}^{p}
\label{eq:appendix-admm-mixed}
\end{equation}%
where $p\neq 2$. The constraints are the same than those specified for the
Tikhonov problem. We define:
\begin{eqnarray*}
f\left( x\right)  &=&\frac{1}{2}\left\Vert A_{1}x-b_{1}\right\Vert
_{2}^{2}+\frac{1}{2}\varrho _{2}\left\Vert \Gamma _{2}\left( x-x_{0}\right)
\right\Vert _{2}^{2}+ \\
&&\mathbf{1}_{\Omega _{1}}\left( x\right) +\mathbf{1}_{\Omega _{2}}\left(
x\right) +\mathbf{1}_{\Omega _{3}}\left( x\right) +\mathbf{1}_{\Omega
_{4}}\left( x\right)
\end{eqnarray*}%
and:%
\begin{equation*}
g\left( z\right) =\frac{1}{p}\varrho _{p}\left\Vert z\right\Vert _{p}^{p}
\end{equation*}%
The $L_{2}-L_{p}$ problem becomes:%
\begin{eqnarray*}
\left\{ x^{\star },z^{\star }\right\}  &=&\arg \min f\left( x\right)
+g\left( z\right)  \\
&\text{s.t.}&\Gamma _{p}\left( x-x_{0}\right) -z=\mathbf{0}
\end{eqnarray*}%
With this specification, the ADMM algorithm is:%
\begin{eqnarray*}
x^{\left( k+1\right) } &=&\arg \min \left\{ f\left( x\right) +\frac{\varphi
^{\left( k\right) }}{2}\left\Vert \Gamma _{p}x-z^{\left( k\right) }-\Gamma
_{p}x_{0}+u^{\left( k\right) }\right\Vert _{2}^{2}\right\}  \\
z^{\left( k+1\right) } &=&\arg \min \left\{ g\left( z\right) +\frac{\varphi
^{\left( k\right) }}{2}\left\Vert \Gamma _{p}x^{\left( k+1\right) }-z-\Gamma
_{p}x_{0}+u^{\left( k\right) }\right\Vert _{2}^{2}\right\}  \\
u^{\left( k+1\right) } &=&u^{\left( k\right) }+\left( \Gamma _{p}x^{\left(
k+1\right) }-z^{\left( k+1\right) }-\Gamma _{p}x_{0}\right)
\end{eqnarray*}%
The $x$-step consists in minimizing a quadratic constrained problem. It can be
carried out explicitly if no inequality constraint is imposed. Otherwise, the $x$-step
can be performed by another ADMM. The $z$-step consists in
computing the proximal operator of $\lambda \left\Vert z\right\Vert _{p}^{p}$
at the point $z=\Gamma _{p}x^{\left( k+1\right) }-\Gamma _{p}x_{0}+u^{\left(
k\right) }$ with $\lambda =\varrho _{p}/\left( p\varphi ^{\left( k\right)
}\right) $. Other choices for the functions $f\left( x\right) $ and $g\left(
z\right) $ give rise to computing constrained proximal operators or the proximal
operator of $x\mapsto \left\Vert \Gamma _{p}x\right\Vert _{p}^{p}$. No explicit
formula is known for the latter, unless a positive multiple of $\Gamma _{p}$
is orthogonal (Beck, 2017). Our choice makes the $z$-step explicit for $p\in
\left\{ 1,2,3,4,5\right\}$, and easily computable for any $p>1$.

\subsubsection{Cardinality constraints}

The ADMM algorithm can also be used to find a portfolio with at most $n_{1}$
non-zero weights. Let us introduce the set $\mathcal{Z}$ of $n_{1}$-sparse
vectors:
\begin{equation}
\mathcal{Z}=\left\{ x\in \mathbb{R}^{n}\mid \limfunc{card}x\leqslant
n_{1},x^{-}\leqslant x\leqslant x^{+}\right\}   \label{eq:appendix-sparse1}
\end{equation}%
We consider the augmented Tikohnov problem:%
\begin{eqnarray}
x^{\star } &=&\arg \min \frac{1}{2}\left\Vert A_{1}x-b_{1}\right\Vert
_{2}^{2}+\frac{1}{2}\varrho _{2}\left\Vert \Gamma _{2}\left( x-x_{0}\right)
\right\Vert _{2}^{2}  \label{eq:appendix-sparse2} \\
&\text{s.t.}&\left\{
\begin{array}{l}
x\in \Omega _{1}\cap \Omega _{2}\cap \Omega _{3}\cap \Omega _{4} \\
\Gamma_1 \left(x - x_0\right) \in \mathcal{Z}%
\end{array}%
\right.   \notag
\end{eqnarray}%
Zou and Hastie (2005) have been introduced Problem
(\ref{eq:appendix-admm-mixed}) with $p=1$ as a convex relaxation to problem
(\ref{eq:appendix-sparse2}). The constraint $x\in \mathcal{Z}$ is forced by the
penalty $\varrho _{1}\left\Vert \Gamma _{1}\left( x-x_{0}\right) \right\Vert
_{1}$ and the strength of the penalty parameter $\varrho _{1}$ must be chosen
as the smallest value that satisfies the constraint $\limfunc{card}x\leqslant
n_{1}$ (Hastie et al., 2009).\smallskip

The projection onto the non-convex set $\mathcal{Z}$ exists and is explicit
(but may not be unique). Diamond \textsl{et al.} (2018) show that:%
\begin{equation*}
\mathcal{P}_{\mathcal{Z}}\left( v\right) =\mathcal{P}_{\Omega_{4}}%
\left( v\left( n_{1}\right) \right)
\end{equation*}%
where $v\left( n_{1}\right) _{i}=v_{i}$ if $i\in \mathcal{I}$, $v\left(
n_{1}\right) _{i}=0$ if $i\not\in \mathcal{I}$, $\mathcal{I}$ is a set of
indices of the $n_{1}$ largest values of $\left\vert v_{i}\right\vert $, and
$\mathcal{P}_{\Omega_{4}}$ is the projection onto $\Omega
_{4}=\left\{ x\in \mathbb{R}^{n}:x^{-}\leqslant x\leqslant x^{+}\right\} $.
As previously, we have:%
\begin{eqnarray*}
f\left( x\right)  &=&\frac{1}{2}\left\Vert A_{1}x-b_{1}\right\Vert
_{2}^{2}+\frac{1}{2}\varrho _{2}\left\Vert \Gamma _{2}\left( x-x_{0}\right)
\right\Vert _{2}^{2}+ \\
&&\mathbf{1}_{\Omega _{1}}\left( x\right) +\mathbf{1}_{\Omega _{2}}\left(
x\right) +\mathbf{1}_{\Omega _{3}}\left( x\right) +\mathbf{1}_{\Omega
_{4}}\left( x\right)
\end{eqnarray*}%
and:%
\begin{equation*}
g\left( z\right) =\mathbf{1}_{\mathcal{Z}}\left( z\right)
\end{equation*}%
with the constraint $\Gamma _{1}\left( x-x_{0}\right) =z$. With this
specification, the ADMM algorithm is:%
\begin{eqnarray*}
x^{\left( k+1\right) } &=&\arg \min \left\{ f\left( x\right) +\frac{\varphi
^{\left( k\right) }}{2}\left\Vert \Gamma _{p}x-z^{\left( k\right) }-\Gamma
_{p}x_{0}+u^{\left( k\right) }\right\Vert _{2}^{2}\right\}  \\
z^{\left( k+1\right) } &=&\mathcal{P}_{\mathcal{Z}}\left( \Gamma
_{1}x^{\left( k+1\right) }-\Gamma _{1}x_{0}+u^{\left( k\right) }\right)  \\
u^{\left( k+1\right) } &=&u^{\left( k\right) }+\left( \Gamma _{1}x^{\left(
k+1\right) }-z^{\left( k+1\right) }-\Gamma _{1}x_{0}\right)
\end{eqnarray*}
Hence, the $z$-step is explicit. The ADMM does not necessarily converge, and
when it does, it does not necessarily converge to an optimal point. Contrary to
the convex case, the possible convergence of the algorithm depends on the
initial values of $x^{0}$ and the penalization parameter $\varphi ^{\left(
k\right) }$. In the non-convex setting, the ADMM may be considered as a local
optimization method, and local neighbor search method with convex relaxation
and restrictions may be used to obtain the convergence of the algorithm
(Diamond \textsl{et al.}, 2018).

\subsection{Proximal operators and projections}
\label{appendix:section-proximal}

As shown previously, the $z$-step of the ADMM algorithm generally
computes the proximal operator of a norm or the projection onto the intersection
of simple convex sets. We review the most useful cases in active asset
management and we refer the reader to Parikh and Boyd (2014), Beck (2017), and
Combettes and M\"uller (2018) for further examples. In most of these cases, the
proximal operators are explicit or consists in determining the zero of a
real-valued function.

\subsubsection{Definition of the proximal operator}

Let $f:\mathbb{R}^{n}\rightarrow \mathbb{R}\cup \left\{ +\infty \right\} $
be a proper closed convex function. The proximal operator $\mathbf{prox}%
_{f}\left( v\right) :\mathbb{R}^{n}\rightarrow \mathbb{R}^{n}$ is defined by:%
\begin{equation}
\mathbf{prox}_{f}\left( v\right) = x^{\star} = \arg \min\nolimits_{x}\left\{ f\left(
x\right) +\frac{1}{2}\left\Vert x-v\right\Vert _{2}^{2}\right\}
\label{eq:appendix-prox1}
\end{equation}%
Since the function $f_{v}\left( x\right) =f\left( x\right) +\dfrac{1}{2}%
\left\Vert x-v\right\Vert _{2}^{2}$ is strongly convex, it has a unique minimum
for every $v\in \mathbb{R}^{n}$ (Beck, 2017; Parikh and Boyd, 2014).\smallskip

If we would like to compute the proximal operator of $\lambda f\left( x\right)
+\mathds{1}_{\Omega }\left( x\right) $ for $\lambda \geqslant 0$,
one has to solve:%
\begin{eqnarray*}
x^{\star } &=&\arg \min\nolimits_{x}\left\{ \lambda f\left( x\right) +\frac{1%
}{2}\left\Vert x-v\right\Vert _{2}^{2}\right\} \\
\text{} &\text{s.t.}&x\in \Omega
\end{eqnarray*}%
In the case $\lambda =0$, we have to determine the orthogonal projection $%
\mathcal{P}_{\Omega }\left( v\right) $ of $v$ onto the set $\Omega $. In the
case $\lambda >0$, we may use different optimization algorithms depending on
the regularity of $f\left( x\right) $ and the presence/absence of the set of
constraints $\Omega $ (Nocedal and Wright, 2006).

\subsubsection{The $L_{p}$ norm}

To compute the proximal operator of $f\left( x\right) =\lambda \dfrac{1}{p}%
\left\Vert x\right\Vert _{p}^{p}$, we may assume that the dimension is $n=1$
as $x\mapsto \left\Vert x\right\Vert _{p}^{p}$ is fully separable:%
\begin{equation*}
f_{v}\left( x\right) =\lambda \frac{1}{p}\left\vert x\right\vert ^{p}+\frac{1%
}{2}\left( x-v\right) ^{2}
\end{equation*}%
The case $p=1$ is standard. When $p>1$ and $\lambda >0$, the derivative of $%
f_{v}\left( x\right) $ is:%
\begin{equation*}
f_{v}^{\prime }\left( x\right) =\lambda \limfunc{sign}\left( x\right)
\left\vert x\right\vert ^{p-1}+x-v
\end{equation*}%
Since $f_{v}^{\prime }\left( x\right) $ is an increasing function with
respect to $x$, we obtain a unique minimum. We deduce the following results:%
\begin{equation*}
\begin{tabular}{cc}
\hline
$f\left( x\right) $ & $\mathbf{prox}_{f}\left( v\right) $ \\ \hline
$\lambda \left\Vert x\right\Vert _{1}$ & $S_{\lambda }\left( v\right)
=\left( \left\vert v\right\vert -\lambda \mathbf{1}\right) \odot \limfunc{%
sign}\left( v\right) $ \\
$\lambda \dfrac{1}{p}\left\Vert x\right\Vert _{p}^{p}$ & $f_{\lambda
,p}^{-1}\left( v\right) $ \\ \hline
\end{tabular}%
\end{equation*}%
where $f_{\lambda ,p}:\mathbb{R}\rightarrow \mathbb{R}$ is the odd and
bijective function defined by:
\begin{equation*}
\forall x\geqslant 0\qquad f_{\lambda ,p}\left( x\right) =\lambda x^{p-1}+x
\end{equation*}%
Explicit computations can be carried out for $p\in \left\{ 2,3,4,5\right\} $%
. In particular, we have:%
\begin{equation*}
f_{\lambda ,2}^{-1}\left( v\right) =\frac{1}{1+\lambda }v\qquad \forall
\,v\in \mathbb{R}
\end{equation*}%
and:%
\begin{equation*}
f_{\lambda ,3}^{-1}\left( v\right) =\frac{1}{\lambda}\left(-\frac{1}{2}+\sqrt{\frac{1}{4}+\lambda v}\right)%
\qquad \forall \,v\geqslant 0
\end{equation*}%
Explicit formulas for cubic and quartic equations are known, so that
explicit expressions for $f_{\lambda ,4}^{-1}\left( v\right) $ and $%
f_{\lambda ,5}^{-1}\left( v\right) $ may be written\footnote{%
As the Galois group of $P\left( X\right) =X^{q}+X-c$ for $c\in \mathbb{Q}$ and
$q\geqslant 5$, may be not solvable, no explicit formula can be provided for
$f_{\lambda ,p}^{-1}\left( v\right) $ when $p\geqslant 6$. However, bisection
method can always be implemented to compute $f_{\lambda ,p}^{-1}\left( v\right)
$ for any $p>1$ and Newton algorithm for $p > 2$.}.\smallskip

\begin{figure}[tbph]
\centering
\caption{Proximal operator of $\dfrac{1}{p}\left\Vert x\right\Vert _{p}^{p}$}
\label{fig:prox2}
\figureskip
\includegraphics[width = \figurewidth, height = \figureheight]{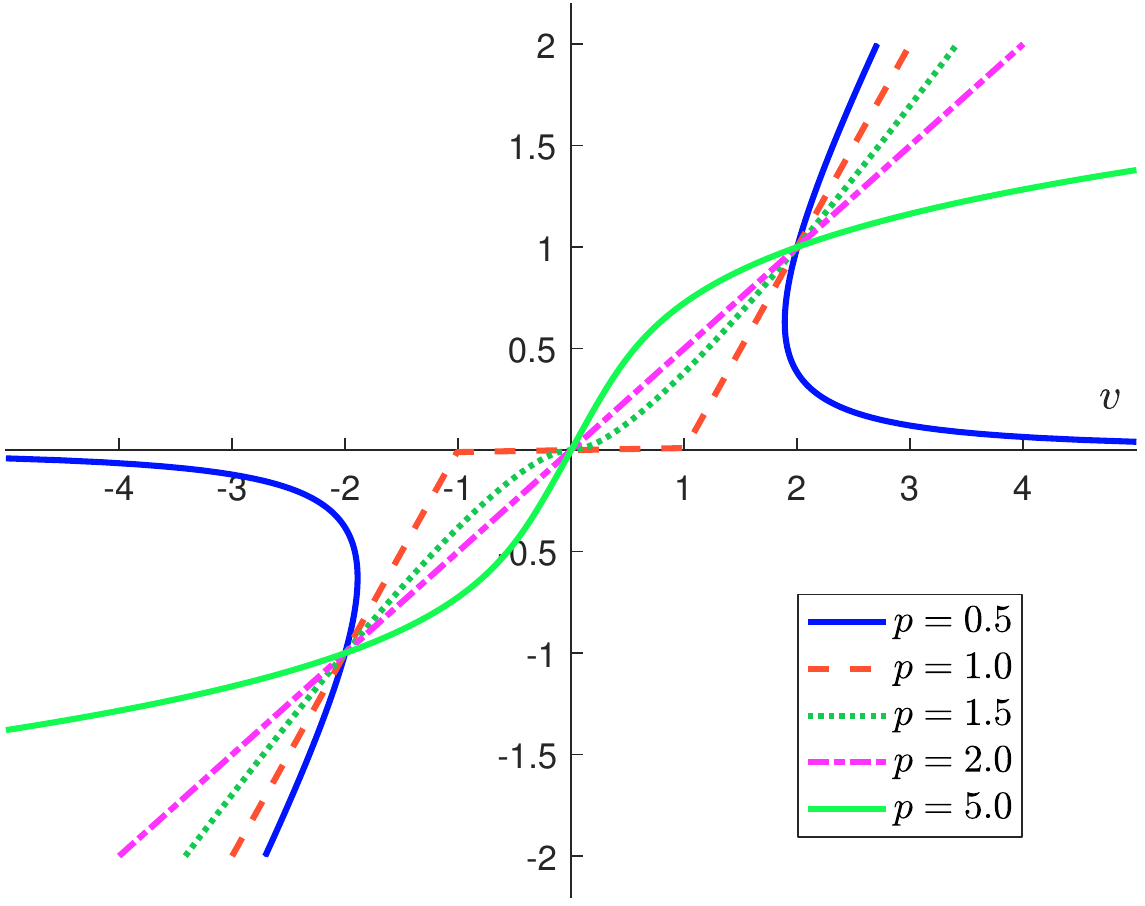}
\end{figure}

In Figure \ref{fig:prox2}, we have reported the proximal operator of $x\mapsto
\lambda \dfrac{1}{p}\left\Vert x\right\Vert _{p}^{p}$ in the one dimension for
several values of $p$ and $\lambda =1$. We verify that $f_{\lambda
,p}^{-1}\left( v\right) $ is an odd function. The proximal operator $S_{\lambda
}\left( v\right) =f_{\lambda ,1}^{-1}\left( v\right) $ is known as the soft
thresholding operator. The proximal map is not uniquely valued for the
non-convex case ($p<1$). The proximal operator for $p=2$ is a line with slope
$1/2$. We also notice that the convexity of the proximal operator is different
for $p<2$ and $p>2$ at $v=1$.

\subsubsection{The case $f\left( x\right) =\mathds{1}_{\Omega }\left(
x\right) $}

If we assume that $f\left( x\right) =\mathds{1}_{\Omega }\left( x\right) $
where $\Omega $ is a (convex) set, we have:%
\begin{eqnarray*}
\mathbf{prox}_{f}\left( v\right)  &=&\arg \min\nolimits_{x}\left\{ \mathds{1}%
_{\Omega }\left( x\right) +\frac{1}{2}\left\Vert x-v\right\Vert
_{2}^{2}\right\}  \\
&=&\mathcal{P}_{\Omega }\left( v\right)
\end{eqnarray*}%
where $\mathcal{P}_{\Omega }\left( v\right) $ is the standard projection. We
give here the results\footnote{%
See Parikh and Boyd (2014), and Beck (2017).} for some polyhedra that are
used in portfolio optimization:%
\begin{equation*}
\begin{tabular}{cc}
\hline
$\Omega $ & $\mathcal{P}_{\Omega }\left( v\right) $ \\ \hline
$A_2 x = b_2$ & $v-A_2^{\dagger }\left( A_2 v - b_2\right) $ \\
$a^{\top }x=b$ & $v-\dfrac{\left( a^{\top }v-b\right) }{\left\Vert
a\right\Vert _{2}^{2}}a$ \\
$a^{\top }x\leqslant b$ & $v-\dfrac{\left( a^{\top }v-b\right) _{+}}{%
\left\Vert a\right\Vert _{2}^{2}}a$ \\
$x^{-}\leqslant x\leqslant x^{+}$ & $v\odot \mathds{1}\left\{ x^{-}\leqslant
v\leqslant x^{+}\right\} +$ \\
& $x^{-}\odot \mathds{1}\left\{ v<x^{-}\right\} +x^{+}\odot \mathds{1}%
\left\{ v>x^{+}\right\} $ \\ \hline
\end{tabular}%
\end{equation*}

If $f$ is a norm, then $f^{\ast }\left( x\right) =\mathds{1}_{\mathcal{B}%
}\left( x\right) $ where $\mathcal{B}$ is the unit ball of the dual norm\footnote{%
The norms $L_{p}$ and $L_{q}$ are dual if and only if the exponents $\left\{
p,q\right\} \in \left[ 1,\infty \right) $ are H\"older conjugates ($%
p^{-1}+q^{-1}=1$).} of $f$. Thus, Moreau decomposition yields:%
\begin{equation*}
\mathbf{prox}_{\lambda f}\left( v\right) =v-\lambda \mathcal{P}_{\mathcal{B}%
}\left( \frac{1}{\lambda }v\right)
\end{equation*}%
meaning that we only use projections onto norm balls.\smallskip

A ball for the $L_{\infty }$ norm is a particular case of box constraint.
The orthogonal projection onto the unit ball for the $L_{2}$ norm is:%
\begin{equation*}
\mathcal{P}_{\mathcal{B}}\left( v\right) =\left\{
\begin{array}{ll}
\dfrac{v}{\left\Vert v\right\Vert _{2}} & \text{for }\left\Vert v\right\Vert
_{2}>1 \\
v & \text{for }\left\Vert v\right\Vert _{2}\leqslant 1%
\end{array}%
\right.
\end{equation*}%
The projection on the unit ball for the $L_{1}$ norm is less straightforward.
It is given by:
\begin{equation*}
\mathcal{P}_{\mathcal{B}}\left( v\right) =\mathrm{sign}\left( v\right) \odot
\left( \left\vert v\right\vert -\lambda \mathbf{1}\right)
\end{equation*}%
where $\lambda $ satisfies:
\begin{equation}
\left\Vert \left\vert v\right\vert -\lambda \mathbf{1}\right\Vert _{1}=1
\label{eq:appendix-prox3}
\end{equation}%
Equation (\ref{eq:appendix-prox3}) can be solved by the bi-section
algorithm\footnote{If the vector $v$ has ordered components, the
value of $\lambda $ is explicit.} or projected subgradient methods
(Duchi \textsl{et al.}, 2008).

\begin{remark}
Note also that the projection onto an $L_{1}$ ball and a simplex are equivalent
problems, applying twice the symmetry $x\mapsto -x$.
\end{remark}

Projections onto intersections of convex sets are examples in which the
computation of the proximal operator reduces to determining a zero of a
real-valued function. For instance, the projection onto the intersection of two
balls $\mathcal{B}_{p}\cap \mathcal{B}_{q}$ is a particular case of projection
onto a sublevel set that is defined by $\left\{ x:f\left( x\right) \leqslant
R\right\} $ where $f\left( x\right) =\left\Vert x\right\Vert
_{q}+\mathds{1}_{\mathcal{B}_{p}}\left( x\right) $. Indeed, we consider a
non-empty $L_{p}$ ball $\mathcal{B}_{p}$ and a non-empty $L_{q}$ ball
$\mathcal{B}_{q}$. The orthogonal projection $\mathcal{P}_{\Omega }$ onto the
intersection $\Omega =\mathcal{B}_{p}\cap \mathcal{B}_{q}$ is given
by:%
\begin{equation*}
\mathcal{P}_{\Omega }\left( v\right) =\left\{
\begin{array}{ll}
\mathcal{P}_{\mathcal{B}_{p}}\left( v\right)  & \text{if }\mathcal{P}_{%
\mathcal{B}_{p}}\left( v\right) \in \mathcal{B}_{q} \\
\mathbf{prox}_{f}\left( v\right)  & \text{if }\mathcal{P}_{\mathcal{B}%
_{p}}\left( v\right) \notin \mathcal{B}_{q}%
\end{array}%
\right.
\end{equation*}%
where $f\left( x\right) =\lambda ^{\star }\left\Vert x\right\Vert _{p}$ and
$\lambda ^{\star }$ is a scalar such that $\mathbf{prox}_{f}\left( v\right) \in
\partial \mathcal{B}_{q}$ where $\partial \mathcal{B}_{q}$ is the boundary of
$\mathcal{B}_{q}$.\smallskip

We now consider the projection of $v$ on the intersection of a convex set
$\Omega $ and a hyperplane $\mathcal{H=}\left\{ x\in \mathbb{R}^{n},a\in
\mathbb{R}^{n}\setminus \left\{ \mathbf{0}\right\} \mid a^{\top }x=b\right\} $.
We have:
\begin{eqnarray*}
x^{\star } &=&\mathcal{P}_{\mathcal{H}\cap \Omega }\left( v\right)  \\
&=&\arg \min_{x\in \mathcal{H}\cap \Omega }\frac{1}{2}\left\Vert
x-v\right\Vert _{2}^{2}
\end{eqnarray*}%
Leaving the constraint $x\in \Omega $ implicit, we can write the partial
Lagrange function for this problem:%
\begin{eqnarray}
\mathcal{L}\left( x,\lambda \right)  &=&\frac{1}{2}\left\Vert x-v\right\Vert
_{2}^{2}+\lambda \left( a^{\top }x-b\right)   \notag \\
&=&\frac{1}{2}\left\Vert x-\left( v-\lambda a\right) \right\Vert
_{2}^{2}+\lambda \left( a^{\top }v-b\right) -\frac{1}{2}\lambda
^{2}\left\Vert a\right\Vert _{2}^{2}  \label{eq:appendix-prox2}
\end{eqnarray}%
As strong duality holds, $x^{\star }$ is the optimal solution if, and only if,
there exists a scalar $\lambda ^{\star }\in \mathbb{R}$ satisfying:
\begin{equation*}
x^{\star }\in \arg \min_{x\in \Omega }\mathcal{L}\left( x,\lambda ^{\star
}\right) \qquad \text{and}\qquad x^{\star }\in \mathcal{H}
\end{equation*}%
Using Equation (\ref{eq:appendix-prox2}), we obtain:%
\begin{equation*}
x^{\star }=\mathcal{P}_{\Omega }\left( v-\lambda ^{\star }a\right) \qquad
\text{and}\qquad x^{\star }\in \mathcal{H}
\end{equation*}%
where $\lambda ^{\star }$ is the solution to the equation:
\begin{equation*}
a^{\top }\mathcal{P}_{\Omega }\left( v-\lambda ^{\star }a\right) =b
\end{equation*}%
Particular cases of the last formula are projections onto the standard simplex
$\Omega =\mathbb{R}_{+}^{n}$, the intersection of two non-empty
balls $\Omega =\mathcal{B}_{p}\cap \mathcal{B}_{q}$ and the hyperplane $%
\Omega =\left\{ x\in \mathbb{R}^{n}\mid \mathbf{1}^{\top }x=0\right\} $.%
\label{appendix:section-proximal-end}

\subsection{Derivation of the PRESS statistic for the Tikhonov regularization}
\label{appendix:press}

We have:%
\begin{equation*}
X^{\top }X=X_{-t}^{\top }X_{-t}+x_{t}x_{t}^{\top }
\end{equation*}%
and:%
\begin{equation*}
X^{\top }Y=X_{-t}^{\top }Y_{-t}+x_{t}y_{t}
\end{equation*}%
The Sherman-Morrison-Woodbury formula\footnote{Suppose $u$ and $v$ are two
vectors and $A$ is an invertible square matrix. It follows that:
\begin{equation*}
\left( A+uv^{\top }\right) ^{-1}=A^{-1}-\frac{1}{1+v^{\top }A^{-1}u}%
A^{-1}uv^{\top }A^{-1}
\end{equation*}%
} leads to:
\begin{eqnarray*}
\hat{\beta}_{-t} &=&\left( X_{-t}^{\top }X_{-t}+\varrho _{2}\Gamma _{2}\Gamma
_{2}^{\top }\right) ^{-1}X_{-t}^{\top }Y_{-t} \\
&=&\left( X^{\top }X+\varrho _{2}\Gamma _{2}\Gamma _{2}^{\top
}-x_{t}x_{t}^{\top }\right) ^{-1}\left( X^{\top }Y-x_{t}y_{t}\right)  \\
&=&\left( S\left( \varrho _{2}\right) ^{-1}-x_{t}x_{t}^{\top }\right)
^{-1}\left( X^{\top }Y-x_{t}y_{t}\right)  \\
&=&\left( S\left( \varrho _{2}\right) +\frac{S\left( \varrho _{2}\right)
x_{t}x_{t}^{\top }S\left( \varrho _{2}\right) }{1-x_{t}^{\top }S\left(
\varrho _{2}\right) x_{t}}\right) \left( X^{\top }Y-x_{t}y_{t}\right)  \\
&=&S\left( \varrho _{2}\right) X^{\top }Y-S\left( \varrho _{2}\right)
x_{t}y_{t}+ \\
&&\frac{S\left( \varrho _{2}\right) x_{t}x_{t}^{\top }S\left( \varrho
_{2}\right) }{1-x_{t}^{\top }S\left( \varrho _{2}\right) x_{t}}X^{\top }Y-%
\frac{S\left( \varrho _{2}\right) x_{t}x_{t}^{\top }S\left( \varrho
_{2}\right) }{1-x_{t}^{\top }S\left( \varrho _{2}\right) x_{t}}x_{t}y_{t}
\end{eqnarray*}%
We denote $z_{t}=x_{t}^{\top }S\left( \varrho _{2}\right) x_{t}$. Since $%
\hat{\beta}=S\left( \varrho _{2}\right) X^{\top }Y$, we get:%
\begin{equation*}
x_{t}^{\top }\hat{\beta}_{-t}=x_{t}^{\top }\hat{\beta}-z_{t}y_{t}+\frac{z_{t}%
}{1-z_{t}}x_{t}^{\top }\hat{\beta}-\frac{z_{t}^{2}}{1-z_{t}}y_{t}
\end{equation*}
Finally, we obtain:
\begin{eqnarray*}
y_{t}-x_{t}^{\top }\hat{\beta}_{-t} &=&y_{t}\left( 1+z_{t}+\frac{z_{t}^{2}}{%
1-z_{t}}\right) -x_{t}^{\top }\hat{\beta}\left( 1+\frac{z_{t}}{1-z_{t}}%
\right)  \\
&=&y_{t}\left( \frac{1}{1-z_{t}}\right) -x_{t}^{\top }\hat{\beta}\left(
\frac{1}{1-z_{t}}\right)  \\
&=&\frac{1}{1-x_{t}^{\top }S\left( \varrho _{2}\right) x_{t}}%
(y_{t}-x_{t}^{\top }\hat{\beta})
\end{eqnarray*}%
It follows that the PRESS statistic is equal to:
\begin{eqnarray*}
\mathcal{P}\mathrm{ress}\left( \varrho _{2}\right)  &=&\sum_{t=1}^{T}\left(
y_{t}-x_{t}^{\top }\hat{\beta}_{-t}\right) ^{2} \\
&=&\sum_{t=1}^{T}\frac{\left( y_{t}-x_{t}^{\top }\hat{\beta}\right) ^{2}}{%
\left( 1-x_{t}^{\top }S\left( \varrho _{2}\right) x_{t}\right) ^{2}}
\end{eqnarray*}

\section{The Black-Litterman model}
\label{appendix:section-bl}

\subsection{Computing the implied risk premia}

Let us consider the following optimization problem:
\begin{eqnarray*}
x^{\star }\left( \gamma \right)  &=&\arg \min \frac{1}{2}x^{\top }\Sigma
x-\gamma x^{\top }\left( \mu -r\mathbf{1}\right)  \\
&\text{s.t.}&\mathbf{1}^{\top }x=1
\end{eqnarray*}%
The unscaled solution is:
\begin{equation*}
x^{\star }=\gamma \Sigma ^{-1}\left( \mu -r\mathbf{1}\right)
\end{equation*}%
Given an initial allocation $x_{0}$, we deduce that this portfolio is
optimal if the vector of expected returns is defined by:
\begin{equation*}
\tilde{\mu}=r+\frac{1}{\gamma }\Sigma x_{0}
\end{equation*}%
By assuming that we know the Sharpe ratio of the initial allocation, we
deduce that:
\begin{equation}
\tilde{\mu}=r+\limfunc{SR}\left( x_{0}\mid r\right) \frac{\Sigma x_{0}}{%
\sqrt{x_{0}^{\top }\Sigma x_{0}}}  \label{eq:appendix-bl1}
\end{equation}%
We retrieve one of the fundamental results from the capital asset pricing
model. At the optimum, risk premia are proportional to marginal risks
(Roncalli, 2013).

\subsection{Conditional distribution of expected returns}

Black and Litterman (1992) state that vector $R_{t}$ of asset returns follow
a Gaussian distribution:
\begin{equation*}
R_{t}\sim \mathcal{N}\left( \tilde{\mu},\Sigma _{m}\right)
\end{equation*}%
where $\tilde{\mu}$ is the implied expected return associated with the
allocation $x_{0}$ and $\Sigma _{m}$ is the market covariance matrix of
asset returns. To specify the portfolio manager's views, they assume that
they are given by this relationship:%
\begin{equation}
PR_{t}=Q+\varepsilon   \label{eq:appendix-bl2}
\end{equation}%
where $P$ is a $\left( k\times n\right) $ matrix, $Q$ is a $\left( k\times
1\right) $ vector and $\varepsilon \sim \mathcal{N}\left( 0,\Sigma
_{\varepsilon }\right) $ is a Gaussian vector of dimension $k$. The $k$
views of the portfolio manager can be expressed in absolute or relative
terms. It follows that the joint distribution of the expected returns $R_{t}$
and the views $\nu _{t}=PR_{t}-\varepsilon $ is given by the following
relationship:
\begin{equation*}
\left(
\begin{array}{c}
R_{t} \\
\nu _{t}%
\end{array}%
\right) \sim \mathcal{N}\left( \left(
\begin{array}{c}
\tilde{\mu} \\
P\tilde{\mu}%
\end{array}%
\right) ,\left(
\begin{array}{cc}
\Sigma _{m} & \Sigma _{m}P^{\top } \\
P\Sigma _{m} & P\Sigma _{m}P^{\top }+\Sigma _{\varepsilon }%
\end{array}%
\right) \right)
\end{equation*}%
By applying the conditional expectation formula\footnote{%
See Appendix \ref{appendix:section-conditional-expectation} on page
\pageref{appendix:section-conditional-expectation}.}, we obtain:
\begin{eqnarray}
\bar{\mu} &=&\mathbb{E}\left[ R_{t}\mid \nu _{t}=Q\right]   \notag \\
&=&\tilde{\mu}+\Sigma _{m}P^{\top }\left( P\Sigma _{m}P^{\top }+\Sigma
_{\varepsilon }\right) ^{-1}\left( Q-P\tilde{\mu}\right)   \notag
\end{eqnarray}%
and:
\begin{eqnarray}
\bar{\Sigma} &=&\mathbb{E}\left[ \left( R_{t}-\bar{\mu}\right) \left( R_{t}-%
\bar{\mu}\right) ^{\top }\mid \nu _{t}=Q\right]   \notag \\
&=&\Sigma _{m}-\Sigma _{m}P^{\top }\left( P\Sigma _{m}P^{\top }+\Sigma
_{\varepsilon }\right) ^{-1}P\Sigma _{m}  \notag
\end{eqnarray}%
The vector of conditional expected returns $\bar{\mu}$ has two components:
\begin{enumerate}
\item The first component corresponds to the vector of implied expected
returns $\tilde{\mu}$.

\item The second component is a correction term which takes into account the
\textit{disequilibrium} $\left( Q-P\tilde{\mu}\right) $ between the manager's
views and the market's views.
\end{enumerate}
In the same way, the conditional covariance matrix has two
components. Indeed, we have\footnote{%
Let $A$, $B$ and $C$ three compatible matrices. We have:
\begin{equation*}
AB^{\top }\left( BAB^{\top }\right) ^{-1}B=I-\left( I+AB^{\top
}C^{-1}B\right) ^{-1}
\end{equation*}%
}:
\begin{eqnarray}
\bar{\Sigma} &=&\left( I_{n}+\Sigma _{m}P^{\top }\Sigma _{\varepsilon
}^{-1}P\right) ^{-1}\Sigma _{m}  \notag \\
&=&\left( \Sigma _{m}^{-1}+P^{\top }\Sigma _{\varepsilon }^{-1}P\right) ^{-1}
\label{eq:appendix-bl3}
\end{eqnarray}%
Again, the conditional covariance matrix is a weighted average of the market
covariance matrix $\Sigma _{m}$ and the covariance matrix $\Sigma
_{\varepsilon }$ of the manager views.

\subsection{The case of absolute views}

If the portfolio manager specifies absolute views, it is equivalent
imposing $P=I_{n}$ and $Q=\breve{\mu}$. We deduce that:
\begin{equation*}
\bar{\mu}=\left( I_{n}-\Sigma _{m}\left( \Sigma _{m}+\Sigma _{\varepsilon
}\right) ^{-1}\right) \tilde{\mu}+\Sigma _{m}\left( \Sigma _{m}+\Sigma
_{\varepsilon }\right) ^{-1}\breve{\mu}
\end{equation*}%
and\footnote{%
We remind that:
\begin{equation*}
A^{-1}+B^{-1}=B^{-1}\left( A+B\right) A^{-1}
\end{equation*}%
}:
\begin{equation*}
\bar{\Sigma}=\Sigma _{m}\left( \Sigma _{m}+\Sigma _{\varepsilon }\right)
^{-1}\Sigma _{\varepsilon }
\end{equation*}%
If we consider the (unscaled) optimal portfolio $\bar{x}$, we obtain:
\begin{eqnarray*}
\bar{x} &=&\gamma \bar{\Sigma}^{-1}\bar{\mu} \\
&=&\gamma \Sigma _{\varepsilon }^{-1}\left( \Sigma _{m}+\Sigma _{\varepsilon
}\right) \Sigma _{m}^{-1}\left( \left( I_{n}-\Sigma _{m}\left( \Sigma
_{m}+\Sigma _{\varepsilon }\right) ^{-1}\right) \tilde{\mu}+\Sigma
_{m}\left( \Sigma _{m}+\Sigma _{\varepsilon }\right) ^{-1}\breve{\mu}\right)
\\
&=&\Sigma _{m}^{-1}\Sigma \tilde{x}+\breve{x}
\end{eqnarray*}%
where $\breve{x}$ is the mean-variance optimized portfolio based on the
manager's views. In particular, if $\Sigma _{m}=\Sigma $, it follows that
the optimal portfolio $\bar{x}$ is simply the sum of the SAA portfolio $\tilde{%
x}$ and the MVO portfolio $\breve{x}$.\smallskip

Let $\hat{\Sigma}$ be the empirical covariance matrix. If we assume that $%
\Sigma _{m}=\tau \hat{\Sigma}$ and $\Sigma _{\varepsilon }=\tau \hat{\Sigma}$%
, we obtain:
\begin{equation*}
\bar{\mu}=\frac{\tilde{\mu}+\breve{\mu}}{2}
\end{equation*}%
and:
\begin{equation*}
\bar{\Sigma}=\frac{\tau }{2}\hat{\Sigma}
\end{equation*}%
The conditional expected returns are therefore an average between the implied
expected returns and the manager's views, whereas the conditional covariance
matrix is proportional to the empirical covariance matrix. In particular, if
$\tau $ is set to $1$, asset volatilities are divided by $\sqrt{2}$. This
type of parametrization is a real problem, because it dramatically reduces
the covariance matrix of asset returns.\smallskip

\begin{figure}[tbph]
\caption{Variance reduction in the Black-Litterman model}
\label{fig:bl1}
\centering
\figureskip
\includegraphics[width = \figurewidth, height = \figureheight]{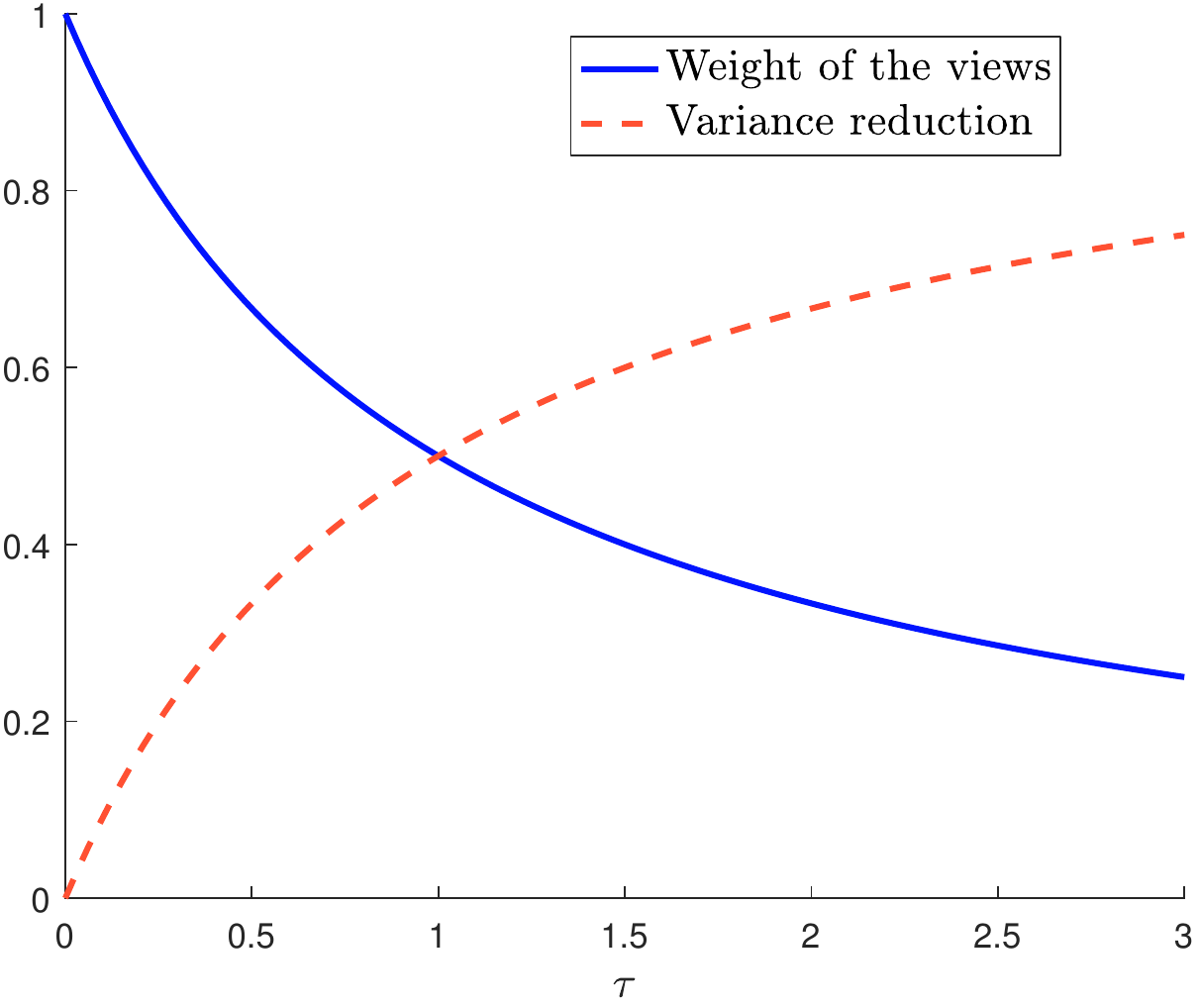}
\end{figure}

We now consider a second approach with $\Sigma _{m}=\hat{\Sigma}$ and $%
\Sigma _{\varepsilon }=\tau \hat{\Sigma}$. It follows that:
\begin{equation}
\bar{\mu}=\frac{\tau }{1+\tau }\tilde{\mu}+\frac{1}{1+\tau }\breve{\mu}
\label{eq:appendix-bl4}
\end{equation}%
and:
\begin{equation*}
\bar{\Sigma}=\frac{\tau }{1+\tau }\hat{\Sigma}
\end{equation*}%
When $\tau \rightarrow 0$, we verify that that the conditional expectation
tends toward the manager's views. However, the covariance matrix also tends towards
the null matrix (see Figure \ref{fig:bl1}). Again, we notice an arbitrage
between the weight of the manager's views and the reduction of the
covariance matrix.\smallskip

In practice, we would like to control the contribution of the manager's views
without modifying necessarily the covariance matrix of asset returns. This is
why we can impose that $\bar{\Sigma}=\hat{\Sigma}$.

\clearpage
\section{Additional results}

\subsection{Tables}
\label{appendix:section-tables}

\begin{table}[tbph]
\centering
\caption{Quality representation of each asset}
\label{tab:mvo2-3}
\tableskip
\begin{tabular}{cc:rrrr}
\hline
\multicolumn{2}{c:}{Factor} & \multicolumn{1}{c}{1} &
\multicolumn{1}{c}{2} & \multicolumn{1}{c}{3} & \multicolumn{1}{c}{4} \\ \hline
\multirow{4}{*}{Asset}
& $\quad 1 \quad$ & $58.35\%$ & $ 0.08\%$ & $ 0.24\%$ & $41.33\%$ \\
& $2$             & $55.18\%$ & $ 5.90\%$ & $38.46\%$ & $ 0.46\%$ \\
& $3$             & $50.25\%$ & $39.36\%$ & $ 9.07\%$ & $ 1.32\%$ \\
& $4$             & $78.91\%$ & $18.87\%$ & $ 0.99\%$ & $ 1.23\%$ \\ \hline
\end{tabular}
\medskip

\centering
\caption{Contribution of each asset}
\label{tab:mvo2-4}
\tableskip
\begin{tabular}{cc:rrrr}
\hline
\multicolumn{2}{c:}{Factor} & \multicolumn{1}{c}{1} &
\multicolumn{1}{c}{2} & \multicolumn{1}{c}{3} & \multicolumn{1}{c}{4} \\ \hline
\multirow{4}{*}{Asset}
& $\quad 1 \quad$ & $13.07\%$ & $ 0.06\%$ & $ 0.33\%$ & $86.54\%$ \\
& $2$             & $17.80\%$ & $ 6.49\%$ & $74.32\%$ & $ 1.38\%$ \\
& $3$             & $20.02\%$ & $53.43\%$ & $21.64\%$ & $ 4.91\%$ \\
& $4$             & $49.11\%$ & $40.02\%$ & $ 3.71\%$ & $ 7.16\%$ \\ \hline
\end{tabular}
\medskip

\centering
\caption{Linear dependence between the four assets ($\mu_1 = 3\%$)}
\label{tab:hedging3a}
\tableskip
\begin{tabular}{c:c:cccc:c} \hline
Asset  & $\alpha_{i}$  & \multicolumn{4}{c:}{$\beta_i$}        & $\mathfrak{R}^2_i$   \\ \hline
1 &         $-2.30\%$ & $     $ & $0.139$ & $0.187$ & $0.250$ & $45.83\%$ \\
2 & ${\TsVIII}2.98\%$ & $0.230$ & $     $ & $0.268$ & $0.191$ & $37.77\%$ \\
3 & ${\TsVIII}4.49\%$ & $0.409$ & $0.354$ & $     $ & $0.045$ & $33.52\%$ \\
4 & ${\TsVIII}4.41\%$ & $0.750$ & $0.347$ & $0.063$ & $     $ & $41.50\%$ \\ \hline
\end{tabular}
\medskip

\centering
\caption{Risk/return analysis of hedging portfolios ($\mu_1 = 3\%$)}
\label{tab:hedging3b}
\tableskip
\begin{tabular}{c:ccc:ccc:c} \hline
Asset  & $\mu_{i}$ & $\hat{\mu}_i$ & $\alpha_i$
       & $\sigma_{i}$ & $\hat{\sigma}_i$ & $s_i$ & $\mathfrak{R}^2_{i}$   \\ \hline
1      & ${\TsV}3.00\%$ & $5.30\%$ &         $-2.30\%$ & $15.00\%$ & $10.16\%$ & $11.04\%$ & $45.83\%$ \\
2      & ${\TsV}8.00\%$ & $5.02\%$ & ${\TsVIII}2.98\%$ & $18.00\%$ & $11.06\%$ & $14.20\%$ & $37.77\%$ \\
3      & ${\TsV}9.00\%$ & $4.51\%$ & ${\TsVIII}4.49\%$ & $20.00\%$ & $11.58\%$ & $16.31\%$ & $33.52\%$ \\
4      &      $10.00\%$ & $5.59\%$ & ${\TsVIII}4.41\%$ & $25.00\%$ & $16.11\%$ & $19.12\%$ & $41.50\%$ \\ \hline
\end{tabular}
\medskip

\centering
\caption{Optimal portfolio ($\mu_1 = 3\%$)}
\label{tab:hedging3c}
\tableskip
\begin{tabular}{c:cccc} \hline
Asset  & $\omega_{i}$ & $y^{\star}_i$ & $z^{\star}_i$ & $x^{\star}_i$  \\ \hline
1      & $84.62\%$ & $53.59\%$ &      $206.52\%$ &         $-75.81\%$ \\
2      & $60.68\%$ & $99.25\%$ &      $164.80\%$ & ${\TsVIII}59.46\%$ \\
3      & $50.43\%$ & $90.44\%$ &      $135.19\%$ & ${\TsVIII}67.87\%$ \\
4      & $70.94\%$ & $64.31\%$ & ${\TsV}86.63\%$ & ${\TsVIII}48.48\%$ \\ \hline
\end{tabular}
\end{table}

\subsection{Figures}
\label{appendix:section-figures}

\clearpage

\begin{figure}[tbph]
\centering
\caption{Mixed regularization with a target portfolio ($x_1^{\star}$)}
\label{fig:mixed1-1}
\figureskip
\includegraphics[width = \figurewidth, height = \figureheight]{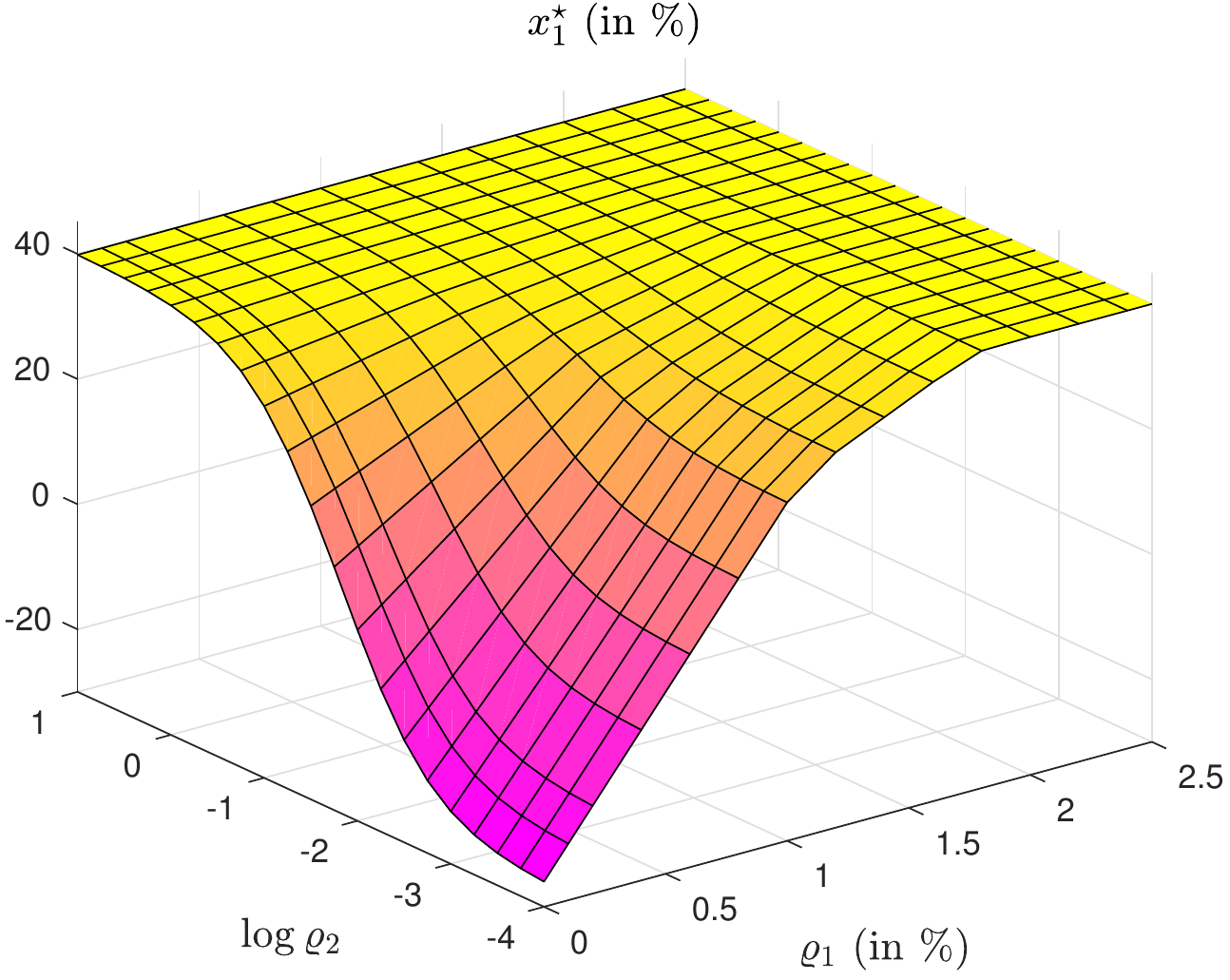}
\vspace*{1.00cm}

\centering
\caption{Mixed regularization with a target portfolio ($x_2^{\star}$)}
\label{fig:mixed1-2}
\figureskip
\includegraphics[width = \figurewidth, height = \figureheight]{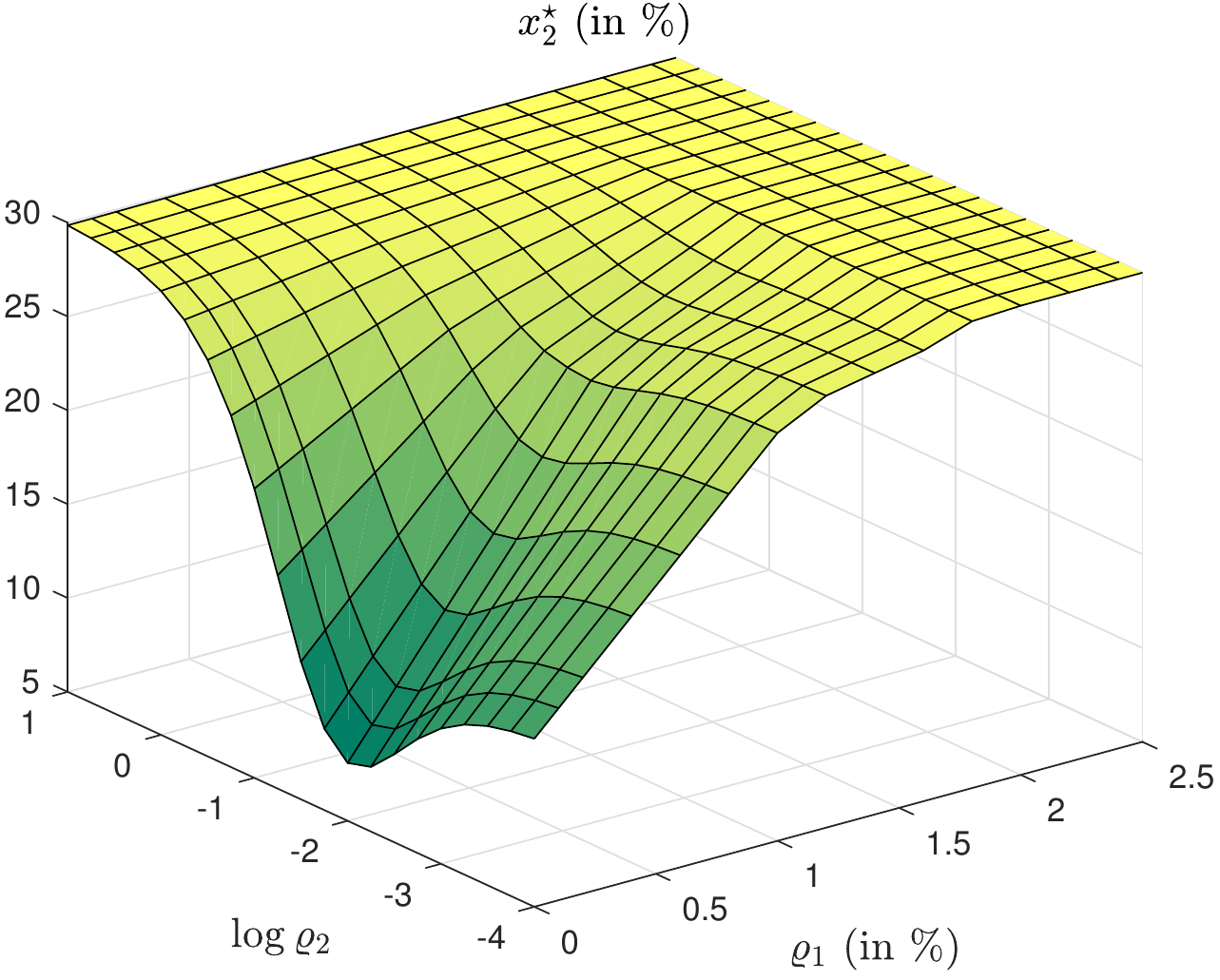}
\end{figure}

\begin{figure}[tbph]
\centering
\caption{Mixed regularization with a target portfolio ($x_3^{\star}$)}
\label{fig:mixed1-3}
\figureskip
\includegraphics[width = \figurewidth, height = \figureheight]{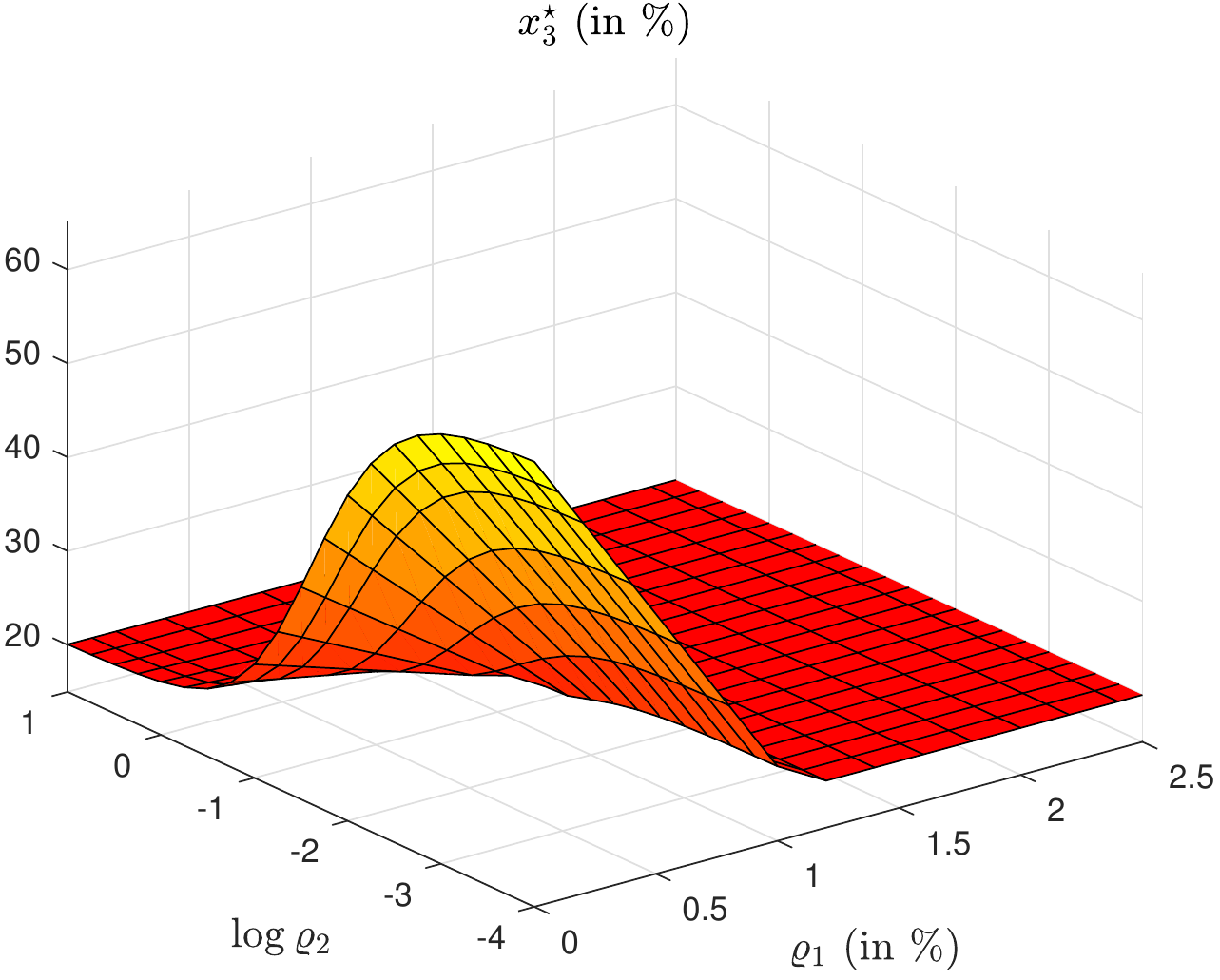}
\vspace*{1.00cm}

\centering
\caption{Mixed regularization with a target portfolio ($x_4^{\star}$)}
\label{fig:mixed1-4}
\figureskip
\includegraphics[width = \figurewidth, height = \figureheight]{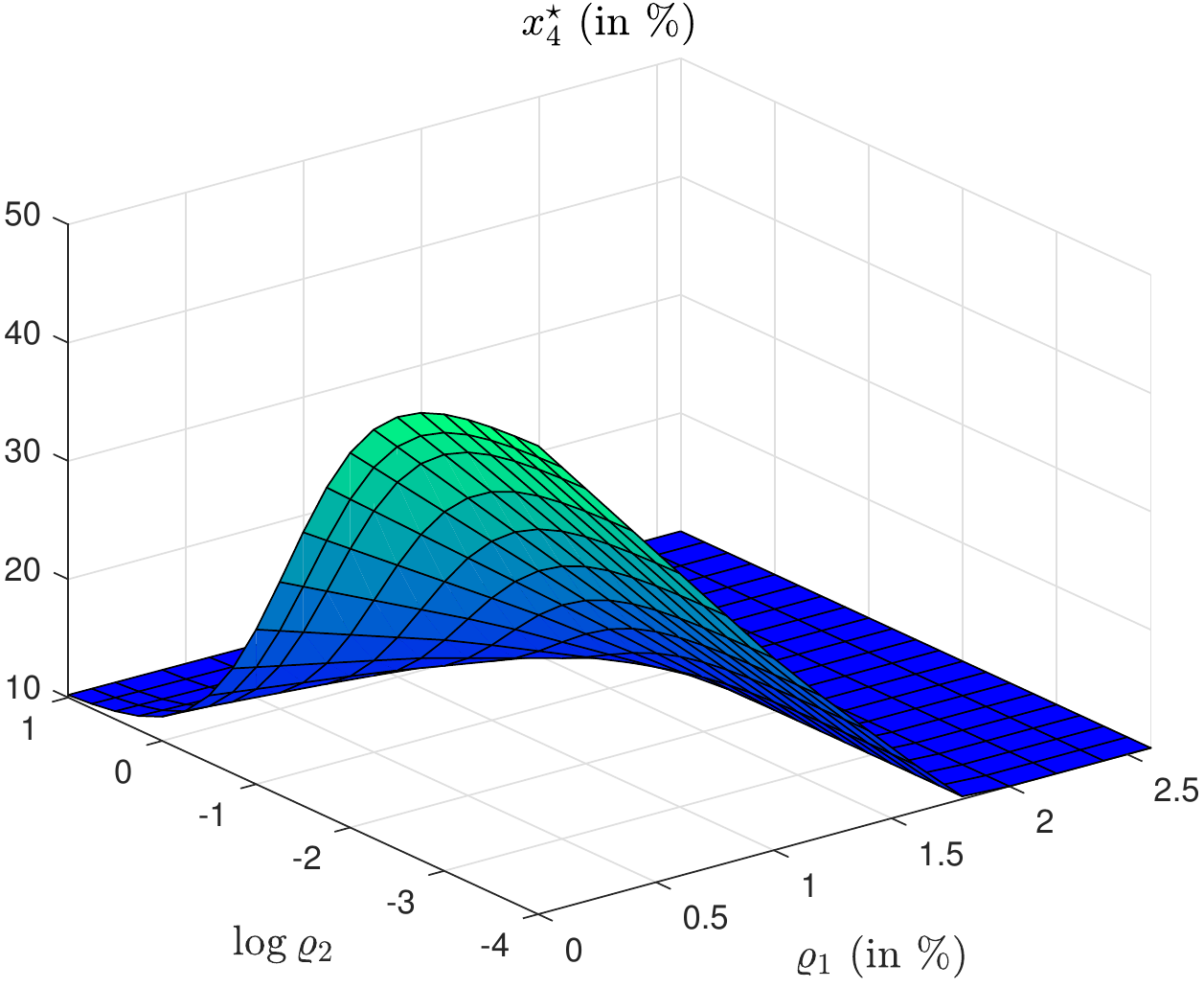}
\end{figure}

\begin{figure}[tbph]
\centering
\caption{Mixed regularization without a target portfolio ($x_1^{\star}$)}
\label{fig:mixed2-1}
\figureskip
\includegraphics[width = \figurewidth, height = \figureheight]{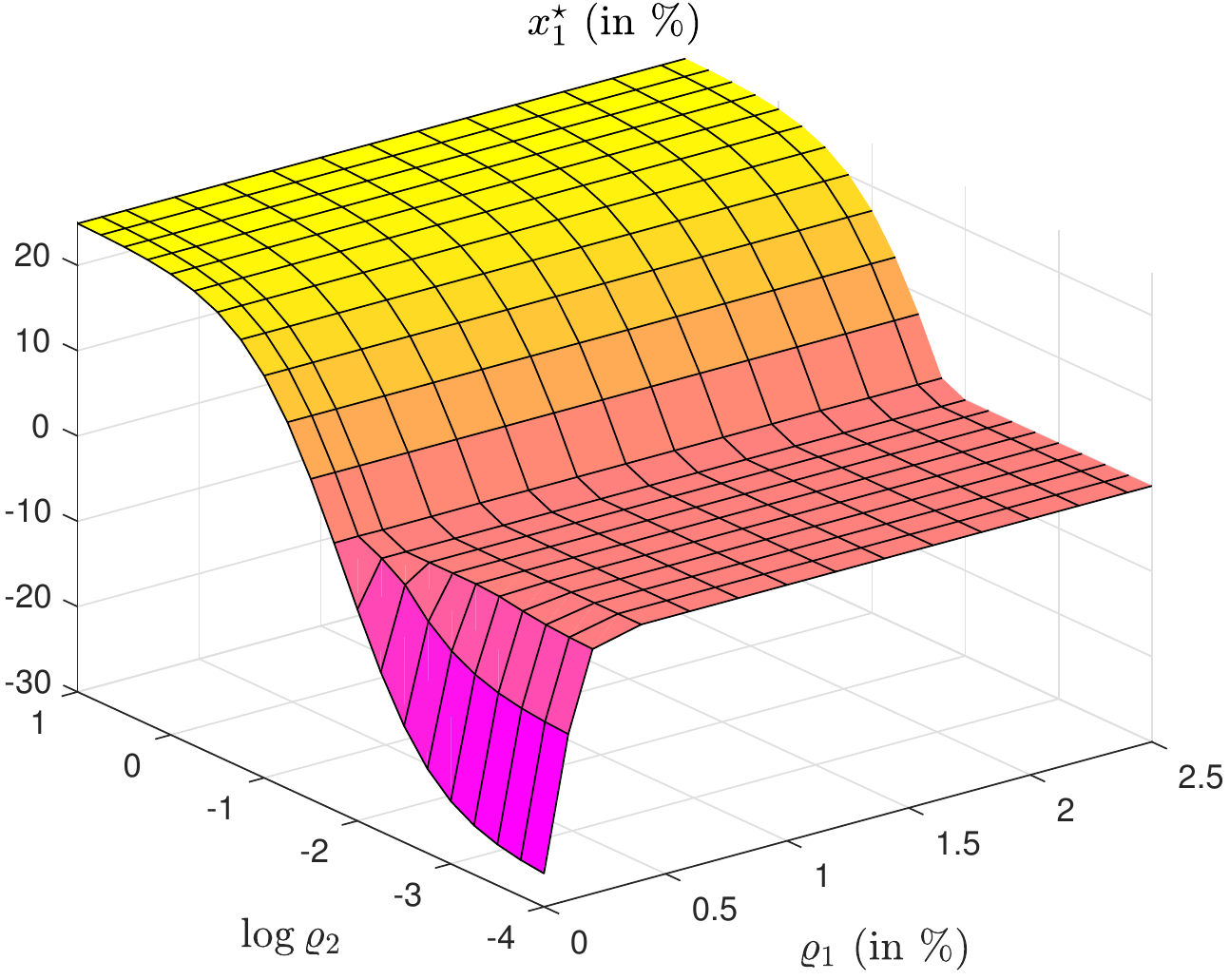}
\vspace*{1.00cm}

\centering
\caption{Mixed regularization without a target portfolio ($x_2^{\star}$)}
\label{fig:mixed2-2}
\figureskip
\includegraphics[width = \figurewidth, height = \figureheight]{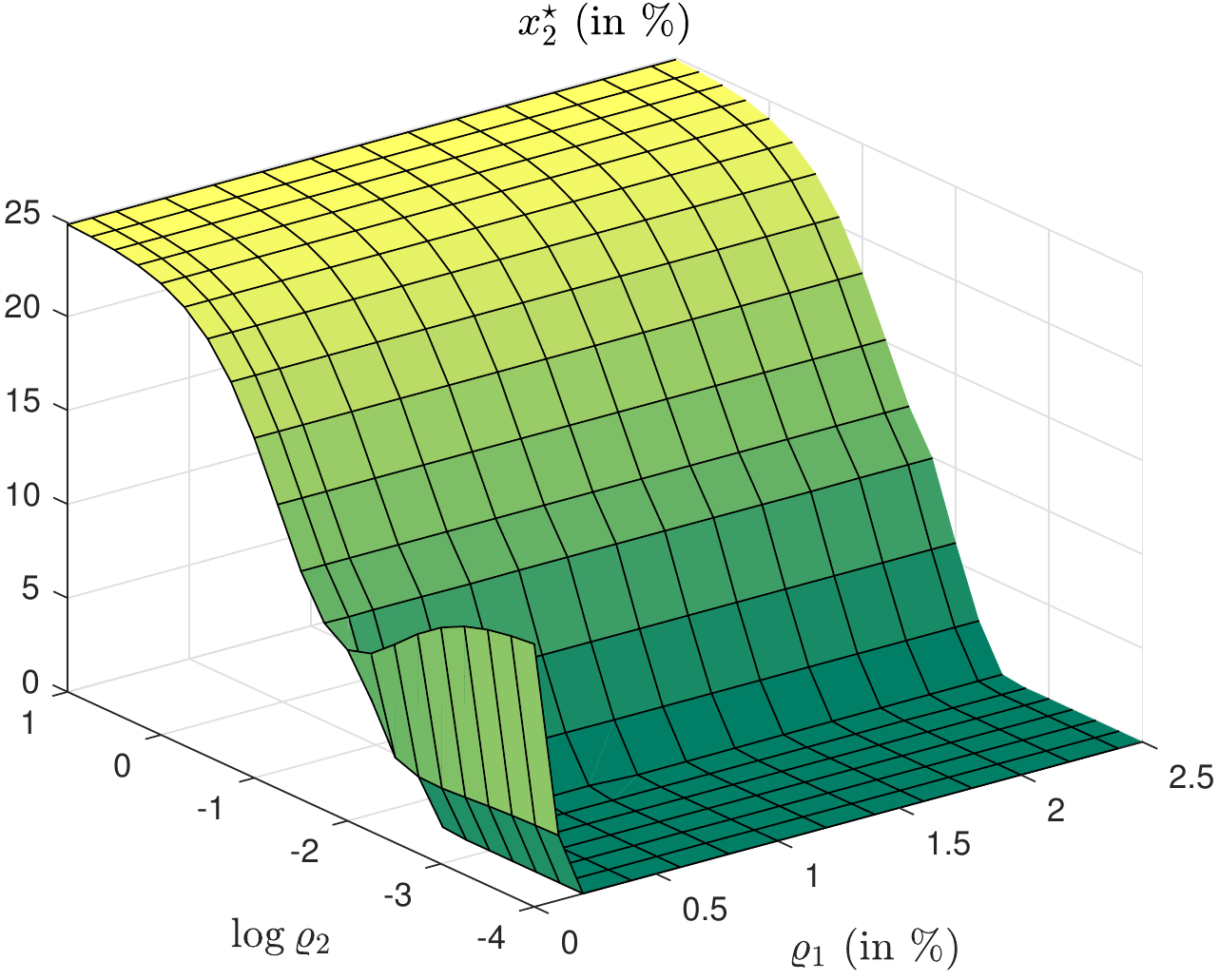}
\end{figure}

\begin{figure}[tbph]
\centering
\caption{Mixed regularization without a target portfolio ($x_3^{\star}$)}
\label{fig:mixed2-3}
\figureskip
\includegraphics[width = \figurewidth, height = \figureheight]{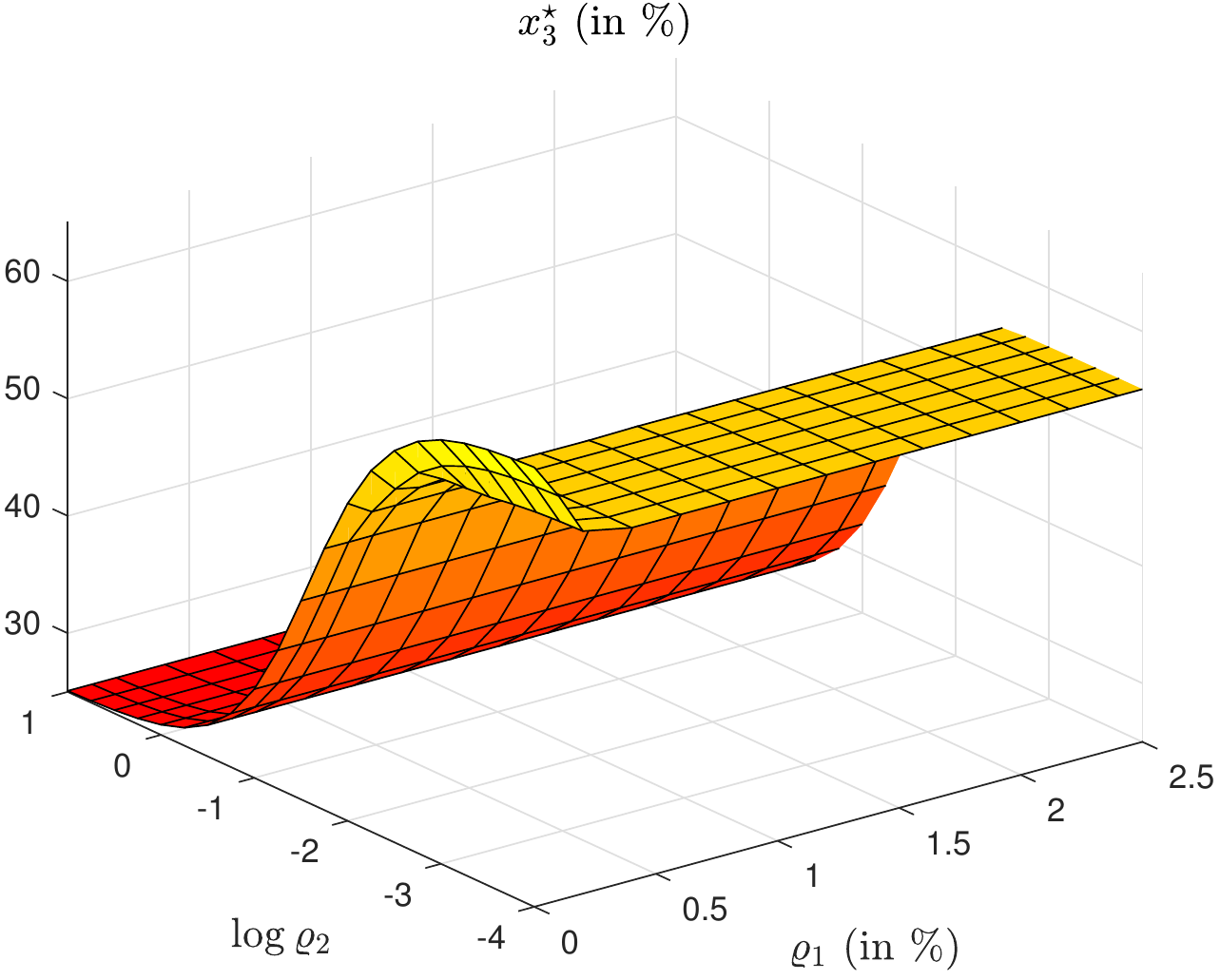}
\vspace*{1.00cm}

\centering
\caption{Mixed regularization without a target portfolio ($x_4^{\star}$)}
\label{fig:mixed2-4}
\figureskip
\includegraphics[width = \figurewidth, height = \figureheight]{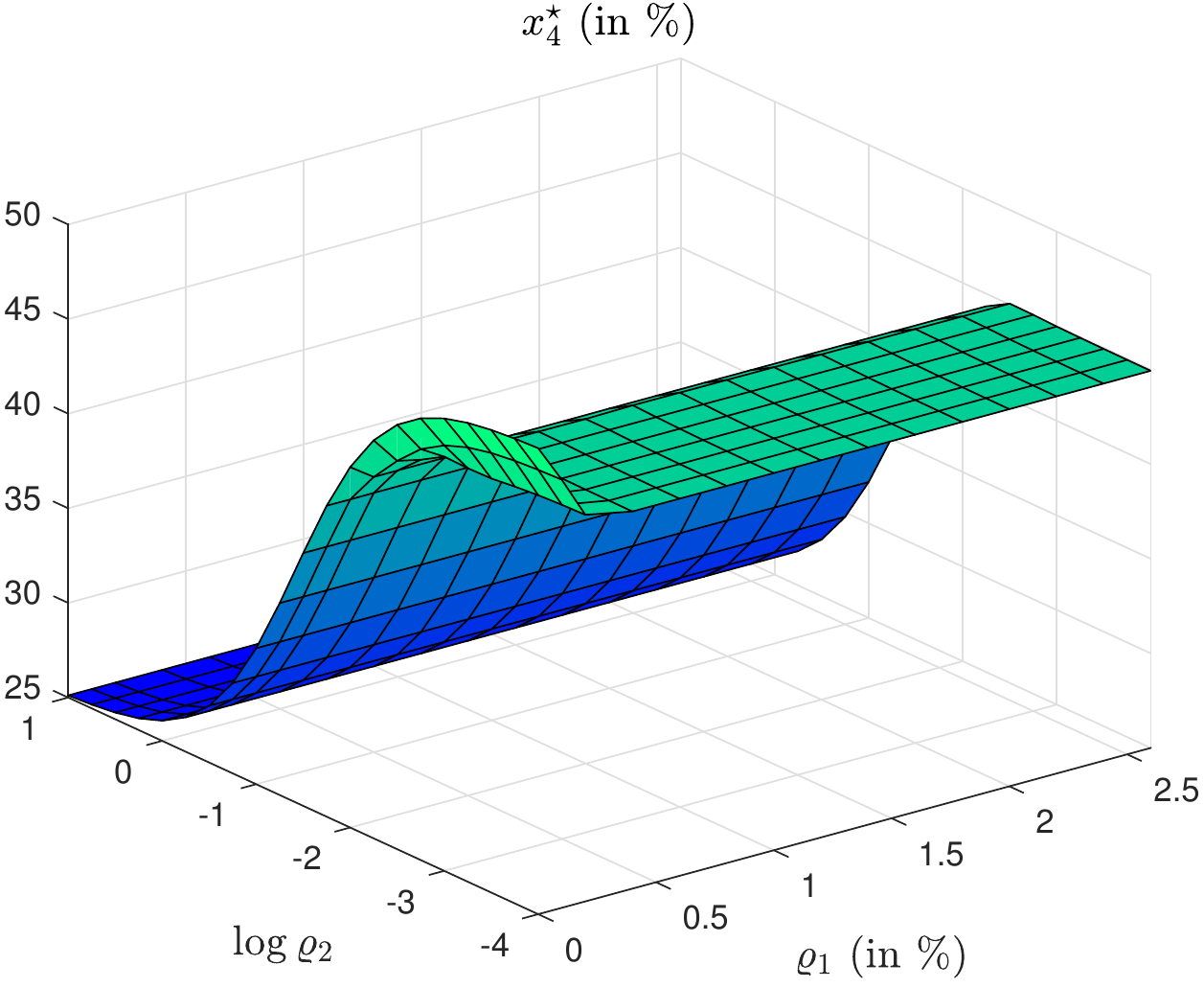}
\end{figure}

\end{document}